\documentclass[a4paper,fleqn,usenatbib]{mnras}
\usepackage[T1]{fontenc}
\usepackage{aecompl}
\usepackage{graphicx}   
\usepackage{amsmath}    
\usepackage{amssymb}    
\usepackage{pdflscape}
\usepackage{subfigure}
\usepackage{booktabs}
\usepackage{longtable}
\usepackage{natbib}
\usepackage{pdflscape}	
\usepackage{multicol}        

\title[NGC~1893]{NGC~1893: a young open cluster rich in multi-type variable stars}

\author[H. F. Xue et al.]{
Hui-Fang Xue $^{1}$,
Jian-Ning Fu $^{1}$\thanks{Corresponding author, e-mail: jnfu@bnu.edu.cn},
Nami Mowlavi $^{1,2}$,
Sophie Saesen $^{2}$,
Fabio Barblan$^{2}$,\newauthor
Yong Yang $^{3}$,
Jia-Shu Niu $^{4}$.\\
$^{1}$Department of Astronomy, Beijing Normal University, Beijing 100875, P.R.China\\
$^{2}$Department of Astronomy, University of Geneva, Ch. des Maillettes 51, CH-1290 Versoix, Switzerland\\
$^{3}$South-Western Institute for Astronomy Research, Yunnan University, Kunming 650500, P.R.China\\
$^{4}$Institute of Theoretical Physics, Shanxi University, Taiyuan, 030006, P.R.China
}

\date{Accepted XXX. Received YYY; in original form ZZZ}

\pubyear{2018}

\begin{document}
\label{firstpage}
\pagerange{\pageref{firstpage}--\pageref{lastpage}}
\maketitle

\begin{abstract}
In this work, we have studied the variable stars in the young open cluster NGC 1893 based on a multi-year photometric survey covering a sky area around the cluster up to $31' \times 31'$ wide. More than 23\,000 images in the $V$ band taken from January 2008 to February 2017 with different telescopes, complemented with 90 images in the $B$ band in 2014 and 2017, were reduced, and light curves were derived in $V$ for 5653 stars. By analyzing these light curves, we detected 147 variable stars (85 of them being new discoveries), including 110 periodic variables, 15 eclipsing binaries and 22 non-periodic variables. Proper motions, radial velocities, color-magnitude and two-color diagrams were used to identify the cluster membership of these variable stars, resulting in 84 members. Periodic variable members were then classified into different variability types, mainly according to their magnitudes and to their periods of variability, as well as to their positions in the Hertzsprung-Russell diagram for the early-type stars. As a result, among main-sequence periodic variable members, we identified five $\beta$ Cep candidates, seven slowly pulsating B-type candidates, and thirteen fast-rotating pulsating B-type (FaRPB) candidates (one of which is a confirmed classical Be star). While most of the FaRPB stars display properties similar to the ones discovered in NGC~3766 by \citet{Mowlavi2013}, five of them have periods below 0.1~d, contrary to expectations. Additional observations, including spectroscopic, are called for to further characterize these stars. We also find a binary candidate harboring a $\delta$-Scuti candidate.
\end{abstract}




\section{Introduction}

Open clusters provide ideal opportunities to simultaneously study a group of stars in gravitationally bound systems. Pulsating stars as members of the same open cluster provide a chance for cluster asteroseismology. When multiple pulsating stars are detected in the same open cluster, the fact that these pulsators should have the same initial chemical compositions, the same ages and almost the same distances to the Earth may bring important additional constraints for the determination of the best-fitted stellar models \citep[e.g.,][]{Arentoft2005}.
Simultaneously, cluster asteroseismology provides a method to estimate the properties of the clusters \citep[e.g.,][]{Saesen2013}.
As far as eclipsing binaries in open clusters, since the stellar parameters of the binary components like the masses, radii, effective temperatures and luminosities can be obtained from very high precision observations, the measurements of the parameters of both the whole cluster and the individual stars get benefited.

In young open clusters (age $\le 10^7$~yr), low-mass stars are still in their pre-main sequence (PMS) stage while intermediate- and high-mass stars have already reached the main sequence (MS) phase of core H burning.
The study of the photometric variability of young open cluster members provides additional key information on the star formation process and the properties of early-type MS stars, knowing that PMS and MS stars display specific variability properties depending on their spectral type and stage of evolution.

T Tauri stars (TTSs) are low-mass PMS stars with either periodic or non-periodic variations. Periodic variability in these stars is thought to be caused by rotational modulation of a star with asymmetric cool or hot spots, while non-periodic variability result from accretion and outflows \citep{Herbst1994}. Compared with TTSs, Herbig Ae/Be stars have larger mass, which evolve towards the zero-age main-sequence (ZAMS) with hot or cool circumstellar dust or both, showing spectral type of A/B with emission lines \citep{Waters1998}. Most Herbig Ae/Be stars show photometric variations which are due to circumstellar patchy dust clouds or evolutionary effect \citep{Ancker1998}. On the MS, early-type stars in young open clusters usually show variable lightness due to pulsations, grouped into different regions of instability strips in the Hertzsprung-Russell (HR) diagram. These include $\beta$ Cep stars, slowly pulsating B (SPB) stars and $\delta$ Scuti stars \citep{Gautschy1996}. Be stars can show non-periodic variability due to the presence of a disk around them. Since they lie in the $\beta$ Cep or SPB instability region in the HR diagram, they can also display periodic variability due to pulsation \citep{Gautschy1996}.

\begin{table*}
\caption[]{Distance and age of NGC~1893.}
\label{distance_age}
\begin{tabular}{lccccc}
\hline \hline
 Photometric bands & $\rm{V_0-M_V}$ (mag) & Reddening (mag) & Distance (kpc) & Age (Myr) &  References \\
\hline
$UBV$                & $13.0\pm0.3$       &  0.44-0.77      &    4           &    -      & \citet{Johnson1961}   \\
$UBV$                & 12.5               &   -             &    -           &    -      & \citet{Humphreys1978} \\
$JHKuvbyH\beta$      & $13.18\pm0.11$     &   -             &    4.3         &   4       & \citet{Tapia1991}     \\
$UBVRIJHKL$          & $12.56\pm0.15$     &  0.4-0.7        &   $3.25\pm0.2$ &   1-5     & \citet{Sharma2007}    \\
$UBVI$               & $12.7\pm0.2$       &  0.4-0.7        &   $3.5\pm0.3$  &   1.5     &  \citet{Lim2014}      \\
\hline
\end{tabular}
\end{table*}

NGC~1893 ($\alpha$=05:22:44, $\delta$=33:24:42) is a Galactic open cluster immersed in the bright diffuse nebulosity of IC 410, which is associated with two pennant nebulae (Sim 190 and Sim 130) \citep{Gaze1952}, and obscured by several conspicuous dust clouds. A substantial number of efforts have been made to determine its distance and age, which are listed in Table~\ref{distance_age}. The distance modulus, reddening, distance and age presented in Table~\ref{distance_age} are broadly consistent with each other. \citet{Sharma2007} presented a comprehensive multiwavelength study of NGC~1893, yielding a spread in reddening between $E(B-V)$=0.4 and 0.7 mag. They estimated the distance modulus $V_0-M_V=12.56\pm0.15$ mag and found that the majority of the young stellar objects have ages between 1 and 5 Myr. \citet{Lim2014} derived a consistent reddening and distance with \citet{Sharma2007}, while they estimated the turn-off age of 1.5 Myr and the median age of 1.9 Myr from the PMS members with a spread of 5 Myr. With such an age, the members with spectral type A and later ones are always in their PMS stages.

$UBV$ photometry and spectroscopy of the cluster were performed by \citet{Massey1995}, who listed the spectral type of 24 stars. \citet{Marco2001} and \citet{Marco2002} identified 50 likely members with the spectral type of O-F in NGC~1893 and confirmed 2 Herbig Be stars, 3 massive T Tauri stars, 2 Herbig A candidates, 1 Herbig B candidate. \citet{Caramazza2008} and \citet{Prisinzano2011} found that NGC~1893 is a very rich cluster with a conspicuous population of pre-main sequence stars and well-studied main sequence cluster population. \citet{Lata2012} and \citet{Lata2014} identified more than one hundred variable stars in 13$\times$13 $arcmin^2$ field of view (FOV) around the center of the cluster by multi-epoch $V$-band photometry taken over 16 nights from 2007, Dec to 2013, Jan. They classified these variable stars as either PMS variables or MS ($\beta$ Cep, SPB, new class) variables.

The present paper provides the most extensive search for and study of the multi-type variable stars in the very young open cluster NGC~1893, and with the largest FOV observed to date (31$\times$31 $arcmin^2$) and time-series photometry that span 9 years.
Section~2 introduces our observations and data reduction. Section~3 describes the method used to search for periodic variables and estimate their cluster membership in Sect.~4. The characteristics of these periodic variables are described in Sect.~5, while Sects.~6 and 7 present the eclipsing binaries and non-periodic variable stars, respectively, found in the cluster. Our results are compared with previous works in Sect.~8. At last, we present a summary in Sect.~9.


\section{Observations and Data Reduction}

\begin{figure}
\centering
\includegraphics[width=0.45\textwidth,height=0.45\textwidth]{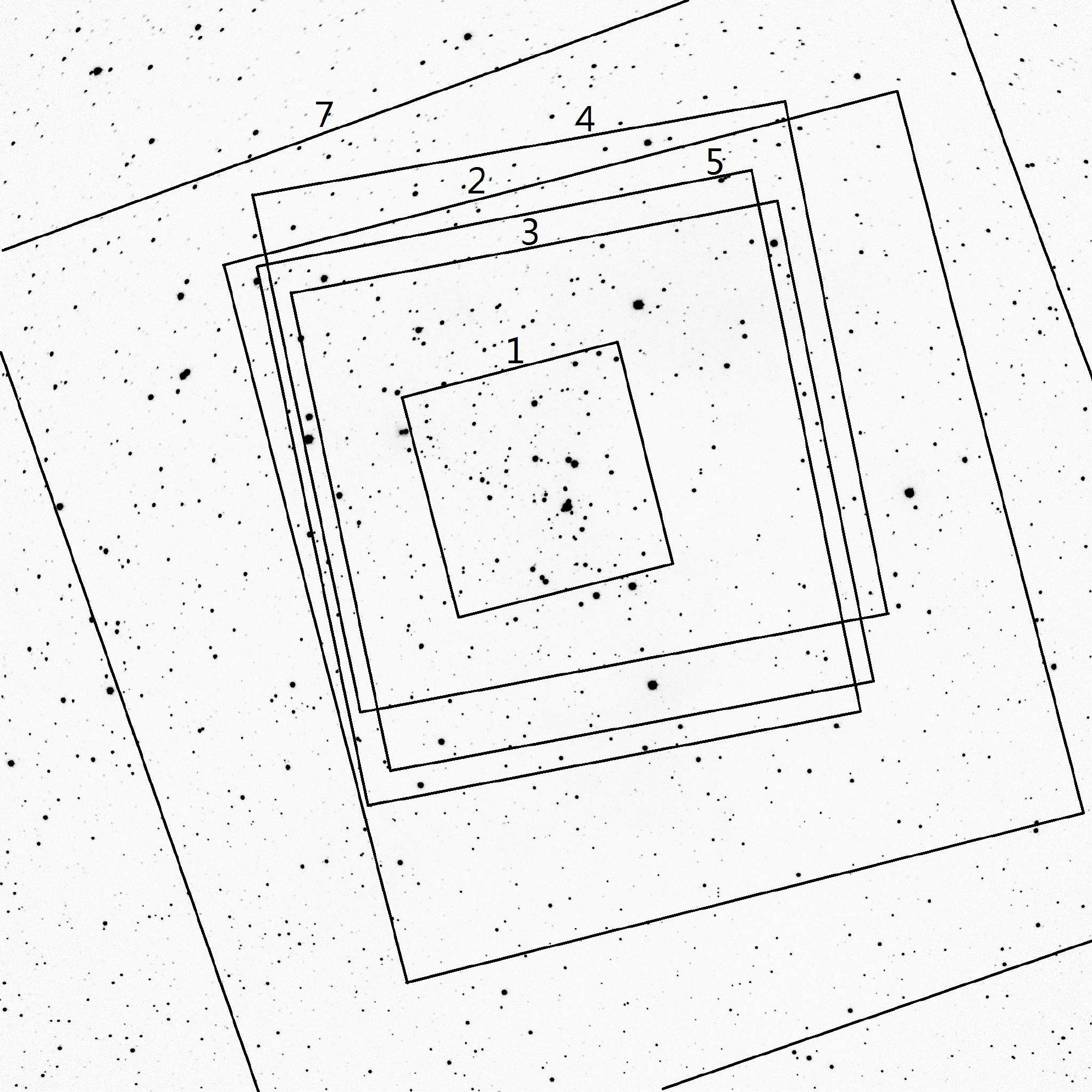}
\caption{CCD image of NGC~1893 collected with the 85-cm telescope in 2014 December, which is numbered as group 6 in Table \ref{tab:obs_log1}. The different FOVs are denoted with boxes and numbers, which correspond to the ID in Table \ref{tab:obs_log1}. North is up and east is to the left.
\label{fig:FOV}}
\end{figure}

\begin{table*}\small
\caption{Log of observation runs in $B$ and $V$ bands. \label{tab:obs_log1}}
\begin{center}
\scalebox{0.85}[0.95]{
\begin{tabular}{ccccccccccc}
\hline
\hline
ID&Date           & Telescope &  CCD  &  FOV  &  Resolution    & Nights & Frames & Exposure Time & Filter & $\sigma_{amp}$     \\
  &               &           &       &($arcmin^2$)&(\arcsec/pixel)&    &        &   (s)         &        &      (mmag)        \\
\hline
1&2008 Jan 05-Apr 05&La Palma 1.2 m (Mercator)&$2158\times2044$&$6.5\times6.5$&0.19& 31 & 2173   & 60-420  &  $V$   & 3.0  \\
2&2008 Oct 23-27&Xinglong 50 cm&$1340\times1300$ & $23\times22$ & 1.00   & 5     & 873    &      90-120   &  $V$   & 2.5        \\
3&2008 Nov 11-20&Xinglong 85 cm&$512\times512$  & $15\times15$  & 1.77   & 8     & 6766   &      15-60    &  $V$   & 3.1        \\
4&2009 Jan 10-15&Xinglong 85 cm&$1024\times1024$& $16\times16$  & 0.96   & 6     & 6602   &      15-25    &  $V$   & 3.9        \\
5&2012 Feb 03-13&Xinglong 85 cm&$1024\times1024$& $16\times16$  & 0.96   & 10    & 3651   &      20-150   &  $V$   & 3.4       \\
6&2014 Dec 11-17&Xinglong 85 cm& $2048\times2048$& $31\times31$  & 0.91   &  6    & 1046   &      8-12     &  $V$   & 4.0       \\
 &                &                &                 &               &        &       & 916    &      130-200  &  $V$   & 2.6      \\
 &                &                &                 &               &        &       & 31     &      15       &  $B$   &          \\
 &                &                &                 &               &        &       & 34     &      200      &  $B$   &          \\
7&2017 Feb 17-23&Xinglong 85 cm&$2048\times2048$& $31\times31$  & 0.91   &  6    & 743    &      6-150    &  $V$   & 3.9      \\
 &                &                &                 &               &        &       & 677    &      60-180   &  $V$   & 2.7    \\
 &                &                &                 &               &        &       & 14     &      15       &  $B$   &        \\
 &                &                &                 &               &        &       & 11     &      200      &  $B$   &    \\
\hline
\end{tabular}
}
\leftline{\small{\ \ \ \ \ \ \ {\bf Note}: $\sigma_{amp,V}=\sigma_{V} \sqrt{\pi/N_V}$ denotes the noise in the amplitude spectrum of the $V$ light curve.}}
\end{center}
\end{table*}

\begin{figure}
  \centering
  \subfigure[]{
    \label{fig:subfig:a}
    \includegraphics[width=0.225\textwidth]{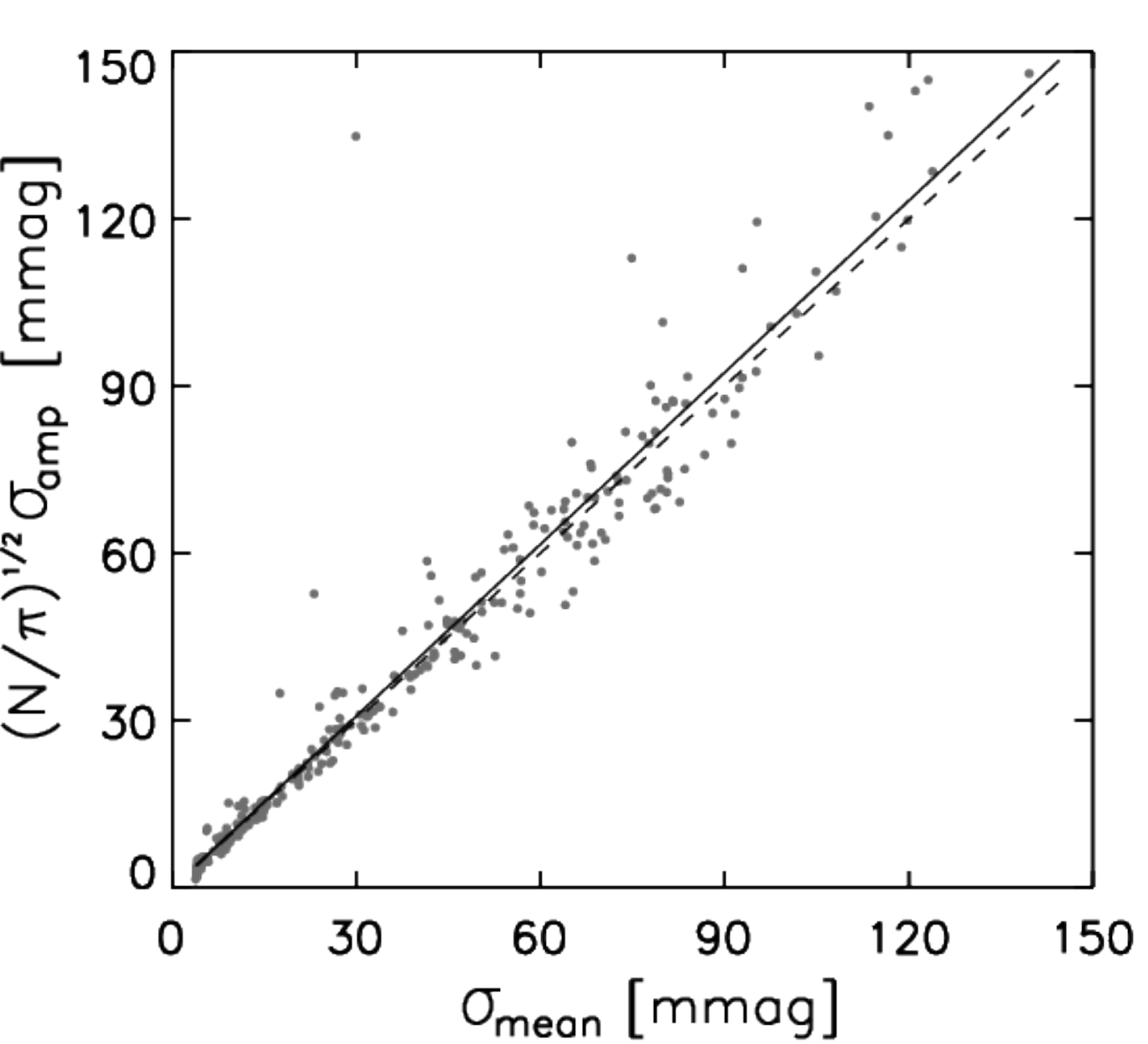}}
  \subfigure[]{
    \label{fig:subfig:b}
    \includegraphics[width=0.225\textwidth]{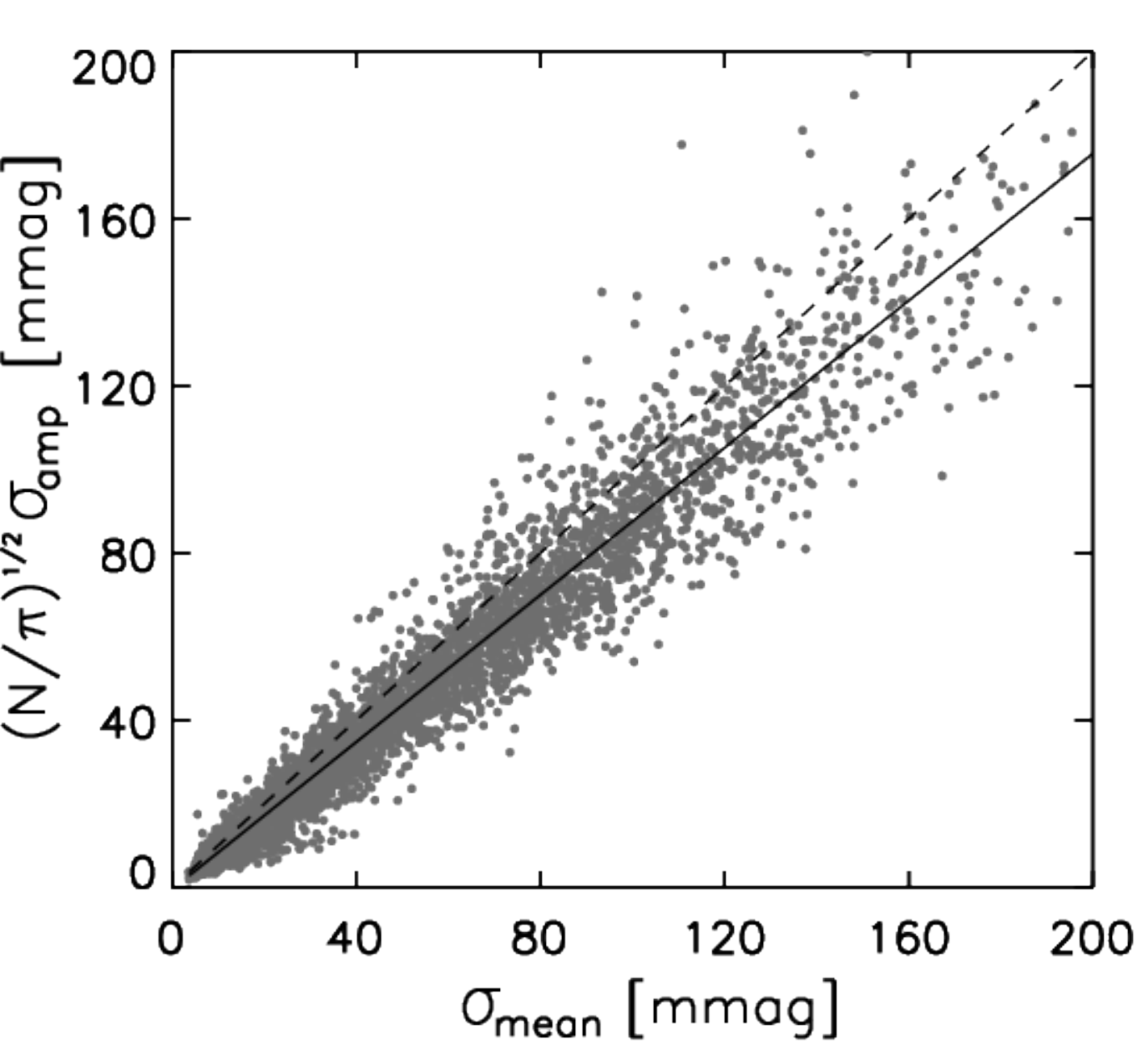}}
  \caption{Comparing the mean measured error and the amplitude noise in the periodogram. The solid lines represent the results of least-square fit. The dashed lines represent the diagonal lines. The subfigure (a) is based on the observations taken from Mercator in 2008. The subfigure (b) the observations taken from Xinglong 85 cm telescope in 2014 in long-exposure mode.}
  \label{error/amp_noise}
\end{figure}

\begin{figure*}
\centering
\subfigure[]{\includegraphics[width=0.3\textwidth]{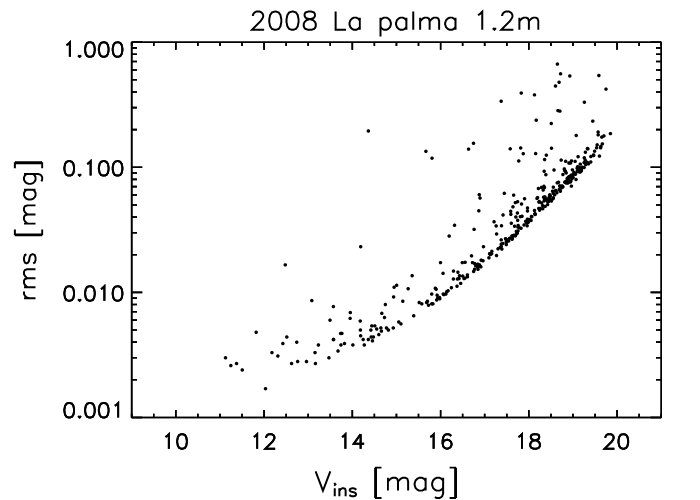}}
\subfigure[]{\includegraphics[width=0.3\textwidth]{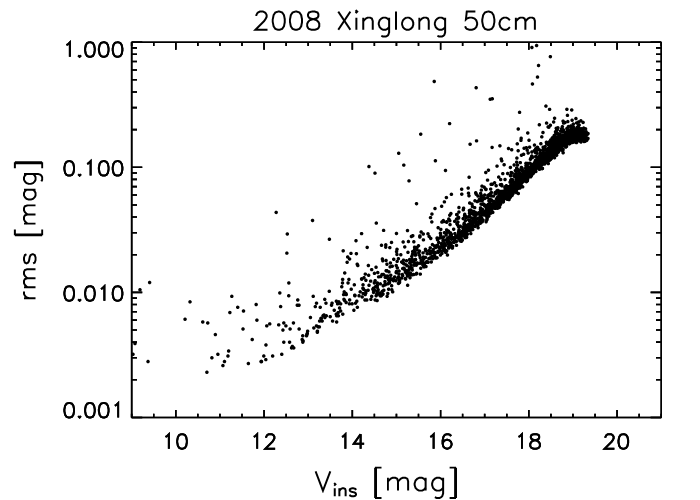}}
\subfigure[]{\includegraphics[width=0.3\textwidth]{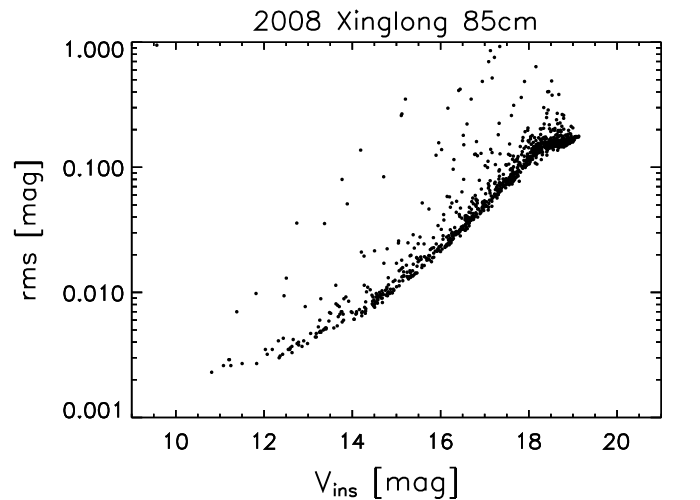}}
\subfigure[]{\includegraphics[width=0.3\textwidth]{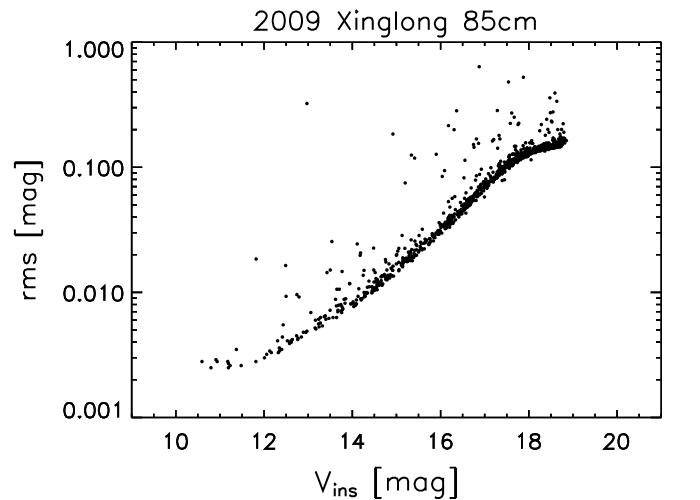}}
\subfigure[]{\includegraphics[width=0.3\textwidth]{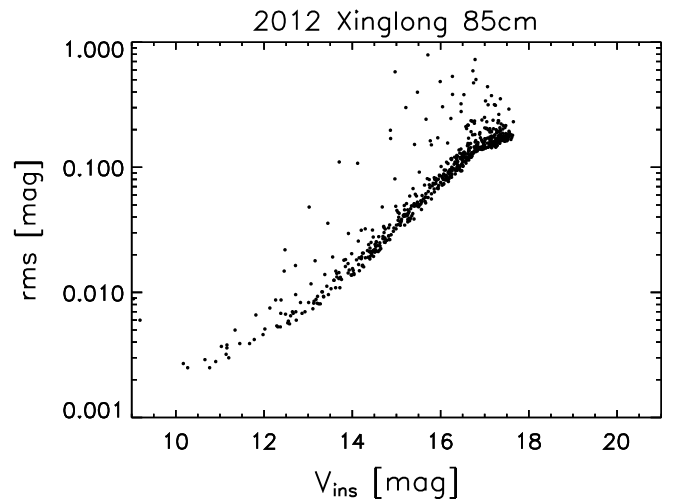}}
\subfigure[]{\includegraphics[width=0.3\textwidth]{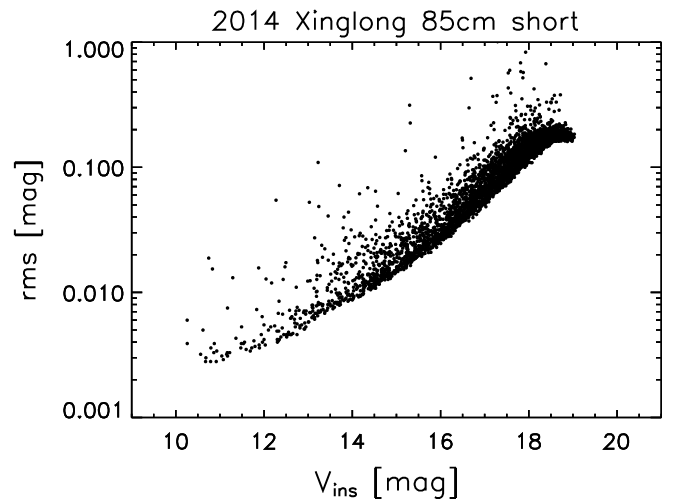}}
\subfigure[]{\includegraphics[width=0.3\textwidth]{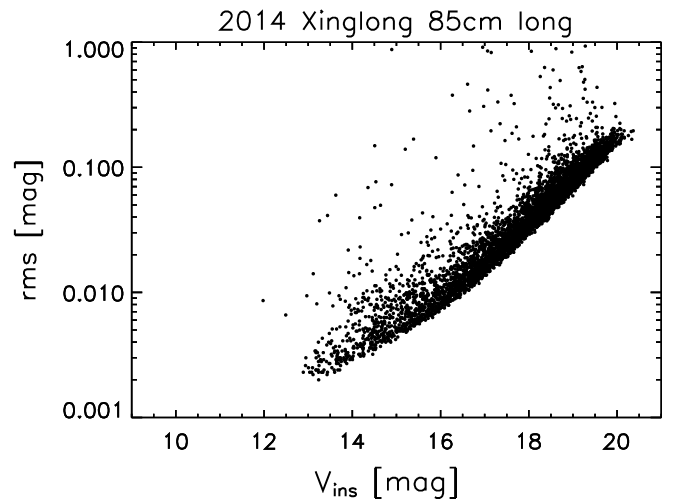}}
\subfigure[]{\includegraphics[width=0.3\textwidth]{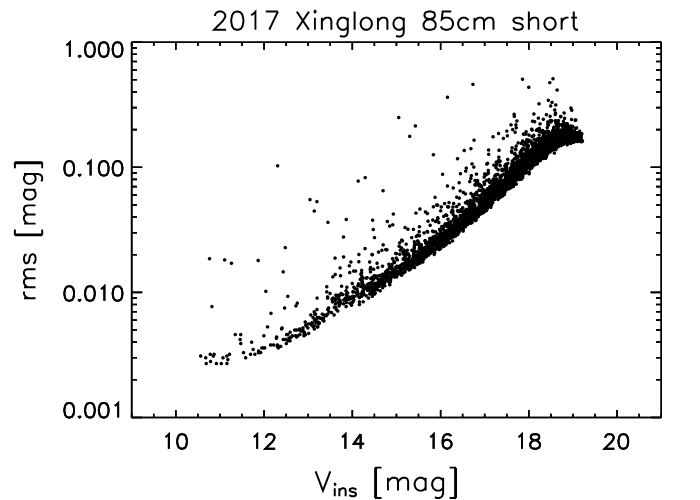}}
\subfigure[]{\includegraphics[width=0.3\textwidth]{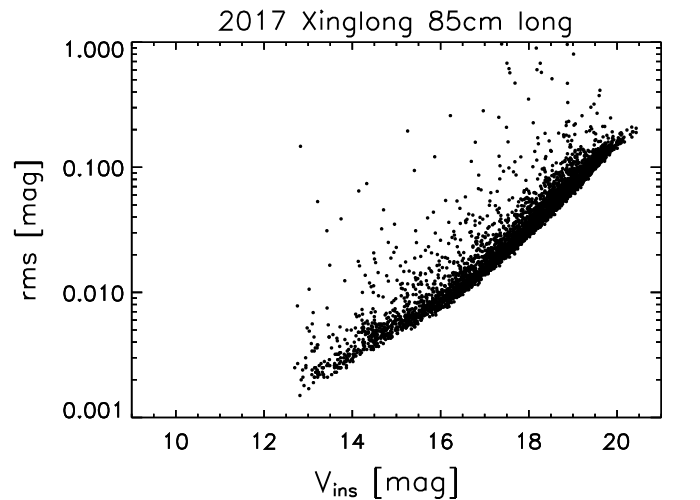}}
\caption{The rms dispersion as a function of the instrumental magnitude of different instruments.
         Subfigures (f) and (h) are based on the short-exposure observations of groups 6 and 7, respectively.
         Subfigures (g) and (i) are based on the long-exposure observations of groups 6 and 7, respectively.
}
\label{rms}
\end{figure*}

Time-series photometry of NGC~1893 were taken using $V$ filter with three telescopes from January 05, 2008 to February 23, 2017, for a total of 72 nights and 23537 frames. The detailed log of the observations is given in Table~\ref{tab:obs_log1}. The observations from January 05, 2008 to April 05, 2008 were done with the 1.2 m Flemish Mercator telescope located at La Palma, Canary Islands, Spain.
All subsequent observations were carried out at the 50 cm or 85 cm telescope located at the Xinglong station, National Astronomical Observatories of China. Most of our observations were collected using the 85 cm telescope.
It must be noted that three different CCD cameras were used on the 85 cm telescope between 2008 to 2017, as indicated in Table~\ref{tab:obs_log1}. The exposure times were adjusted according to weather condition and height of the target on the sky.
In the 2014 and 2017 campaigns, we alternated short and long exposures to focus on bright stars and faint stars, respectively, and we alternately took a small number of images in $B$ and $V$ bands for the color index analysis of the stars in the cluster. Figure~\ref{fig:FOV} shows a CCD image with the different FOVs of the telescopes.

Data pre-processing consisted in bias subtraction and flat field correction, using the IRAF\footnote{Image Reduction and Analysis Facility is developed and distributed by the National Optical Astronomy Observatories, which is operated by the Association of Universities for Research in Astronomy in Tucson, Arizona, USA, under operative agreement with the National Science Foundation.}.
For the Mercator run in 2008, bias images were taken in 8 nights among the 31 observation nights, while flat fields were taken every night. For the Xinglong observations, biases and flat fields were taken every night. The procedure of bias subtraction and flat field correction follows the method described in Sect.~4 of \citet{Saesen2010}.

The detailed procedure of data reduction was presented in \citet{Saesen2010} and \citet{Mowlavi2013}.
Here, we only summarize the proceeding scheme.
We use DAOPHOT $\amalg$ \citep{Stetson1987} and ALLSTAR \citep{Stetson1988} packages to extract the magnitudes of different stars on the frames.
First, we use a template image with the largest FOV and the best seeing to make a master star list with the routine FIND.
As some stars overlap in the template image but are distinguished in frames with higher resolution, they are added to the master star list.
The total number of stars in the list amounts to 5653.
Second, we convert the list to an instrument star list of a given telescope by transferring the coordinates of the list to the CCD images, providing thereby an exclusive number for every star in our observations.
Third, the magnitudes of the stars in each image are calculated using a combination of aperture and point spread function (PSF) photometry.
We choose the most appropriate number of reference stars for each group of observations listed in Table~\ref{tab:obs_log1}, by comparing the photometric precision of stars based on the different sets of reference stars.
Multi-differential photometry is performed to quantify the relative light variations of the measured stars and get an error estimate for each data point, using the chosen reference stars.
The estimated error will be used as the weight in the subsequent frequency analysis to improve the signal to noise level.
However, the errors on the data points from different instruments are not comparable due to the characteristics of the instruments.
To calibrate the errors of data points, we use the relation described in Sect.~5.3 of \citet{Saesen2010}.
For a given star $i$ and image $j$, we have
\begin{eqnarray}
  <noise_{i,j}>=\sigma_{mean,i}=\sqrt{N/\pi}\sigma_{amp,i}
\label{Eq:noise}
\end{eqnarray}
where $\sigma_{mean,i}$ is the noise in the time domain, i.e., the mean measured error of star $i$, and $\sigma_{amp,i}$ is the noise in the frequency domain, i.e., the amplitude noise in the periodogram of star $i$. $N$ is the number of measurements of star $i$.
For each group of observations, we fit $\sigma_{mean}$ and $\sigma_{amp}$ by the least square method and use the fit coefficient as the scaling factor to calibrate the measured errors such as to satisfy Eq.~\ref{Eq:noise}.
This procedure was applied to the Xinglong 85 cm observations, but was not needed for the Mercator and Xinglong 50 cm telescopes because their scaling factors derived from Eq.~\ref{Eq:noise} were already close to unity (1.03 and 1.04, respectively).
The comparison of the mean measured error and the amplitude noise in the periodogram for observations from Mercator is shown in Fig.~\ref{error/amp_noise}~(a) as an example.
For the other observation groups, the scaling factors range from 0.66 to 0.90.
An example is shown in Fig.~\ref{error/amp_noise}~(b), that compares the mean measured error and the noise amplitude in the periodogram for the observations performed with the Xinglong 85 cm telescope in 2014, where we calibrated the measured error by applying the scaling factor of 0.88.
The resulting error estimates of the observations performed with the different instruments become inter-comparable, and can be used as the weights of the data points.
Fourth, the Sys-Rem algorithm developed by \citet{Tamuz2005} is used to remove the linear systematic effects for a large number of stars, in a way similar to \citet[Sect. 5.4]{Saesen2010}.
Finally, we follow the process of \citet[Sect. 2.2]{Mowlavi2013} to do star selection and light curve cleaning.
The rms dispersion as a function of instrumental magnitude of different instruments is shown in Figure~\ref{rms}.

The instrumental magnitudes are transformed to the standard magnitudes using the catalogue provided by \citet{Lim2014}. We match our stars with the secondary standard stars from the catalogue. Then, the transformed relations are derived by linear fitting.
For the instrumental magnitudes taken in Dec 11-17, 2014 with short-exposure mode, the relations are
\begin{eqnarray}
B=-3.24+1.051B_{ins}
\label{Eq:trans_Bshort}
\end{eqnarray}
\begin{eqnarray}
V=-2.48+1.002V_{ins}
\label{Eq:trans_Bshort}
\end{eqnarray}
while the relations of long-exposure mode are
\begin{eqnarray}
B=-2.74+1.033B_{ins}
\label{Eq:trans_Bshort}
\end{eqnarray}
\begin{eqnarray}
V=-2.25+0.996V_{ins}
\label{Eq:trans_Bshort}
\end{eqnarray}
We apply these relations to all stars in our photometric region. Most of the stars are transformed to standard magnitudes by using the observations taken in Dec 11-17, 2014, while the stars which do not locate in the FOV in Dec 11-17, 2014 are transformed with the observations taken in Feb 17-23, 2017. Columns 6 and 7 of Table~\ref{tab:cat_variables} list the transformed magnitudes in $B$ and $V$ bands of the variable stars.


\section{Periodic variability analysis}

\subsection{Variable star detection}\label{Sec:vari_dect}

We use the observations of groups 1, 6 and 7 (cf. Table 2) to search for variable stars.
The observations from Mercator in 2008 (group~1) has the best sky resolution. The observations from Xinglong with the 85 cm telescope in 2014 (group~6) and 2017 (group~7) have the largest FOV, of $31^\prime \times 31^\prime$, which is helpful to search for variable stars in the outer regions of the cluster.
Besides, these observations were taken in both short-exposure and long-exposure modes, which are beneficial to search both bright and faint variable stars, respectively.
The groups 1, 6 and 7 also have high photometric accuracies, which are 0.003, 0.006 and 0.006 mag, respectively\footnote{Here, the photometric accuracies of groups 6 and 7 are derived from the observations of short-exposure mode.}.

Our procedure to search for variable stars follows the one described in \citet{Saesen2010} and \citet{Mowlavi2013}.
In brief, we calculate generalized Lomb-Scargle periodograms \citep{Zechmeister2009} in 0-50 $\rm d^{-1}$, where the weight of each data point is set as the square of the error of the measurement.
To calculate the signal-to-noise (S/N) of a frequency, the amplitude is taken as the signal and the mean amplitude after prewhitening the frequency as the noise.
In order to account for the increasing noise level at lower frequencies, the width of the frequency interval used to compute the noise is taken equal to 1 $\rm d^{-1}$ for $f \in \rm[0-3]\ d^{-1}$, 1.9 $\rm d^{-1}$ for $f \in \rm[3-6]\ d^{-1}$, 3.9 $\rm d^{-1}$ for $f \in \rm[6-11]\ d^{-1}$ and 5 $\rm d^{-1}$ for $f \in \rm[11-50]\ d^{-1}$.
We pick the variable candidates which have at least one significant frequency with S/N above 4.0 \citep{Breger1993}. Among the variable candidates, we disregard frequencies that appear repeatedly in several stars, within the frequency resolution $\delta{f}$, generally resulting from systematic effects. Then, we inspect the light curves of all other stars to identify missing eclipsing binaries and other non-periodic variables.

In total, 147 variable candidates are reported, consisting of 110 periodic variable stars, 15 eclipsing binaries, and 22 non-period variable stars.

\subsection{Light curve combination}

After having reduced independently each of the seven groups of observations listed in Table~\ref{tab:obs_log1}, we have to combine the resulting light curves.
The transparency instability of the atmosphere and/or the different sensitivities of the cameras between the different observation groups result in magnitude shifts that lead to low-frequency peaks in the periodograms.
We remove the magnitude shifts in the following way.
We first calculate the periodogram of every variable candidate using the whole light curve that combines all multi-group observations, but excluding the observations with low photometric precisions, and we compute the residual pre-whitened light curves.
For each group of observations, the mean magnitude difference between the observed and the pre-whitened light curve is then subtracted from the observed data.
We adjust in this way the light curves obtained in each observation group, getting rid of the magnitude shifts.
The adjusted light curves are then combined for each variable candidate.

\subsection{Frequency analysis}

\begin{figure}
\centering
\includegraphics[width=0.46\textwidth]{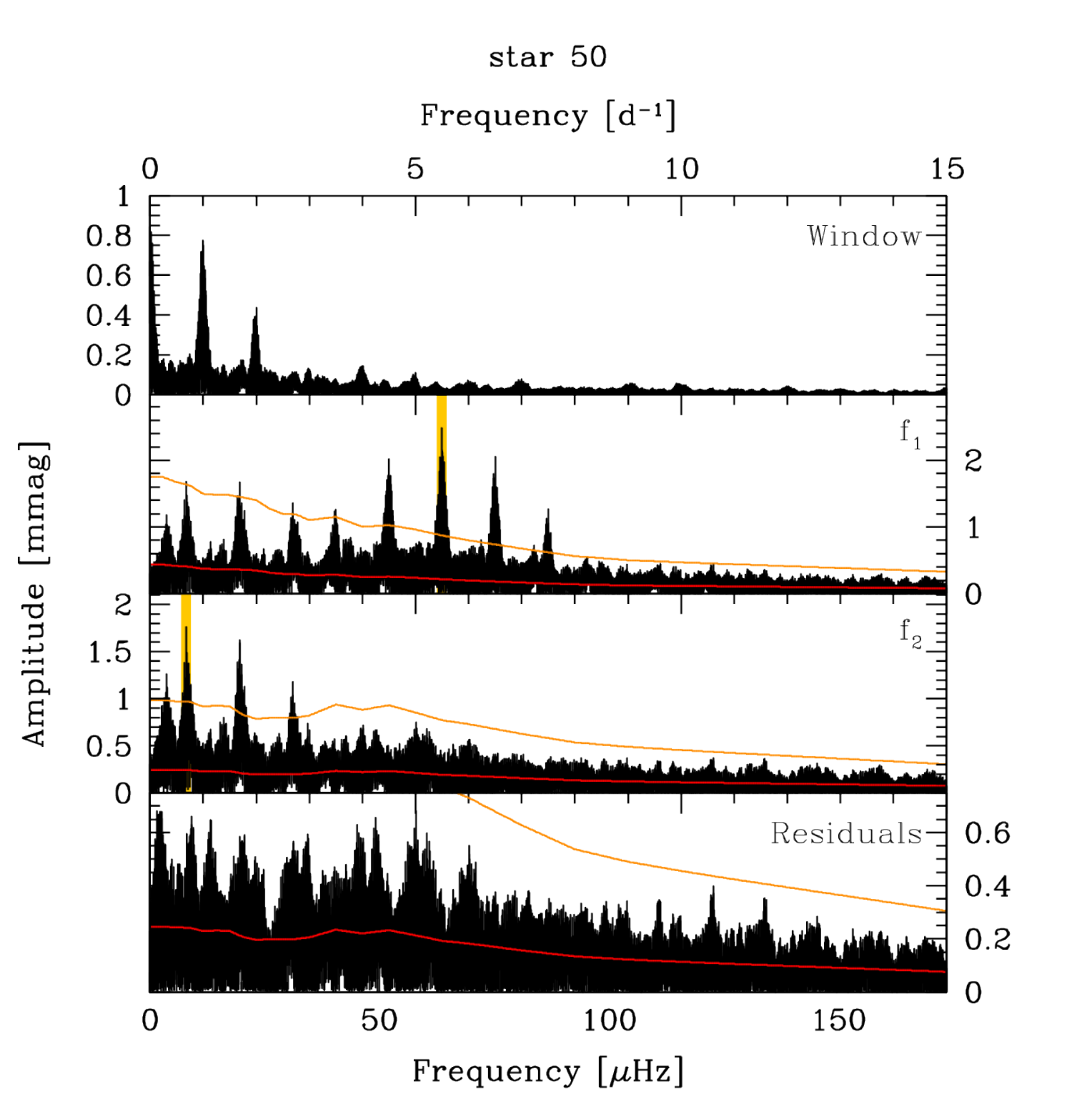}
\caption{Frequency analysis for star 50. The spectral window of the combined light curve is shown in the top panel. The subsequent panels show the generalised Lomb-Scargle periodograms at different steps of prewhitening. The detected frequencies are marked with the yellow paintings. The red and orange lines represent the noise level and the $\rm S/N=4$ level, respectively.
\label{fig:periodogram}
}
\end{figure}

The combined light curve is used to calculate the periodogram as described in Sect.~3.1, and pre-whiten the light curve.
This is applied to all periodic variable stars automatically.
The frequency search is stopped when the S/N value is smaller than 4.0. Figure~\ref{fig:periodogram} shows the spectral window function and the pre-whitening procedure for star 50 as an example of frequency analysis.
The periods, amplitudes and S/N values of the detected periodic variables are listed in Table~\ref{tab:cat_periodic}.\footnote{The eclipsing binary star 2091 hosting a pulsating companion is not reported in Table~\ref{tab:cat_periodic}, but listed in Table~\ref{tab:freq_star2091}.}
Only independent frequencies are reported, thus excluding harmonic and alias frequencies. The errors of the frequencies and amplitudes were estimated with the Monte Carlo simulations, which were based on the simulated light curves produced by an addition of the observed data and a Gaussian distribution random variable obeying $N(0,\sigma_{obs})$ \footnote{Here, $\sigma_{obs}$ indicates the measurement error.} (see Sect.~3 of \citet{Fu2013}).

Among the 110 periodic variable stars, 74, 17, 10, 6, 2 and 1 stars have 1, 2, 3, 4, 5 and 6 independent frequencies, respectively.

\section{Cluster membership of the variable stars}

In the next subsections, we successively consider the proper motions, radial velocities, and multi-band photometry to identify the variable members.
The relevant data collected from the literature are summarized in Table \ref{tab:cat_variables}.

\subsection{Proper motion}

Proper motions (PMs) are available for 3359 stars from a cross-match with the PPMXL catalog \citep{Roeser2010}.
Cluster membership is estimated from these PMs using a maximum likelihood method derived by \citet{Tian1998}.
The method estimates two frequency functions for each star $i$,
$\phi_{c}^{PM}(i;\ \mu_{xc},\ \mu_{yc},\ \sigma_{c})$ for the cluster \citep[Eq.~8 of][]{Tian1998}, and
$\phi_{f}^{PM}(i;\ \mu_{xf},\ \mu_{yf},\ \sigma_{xf},\ \sigma_{yf},\ \gamma)$ for the field \citep[Eq.~9 of][]{Tian1998}.
The parameters $\mu_{xc},\ \mu_{yc},\ \sigma_{c}$ are the PM center and the intrinsic PM dispersion of cluster members, the parameters $\mu_{xf},\ \mu_{yf},\ \sigma_{xf},\ \sigma_{yf}$ are the PM center and the intrinsic PM dispersion of field stars, and the additional $\gamma$ parameter is the correlation coefficient of the distribution of field stars.
The distribution function of the $i$th star in the cluster is given by
\begin{eqnarray}
  \phi^{PM}(i)&=&n_{c}\phi_{c}^{PM}(i)+n_{f}\phi_{f}^{PM}(i)
\end{eqnarray}
where the normalized number of cluster stars $n_c$ and of field stars $n_f$ satisfy the relation $n_c+n_f=1$.

Since the PM centers of the cluster and of field stars are similar, we set the cluster intrinsic dispersion $\sigma_c$ to zero following the discussion in Sect.~3.1 of \citet{Balaguernunez2004}.
A maximum likelihood method is then used to determine the eight remaining unknown parameters \citep[see][]{Tian1998}.
The results are listed in Table~\ref{table:PM_parameters}.
The PM membership probability $P_\mu$ of the $i$th star is then given by
\begin{eqnarray}
  P_{\mu}(i)&=&\frac{\phi_{c}^{PM}(i)}{\phi^{PM}(i)}
\end{eqnarray}
Table~\ref{tab:pm_rv_prob} lists the derived $P_{\mu}$ of 140 variable stars, the remaining variable stars lacking a cross-match in the PPMXL catalog.

\begin{table}
\caption{Distribution parameters of PM frequency function for NGC~1893. The units of $\mu$ and $\sigma$ are both $mas\, yr^{-1}$.}
\label{table:PM_parameters}
\centering
\scalebox{0.9}[0.9]{
\begin{tabular}{ccccccccc}
\hline\hline
$n_c$  &$\gamma$ & $\mu_{xc}$  & $\mu_{yc}$ &$\sigma_c$  & $\mu_{xf}$ &$\mu_{yf}$ &$\sigma_{xf}$  &$\sigma_{yf}$\\
\hline
0.84 &-0.19 & -0.35 & -3.97   & 0.00  & 0.05  & -3.99 & 15.10 & 12.83 \\
\hline
\end{tabular}
}
\end{table}

\subsection{Radial velocity}

Radial velocity (RV) data are found for 333 stars in NGC~1893 from the second release of value-added catalogues of the LAMOST Spectroscopic Survey of the Galactic Anticentre (LSS-GAC DR2) \citep{Xiang2017LAMOST}.
The cross-match is performed using a 3 arc-second match radius, and only stars from the LSS-GAC DR2 catalog that have signal to noise ratios (S/N) in the blue band (4750 \AA) larger than 10 are considered.
We note that \citet{Gieseking1981} has applied the frequency function method to the one-dimensional RV space to determine the RV membership probability.
They assumed that the radial velocities of the field stars and the cluster stars satisfied two one-dimensional Gaussian distributions, respectively. Then the distribution functions of RV for the cluster stars $\phi_{c}^{RV}$ and the field stars $\phi_{f}^{RV}$ can be expressed, respectively, as,
\begin{eqnarray}
\phi_{c}^{RV}(i)=\frac{1}{2\pi(\sigma_c^2+\epsilon_i^2)}
exp\{-\frac{1}{2}[\frac{(v_i-v_c)^2}{\sigma_c^2+\epsilon_i^2}]\} \\
\phi_{f}^{RV}(i)=\frac{1}{2\pi(\sigma_f^2+\epsilon_i^2)}
exp\{-\frac{1}{2}[\frac{(v_i-v_f)^2}{\sigma_f^2+\epsilon_i^2}]\}
\end{eqnarray}
   where $v_i$ is the RV of the $i$th star and $\epsilon_i$ the corresponding error. $(v_c,\ v_f)$ are the RV centers of the cluster and the field stars, respectively, and $(\sigma_c,\ \sigma_f)$ are the intrinsic RV dispersions of the field stars and the cluster stars, respectively. So, the frequency function $\phi^{RV}(i)$ and the RV membership probability  $P_{v}$ of the $i$th star are,
\begin{eqnarray}
\phi^{RV}(i)&=&n_{c}\phi_{c}^{RV}(i)+n_{f}\phi_{f}^{RV}(i)\\
P_{v}(i)&=&\frac{\phi_{c}^{RV}(i)}{\phi^{RV}(i)}
\end{eqnarray}
   The five undetermined parameters ($n_c,\ v_c,\ v_f,\ \sigma_c,\ \sigma_f$) are derived by applying the maximum likelihood method and listed in Table~\ref{tab:RV_parameters}.
   Table~\ref{tab:pm_rv_prob} lists the $P_{v}$ values for 22 variable stars, while the RV data of the other variable stars are absent in the LSS-GAC DR2 catalog.

   We consider a star to be member of the cluster when the kinematic membership probability ${P_\mu}\times{P_{v}} > 50\%$ if both ${P_\mu}$ and ${P_v}$ are available. When ${P_v}$ is not available, we use the criterion ${P_\mu} > 50\%$ to identify the member candidates.

\begin{table}
\caption{Distribution parameters of RV frequency function for NGC~1893. The units of $v$ and $\sigma$ are both $km\,s^{-1}$.}
\label{tab:RV_parameters}
\centering
\begin{tabular}{lllll}
\hline\hline
$n_c$   & $v_c$  & $\sigma_c$ & $v_f$ &$\sigma_f$\\
\hline
0.43&-4.20&2.49&2.36&26.32\\
\hline
\end{tabular}
\end{table}

\begin{table*}
\caption{Membership probability $P_{\mu}$ and $P_{v}$ of the variable stars in the field of view of NGC~1893.}
\label{tab:pm_rv_prob}
\centering
\begin{tabular}{lcc|lcc|lcc|lcc|lcc}
\hline\hline
 Star ID & $P_{\mu}$  & $P_{v}$ &  Star ID & $P_{\mu}$  & $P_{v}$ &  Star ID & $P_{\mu}$  & $P_{v}$ &  Star ID & $P_{\mu}$  & $P_{v}$ &  Star ID & $P_{\mu}$  & $P_{v}$ \\
\hline
22 & 0.01  & $-$ & 269 & 0.09 & 0.74& 861 & 0.99 & $-$ & 1893 & 0.99& $-$ & 3215 & 0.99& $-$ \\
23 & 0.99  & 0.60& 271 & 0.84 & $-$ & 922 & 0.65 & $-$ & 1908 & 0.98& $-$ & 3219 & 0.64& $-$ \\
25 & 0.99  & $-$ & 283 & 0.98 & $-$ & 926 & 0.99 & $-$ & 1913 & 0.99& 0.00& 3343 & 0.99& $-$ \\ 
40 & 0.99  & $-$ & 305 & 0.99 & $-$ & 962 & 0.99 & $-$ & 1921 & 0.98& $-$ & 3349 & 0.99& $-$ \\
43 & 0.99  & 0.91& 313 & 0.92 & $-$ & 986 & 0.96 & $-$ & 1961 & 0.96& $-$ & 3521 & 0.98& $-$ \\
50 & 0.99  & $-$ & 325 & 0.99 & 0.44& 1049 & $-$ & $-$ & 1994 & 0.99& $-$ & 3627 & 0.99& $-$ \\
52 & 0.99  & $-$ & 327 & 0.97 & $-$ & 1069 & 0.97& $-$ & 2005 & 0.97& $-$ & 3680 & 0.94& $-$ \\
54 & 0.99  & $-$ & 338 & 0.99 & $-$ & 1098 & 0.97& $-$ & 2038 & 0.98& 0.00& 3695 & $-$ & $-$ \\
55 & 0.99  & $-$ & 341 & 0.97 & $-$ & 1101 & 0.99& $-$ & 2051 & 0.99& $-$ & 4056 & 0.97& $-$ \\
58 & 0.99  & 0.90& 342 & 0.00 & 0.04& 1114 & 0.99& $-$ & 2074 & 0.01& $-$ & 4082 & 0.83& $-$ \\
62 & 0.98  & $-$ & 360 & 0.98 & $-$ & 1202 & 0.99& $-$ & 2091 & 0.62& $-$ & 4162 & 0.96& $-$ \\
65 & 0.99  & $-$ & 368 & 0.96 & $-$ & 1208 & 0.88& $-$ & 2186 & 0.87& $-$ & 4225 & 0.98& $-$ \\
80 & 0.94  & $-$ & 414 & 0.81 & $-$ & 1277 & 0.99& $-$ & 2302 & $-$ & $-$ & 4495 & $-$ & $-$ \\
96 & 0.89  & $-$ & 441 & 0.98 & 0.93& 1287 & 0.99& $-$ & 2347 & 0.99& $-$ & 4496 & 0.00& $-$ \\
101 & 0.32 & $-$ & 472 & 0.00 & $-$ & 1301 & 0.95& $-$ & 2352 & 0.99& $-$ & 4671 & 0.95& $-$ \\
122 & 0.99 & 0.85& 509 & 0.91 & $-$ & 1326 & 0.99& $-$ & 2450 & 0.94& $-$ & 4716 & 0.97& $-$ \\
127 & 0.53 & $-$ & 528 & 0.96 & $-$ & 1376 & 0.00& $-$ & 2466 & 0.04& $-$ & 4975 & 0.89& $-$ \\
128 & 0.98 & 0.36& 534 & 0.98 & $-$ & 1380 & 0.99& $-$ & 2611 & 0.96& $-$ & 5048 & 0.98& $-$ \\
130 & 0.88 & $-$ & 591 & 0.98 & $-$ & 1388 & 0.96& $-$ & 2640 & 0.98& $-$ & 5130 & 0.99& $-$ \\
149 & 0.93 & $-$ & 599 & 0.99 & $-$ & 1390 & 0.88& $-$ & 2697 & $-$ & $-$ & 5156 & 0.94& $-$ \\
182 & 0.90 & $-$ & 638 & 0.97 & $-$ & 1425 & 0.99& $-$ & 2773 & 0.00& $-$ & 5158 & $-$ & $-$ \\
184 & 0.99 & $-$ & 660 & 0.00 & 0.00& 1572 & 0.98& $-$ & 2777 & 0.87& $-$ & 5262 & 0.92& 0.01\\
190 & 0.99 & 0.87& 666 & 0.90 & 0.00& 1577 & 0.98& $-$ & 2793 & 0.99& $-$ & 5269 & 0.87& $-$ \\
197 & 0.94 & $-$ & 706 & 0.98 & 0.73& 1590 & 0.99& $-$ & 2882 & $-$ & $-$ & 5282 & 0.00& $-$ \\
212 & 0.99 & 0.76& 710 & 0.99 & $-$ & 1620 & 0.99& $-$ & 2906 & 0.00& $-$ & 5283 & 0.18& $-$ \\
218 & 0.49 & 0.00& 721 & 0.99 & $-$ & 1630 & 0.98& $-$ & 2936 & 0.98& $-$ & 5284 & 0.12& $-$ \\
233 & 0.48 & $-$ & 738 & 0.98 & 0.00& 1657 & 0.99& $-$ & 2975 & 0.99& $-$ & 5569 & 0.96& $-$ \\
235 & 0.01 & $-$ & 741 & 0.99 & $-$ & 1664 & 0.99& 0.00& 2986 & 0.00& $-$ &      &     &     \\
240 & 0.99 & $-$ & 824 & 0.98 & $-$ & 1665 & 0.97& 0.00& 3093 & 0.69& $-$ &      &     &     \\
262 & 0.98 & $-$ & 844 & 0.98 & 0.29& 1810 & 0.98& $-$ & 3214 & 0.99& $-$ &      &     &     \\
\hline
\end{tabular}
\end{table*}

\subsection{CMD and TCD}\label{Sect:CMD_TCD}

\begin{figure}
\centering
\includegraphics[width=0.5\textwidth]{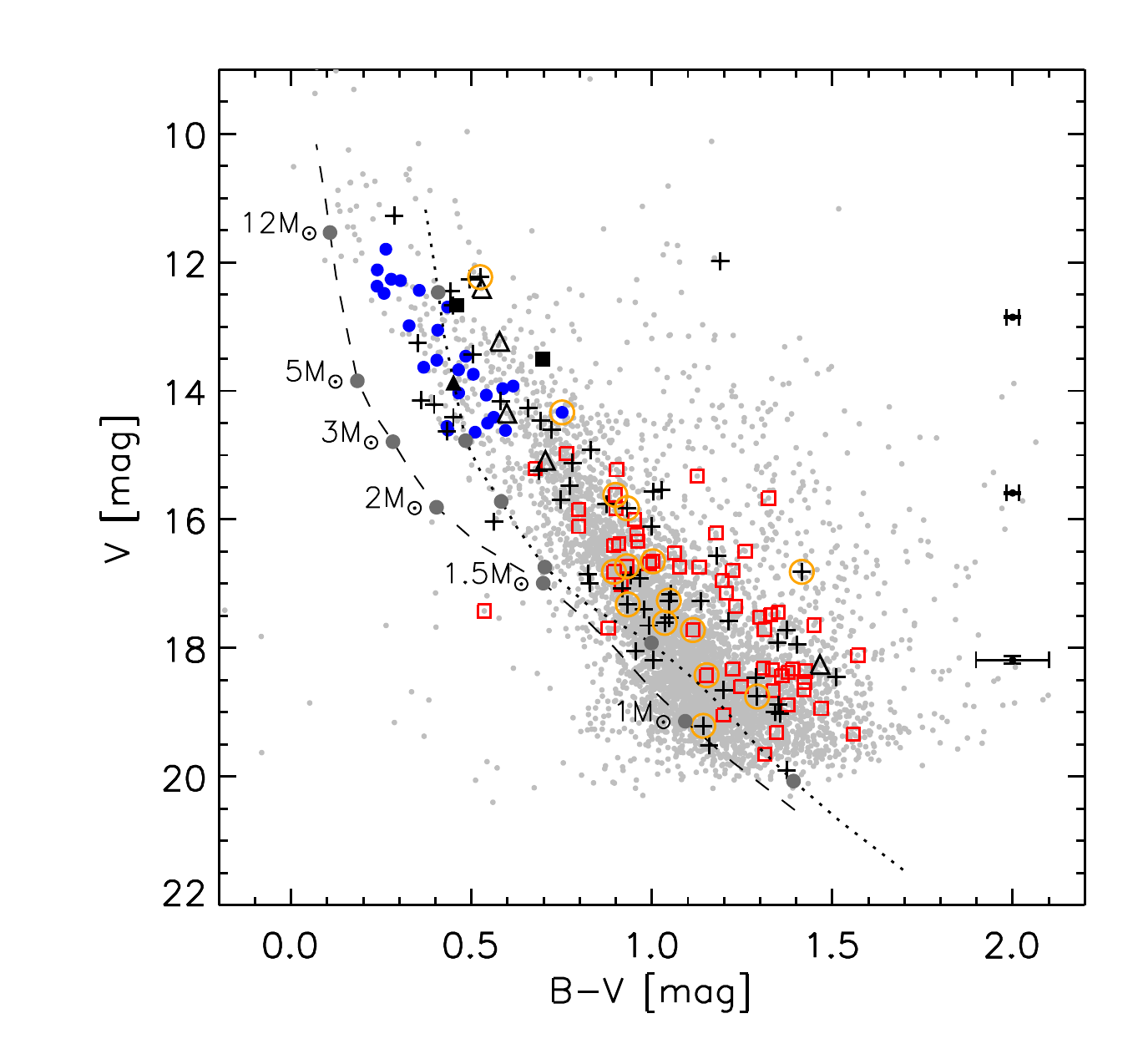}
\caption{$V/B-V$ CMD for the variable stars lying within the field of view of NGC~1893.
         The dash and dotted lines represent the zero-age main-sequence (ZAMS) by \citet{Ekstr2012} for Z=0.014, shifted along the distance modulus of $V_0-M_V=12.7\ \rm{mag}$ \citep{Lim2014} for $E(B-V)_{min}=0.4$~mag (dashed line) and $E(B-V)_{max}=0.7$~mag (dotted line), respectively.
         The dark grey points on the ZAMS lines mark the locations of few specific masses, as labeled next the points on the dashed line.
         The three error bars on the right of the figure show the mean magnitude and colour errors for the variable stars with $V \in \rm[11-14],\ \rm[14-17],\ \rm[17-20]$ mag, respectively.
         The blue dots, red open squares, black pluses and orange circles represent MS variable members, PMS variable members, field variable stars and eclipsing binaries, respectively.
         Open triangles represent variable stars for which membership could not be determined.
         The two stars classified as Herbig Be candidates are marked with filled squares (stars 62 and 130), while the star classified as a classical Be star in literature is marked in black filled triangle (star 182). The grey dots plotted in the background of the figure represent all the identified stars in the field of view of NGC~1893.
         }
\label{fig:V_BV}
\end{figure}

\begin{figure}
\centering
\includegraphics[width=0.5\textwidth]{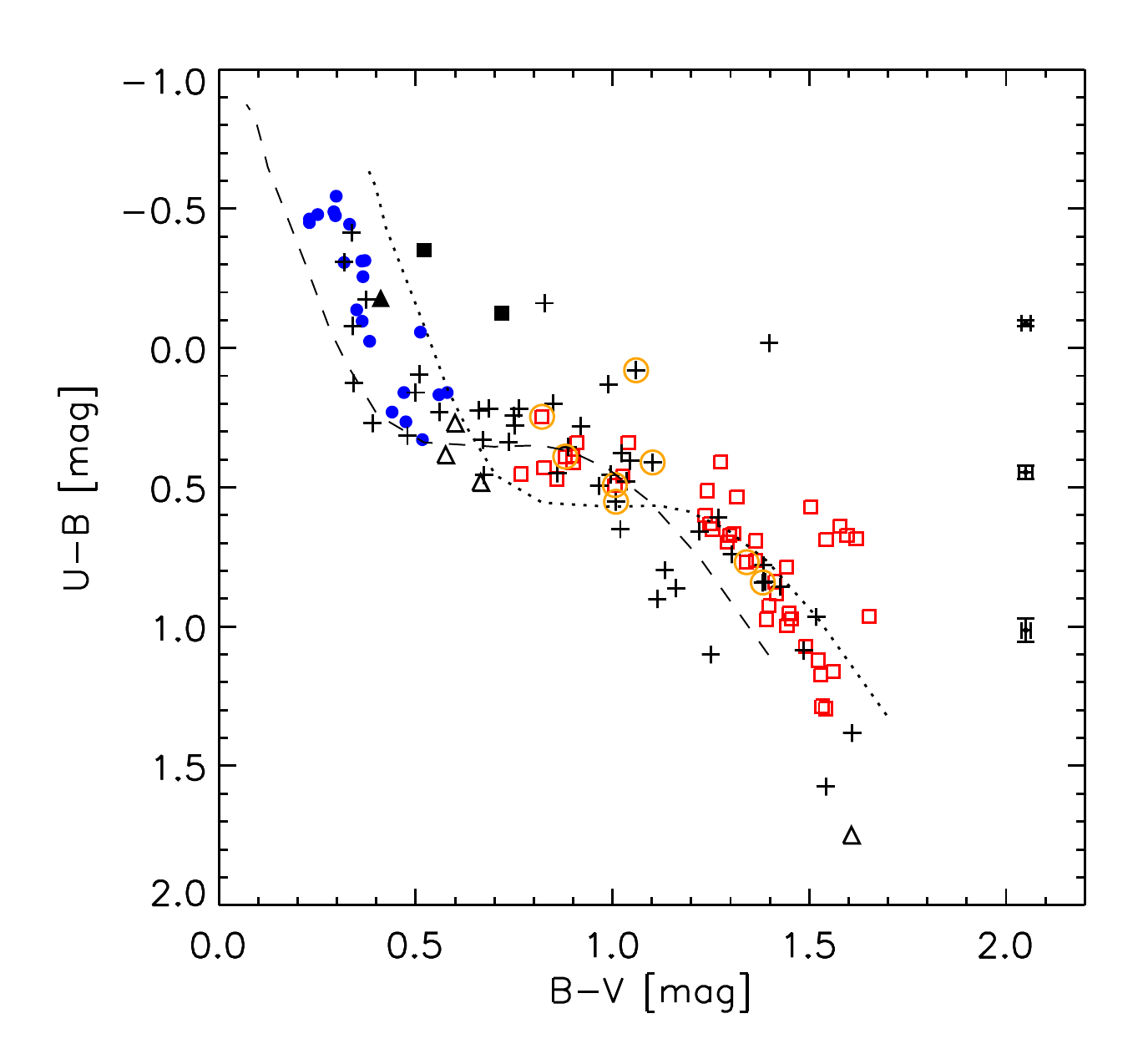}
\caption{$U-B/B-V$ TCD for the variable stars lying within the field of view of NGC~1893.
         The dash and dotted lines represent the zero-age main-sequence (ZAMS) by \citet{Ekstr2012} for Z=0.014, shifted along the reddening vector of 0.72 for $E(B-V)_{min}=0.4$~mag (dashed line) and $E(B-V)_{max}=0.7$~mag (dotted line), respectively.
         The three error bars on the right of the figure show the mean errors of the variable stars for $U-B \in \rm[-0.6-0.2],\ \rm[0.2-0.7],\ \rm[0.7-2.0]$ mag, respectively.
         The interpretation of the symbols is the same as in Fig.~\ref{fig:V_BV}.
         }
\label{fig:UB_BV}
\end{figure}

\begin{figure}
\centering
\includegraphics[width=0.5\textwidth]{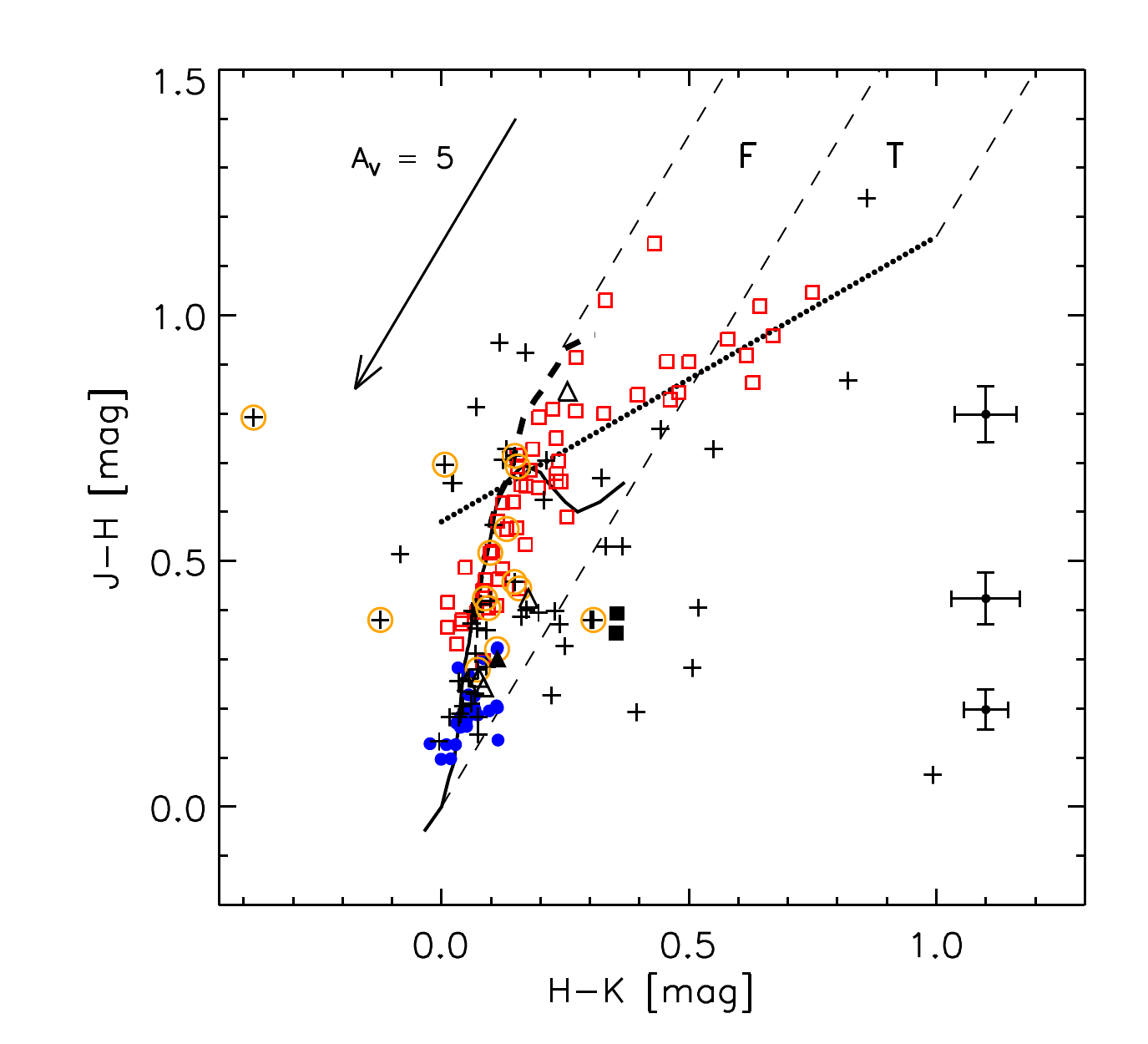}
\caption{$J-H/H-K$ TCD for the variable stars lying within the field of view of NGC~1893, $JHK$ being the 2MASS photometry converted to the California Institute of Technology (CIT) system.
         The sequences for dwarfs (solid curve) and giants (thick dash curve) are from \citet{Bessell1988}.
         The dotted line represents the locus of intrinsic classical T Tauri stars (CTTS) \citep{Meyer1997}.
         The solid and thick dashed curves represent the unreddened MS and giant branch, respectively \citep{Bessell1988}.
         The parallel dashed lines are the locus of reddening.
         The arrow denotes a reddening vector corresponding to a visual extinction of $A_V=5$~mag, based on extinction ratios $A_J/A_V=0.265$, $A_H/A_V=0.155$ and $A_{K_S}/A_V=0.090$ taken from \citet{Cohen1981}.
         `F' sources could be either field stars or
         class \uppercase\expandafter{\romannumeral2}/class \uppercase\expandafter{\romannumeral3} sources with small NIR excesses. `T' sources are considered to be mostly classical T Tauri stars (class \uppercase\expandafter{\romannumeral2} objects) with large NIR excesses.
         The three error bars show the mean errors of the variable stars for $J-H \in \rm[0-0.3],\ \rm[0.3-0.6],\ \rm[0.6-1.2]$ mag, respectively.
         The marker shapes and colors have the same interpretation as in Fig.~\ref{fig:V_BV}.
\label{fig:JH_HK}}
\end{figure}

\begin{figure*}
\centering
\subfigure[]{\includegraphics[width=0.85\textwidth]{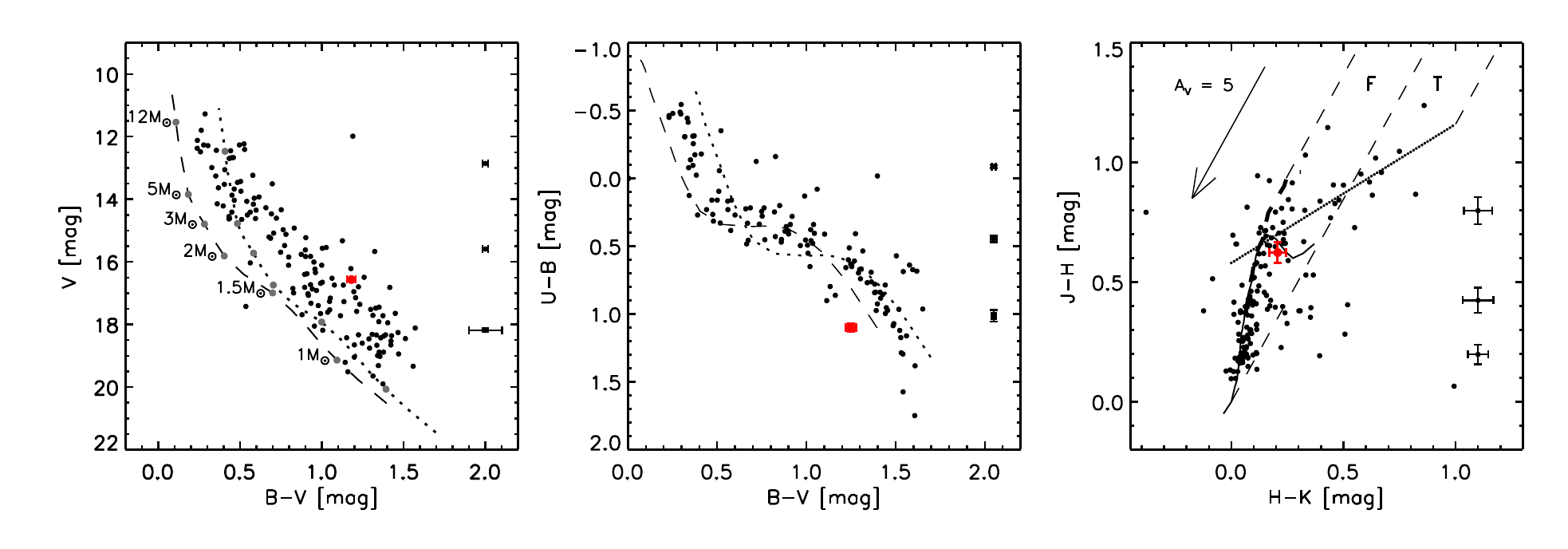}}
\subfigure[]{\includegraphics[width=0.85\textwidth]{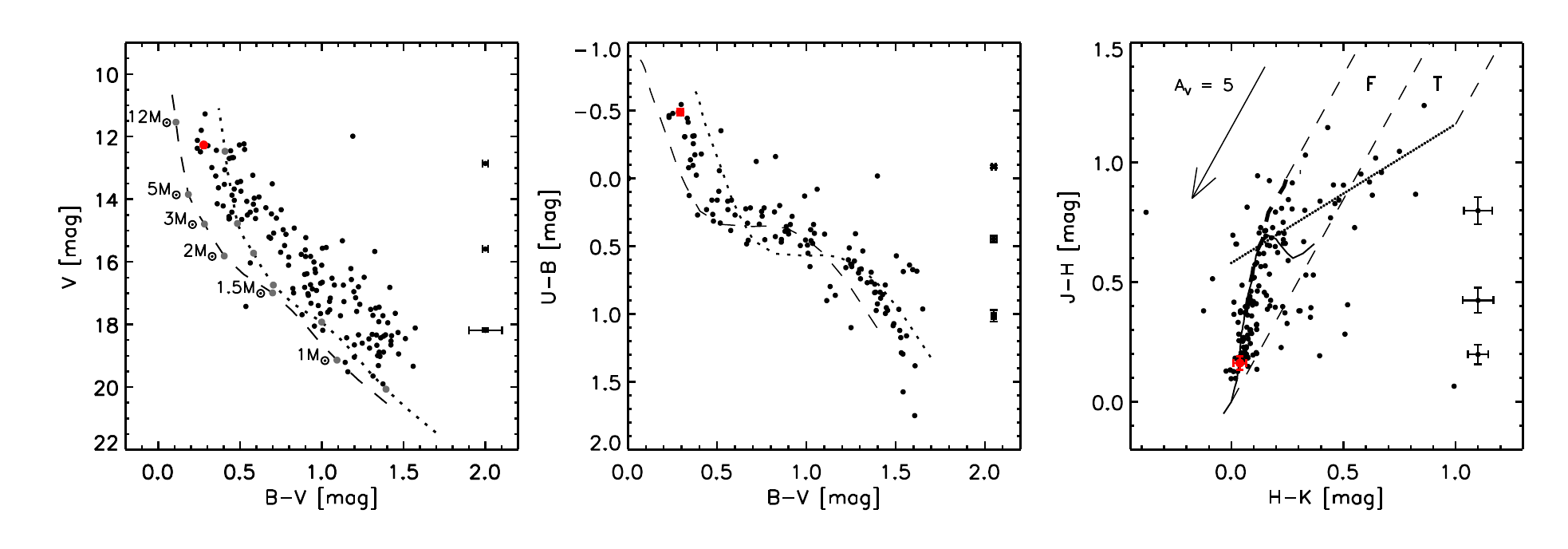}}
\subfigure[]{\includegraphics[width=0.85\textwidth]{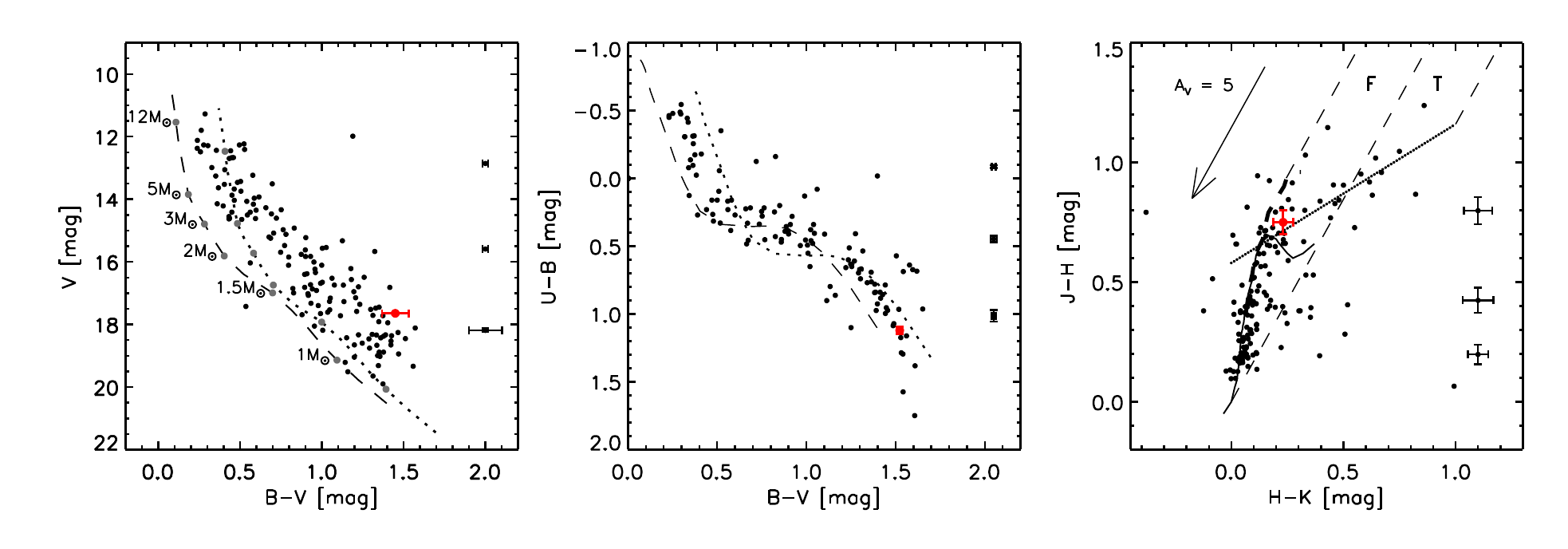}}
\caption{
(a) Position of field star candidate 1390 (red point) in the $V/B-V$ CMD (left), $U-B/B-V$ TCD (middle) and $J-H/H-K$ TCD (right), superposed on the variable stars of the cluster.
         The symbols and lines have the same meaning as in Fig.~\ref{fig:V_BV} for the $V/B-V$ CMD, as in Fig.~\ref{fig:UB_BV} for the $U-B/B-V$ TCD, and as in Fig.~\ref{fig:JH_HK} for the $J-H/H-K$ TCD.
         (b) Same as (a), but for star 50, a MS member candidate from its position in the CMD and TCDs.
         (c) Same as (a), but for star 2186, a PMS member candidate from its position in the CMD and TCDs.
\label{fig:exam_star1390}}
\end{figure*}

We limit the study to the 147 variable stars detected in Sect.~\ref{Sec:vari_dect}.

The $V/B-V$ CMD of the variable stars is shown in Fig.~\ref{fig:V_BV}.
It is constructed using the $V$ and $B$ values from the present work.
For reference, solar metallicity ZAMS lines are plotted in the figure for two reddening values, of 0.4~mag (dashed line) and 0.7~mag (dotted line), in the range of reddening reported by \citet{Lim2014} and \citet{Sharma2007} for the cluster.

The $U-B/B-V$ TCD is shown in Fig.~\ref{fig:UB_BV}.
Since we do not have measurements in $U$ band, we take colours published in the literature, as available.
We use $U-B$ and $B-V$ colours from \citet{Lim2014} for 101 of our variable stars, and from \citet{Massey1995} for 19 additional ones.
Solar metallicity ZAMS lines are also plotted in the figure for two reddening values.

The $J-H/H-K$ TCD is shown in Fig.~\ref{fig:JH_HK}.
The $JHK$ data of the 146 variable stars are taken from the 2MASS catalogue \citep{Cutri2003} and converted to the California Institute of Technology (CIT) system using the relations provided by \citet{Carpenter2001}.

The $V/B-V$ CMD (Fig.~\ref{fig:V_BV}), $U-B/B-V$ TCD (Fig.~\ref{fig:UB_BV}) and $J-H/H-K$ TCD (Fig.~\ref{fig:JH_HK}) are used to identify membership of the stars that lack PM or RV data.
First, the stars who have deflected positions in these CMD and TCDs are excluded from the list of members.
We then follow the criteria suggested by \citet{Lim2014} to distinguish MS and PMS stars: members with $V \leq 16$~mag, 0.0~mag $\leq B-V \leq$ 0.6~mag, and $-1.0$~mag $\leq U-B \leq 0.5$~mag are classified as MS stars, and the other members that locate in Fig.~\ref{fig:JH_HK} close to the upper part of the MS branch (solid curve) or close to the 'T' and 'F' regions are classified as PMS stars.
Three examples are shown in Fig.~\ref{fig:exam_star1390}.
For star 1390 (upper plots in the figure), $P_\mu=0.88$ while no RV data is available.
Its position in the $J-H/H-K$ TCD is compatible with a potential member candidate with NIR excess, but its deflected position in the $U-B/B-V$ TCD suggests it may not be a member.
Star 50 ($P_\mu=0.98$, second line of plots in Fig.~\ref{fig:exam_star1390}) is clearly a MS star of the cluster given its positions in the TCDs.
The third example, star 2186 ($P_\mu=0.87$, bottom plots in Fig.~\ref{fig:exam_star1390}), shows a PMS star due to its positions in the TCDs.

The membership analysis performed above can further be supplemented by spectroscopic data when available and necessary.
\citet{Marco2002} displayed the spectrum of star 130 (their star S3R1N3) with strong Balmer emission lines and a very reddened continuum, and estimated its spectral type to be B0.5\uppercase\expandafter{\romannumeral4}e.
They considered it to be almost certainly a Herbig Be star.
We therefore classify it as a PMS star.
The position of star 62 in the TCD is similar to that of star 130, which suggests it to be a Herbig Be candidate as well.
The two Herbig Be candidates are marked with filled squares in the figures, and their folded light curves shown in Fig.~\ref{fig:HerbigAeBe}.
Finally, we reject star 54 from the list of cluster members following the spectral analysis of \citet{Marco2002}, who find a F0\uppercase\expandafter{\romannumeral3}-\uppercase\expandafter{\romannumeral4} spectral type for this star (their star S3R1N3).

In summary, we identify 27 MS members, 57 PMS members and 58 field stars among the variable stars, while we could not conclude on the membership of five stars (stars 25, 96, 271, 528 and 3219).
They are marked with specific symbols in Figs.~\ref{fig:V_BV} to \ref{fig:JH_HK}, and flagged accordingly in Tables~\ref{tab:cat_periodic} (column 5), \ref{tab:cat_binary} (column 3) and \ref{tab:cat_unknown_P} (column 2) in the Appendix.
Among the members, 25 MS and 37 PMS stars are periodic variable stars.


\section{Periodic variable members}
\label{Sect:PeriodicMembers}

\begin{figure}
\centering
\includegraphics[width=0.45\textwidth]{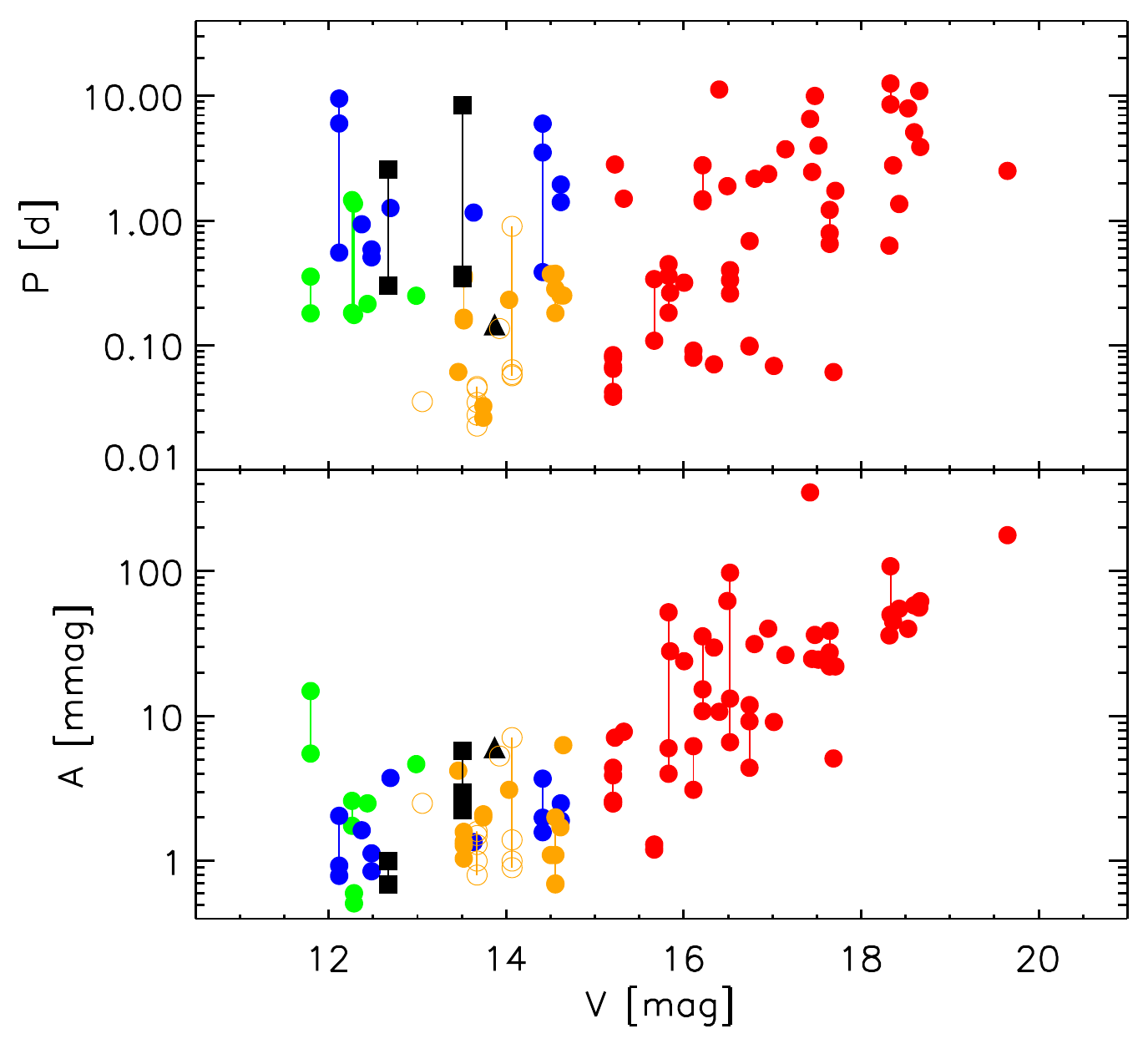}
\caption{Periods (top panel) and associated pulsation amplitudes (bottom panel) of the periodic member candidates of NGC~1893 as a function of their $V$ magnitude.
         Multiple periods of the same object are connected with solid lines.
         The $\beta$~Cep, SPB, FaRPB, and PMS candidates are marked with green, blue, orange, and red points, respectively.
         FaRPB candidates for which the position in the HR diagram (Fig.~\ref{fig:HR_MS}) could not be determined are marked with orange open circles.
         The black filled squares represent two Herbig Be candidates and the black filled triangle represents a classical Be candidate.
         }
\label{fig:A_P_V_members}
\end{figure}

\begin{figure}
\centering
\includegraphics[width=0.45\textwidth]{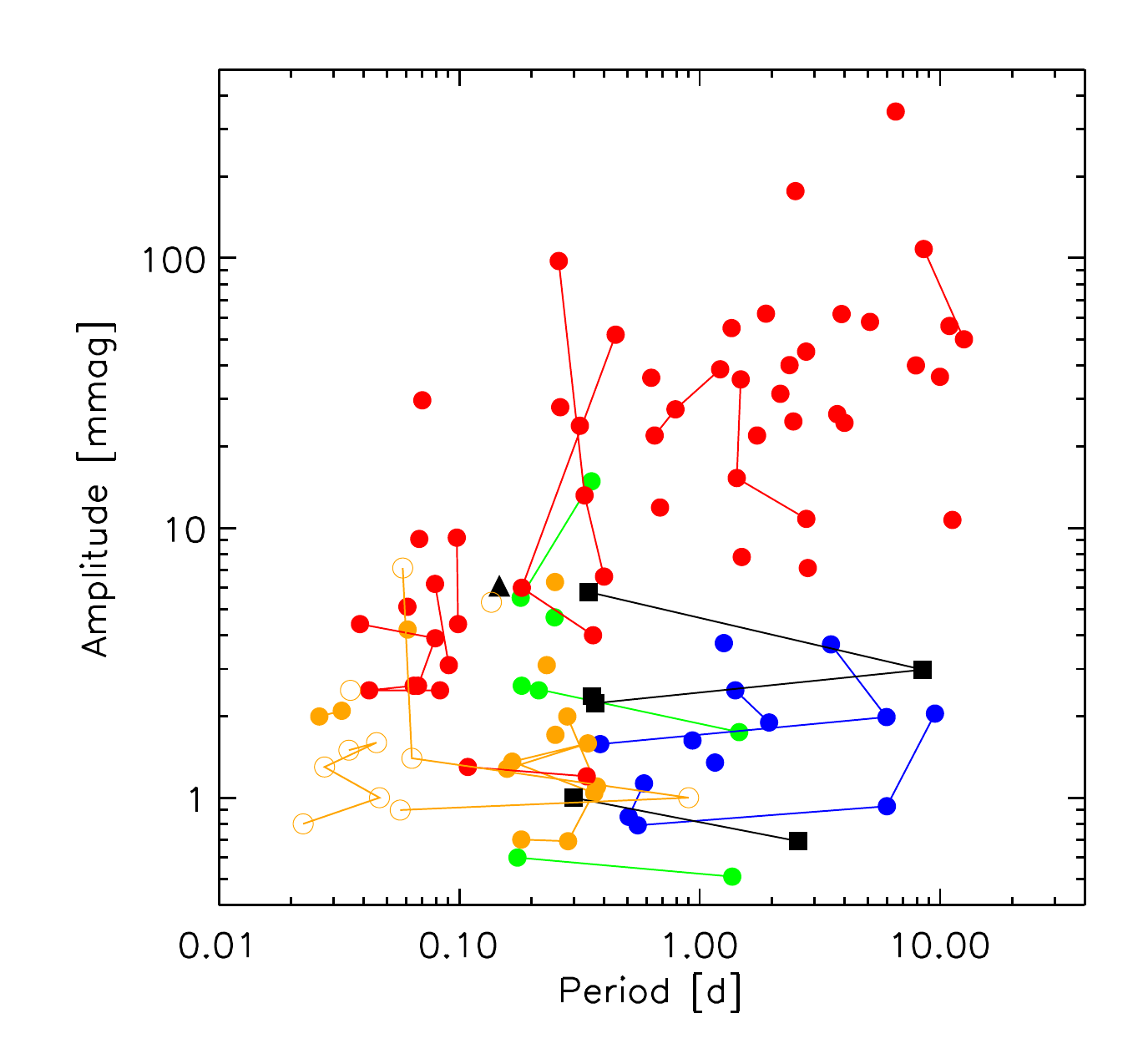}
\caption{Pulsation amplitude versus period of all periodic member candidates of NGC~1893.
         The symbols and colors have the same meaning as in Fig.~\ref{fig:A_P_V_members}.}
\label{fig:A_P_members}
\end{figure}

\begin{figure}
\centering
\includegraphics[width=0.45\textwidth]{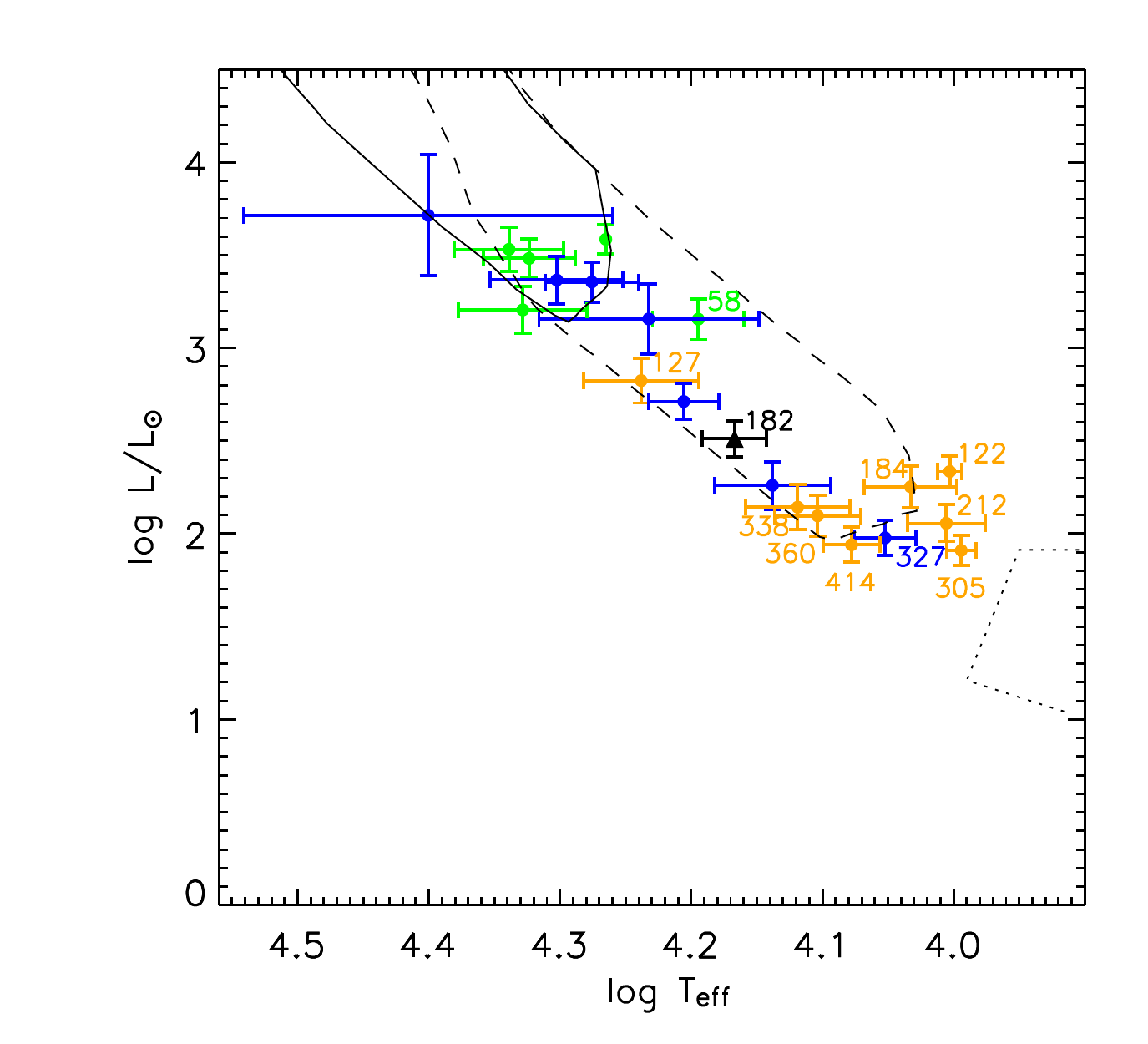}
\caption{HR diagram of the MS periodic member candidates of NGC~1893 having $U$, $B$ and $V$ data from which the effective temperature and luminosity are derived.
         The continuous and dashed lines delimit the theoretical instability strip of $\beta$~Cep and SPB, respectively, taken from \citet{Miglio2007}.
         The dotted line indicates the empirical $\delta$ Scuti instability strip from \citet{Balona2011}.
         The green, blue and orange points identify the $\beta$~Cep, SPB and FaRPB candidates, respectively, identified in this work, and the filled triangle identifies the classical Be star.
         See text for the details of the classification.
\label{fig:HR_MS}}
\end{figure}

Excluding eclipsing binaries which are presented in Sect.~6, 110 stars are found to be periodically variable with significant frequencies.
Based on the work presented in Sect.~4, 62 of them are members, with 25 MS members and 37 PMS members.
From now on, we limit the study to these 62 members.
Their periods and amplitudes are shown in Fig.~\ref{fig:A_P_V_members} as a function of their $V$ magnitudes, while the relation between amplitudes and periods is shown in Fig.~\ref{fig:A_P_members}.
The results in the figures are colour-coded according to the classification performed in this section.
The classification of the MS periodic members is presented in Sects.~\ref{Sect:betaCephei} to \ref{Sect:FaRPB}, and PMS stars are adressed in Sect.~\ref{Sect:PMS}.
The results of the classification are also given in Table~\ref{tab:cat_periodic}.

For the 25 MS periodic variables, classification is performed based on their periods and magnitudes, as well as on their position in the HR diagram when possible.
To estimate the location in the HR diagram, we follow the procedure based on the Q-method \citep{Moreno1975} that enables to estimate the intrinsic color index $(B-V)_0$ of early-type stars from their $U$, $B$ and $V$ magnitudes \citep[also used by][]{Lata2014}.
The method computes the quantity $Q=(U-B)-0.72\,(B-V)$, from  which the intrinsic color $(B-V)_0$ is derived through $(B-V)_0=0.339\,Q-0.007$.
It thus requires the knowledge of $U$, $B$ and $V$ magnitudes, which are available for 21 of our MS periodic variables.
The effective temperature $T_\mathrm{eff}$ is then derived from $(B-V)_0$ using the relations provided by \citet{Torres2010}.
The absolute bolometric magnitude $M_\mathrm{bol}$, on the other hand, is computed from $M_{bol}=M_V+BC$ taking a distance modulus $V_0-M_V=12.7\pm0.2$~mag \citep{Lim2014} and using bolometric corrections BC provided by \citet{Torres2010} based on $(B-V)_0$.
The luminosity then follows using $log(L/L_{\odot})=-0.4\,(M_{bol}-M_{bol, {\odot}})$.
The uncertainties on $T_\mathrm{eff}$ and $log(L/L_{\odot})$ are estimated based on the $U$, $B$ and $V$ uncertainties and on the uncertainty of the distance modulus.
The resulting locations in the HR diagram of the 21 MS periodic variables with available $U$, $B$ and $V$ photometry are shown in Fig.~\ref{fig:HR_MS}.

\subsection{$\beta$ Cephei candidates}
\label{Sect:betaCephei}

$\beta$ Cep stars are pulsating early-type B stars with periods between $\sim$0.08~d and $\sim$0.3~d \citep{Aerts2010}.
The predicted location of their instability strip in the HR diagram is indicated by the solid line in Fig.~\ref{fig:HR_MS}, as predicted by \citet{Miglio2007} for radial modes and non-radial p- and g-modes of degree $1 \le \ell \le 3$.
Four of our periodic candidates in this instability strip have their dominant periods between 0.17~d and 0.35~d (stars 23, 43, 50 and 80,  marked with green points in the figure).
They are consequently classified as $\beta$~Cep stars.
\citet{Marco2001} and \citet{Marco2002} estimated their spectral types to be B4, B3, B2 and B2, respectively. This is consistent with the spectral types of Galactic $\beta$~Cep stars catalogued by \citet{Stankov2005}, in which the majority have B0 to B2 spectral types, with a few $\beta$~Cep stars extending up to B5 type.

In addition to these four stars, we also classify star 58 ($P=0.21407$~d, $A=2.5$~mmag) as a $\beta$ Cep star.
Its effective temperature brings it slightly out of the $\beta$ Cep instability strip (see Fig.~\ref{fig:HR_MS}), but its luminosity and B4 spectral type reported by \citet{Marco2001} are compatible with $\beta$ Cep variables.

The five $\beta$ Cep candidates are marked in green in Figs.~\ref{fig:A_P_V_members} to \ref{fig:HR_MS}, and their folded light curves are shown in Fig.~\ref{fig:Beta_Cep_PD1}.

\subsection{Slowly pulsating B candidates}

Slowly pulsating B stars are pulsating with periods of the order of the day and up to a few days \citep{Stankov2005}.
Their instability strip in the HR diagram is shown by the dashed line in Fig.~\ref{fig:HR_MS}.
It partly overlaps with that of $\beta$~Cephei stars, and extends to fainter luminosities and cooler effective temperatures.

Seven stars (stars 40, 52, 55, 65, 149, 283, 327) locate in or very close to the SPB instability region with periods larger than 0.5~d.
We consequently classify them as SPB candidates.
They are marked with blue points in Figs.~\ref{fig:A_P_V_members} to \ref{fig:HR_MS}, and their folded light curves are shown in Fig.~\ref{fig:SPB}.
\citet{Marco2001} estimated the spectral types of stars 40, 55, 65, 149 and 283 as B2, B3, B5, B5 and B6, respectively, compatible with a SPB classification.

\subsection{Fast-rotating pulsating B candidates}
\label{Sect:FaRPB}

\citet{Degroote2009} and \citet{Mowlavi2013} reported the discovery of a new class of variable stars lying in the HR diagram between the red edge of SPB stars and the blue edge of $\delta$ Scuti stars, where no pulsation is predicted to occur based on standard stellar models.
\citet{Degroote2009} found the new variable stars in CoRoT data and described them as cool, short-term (i.e. $\beta$ Cep-like) candidate pulsators.
\citet{Mowlavi2013}, on the other hand, discovered them in an open cluster, NGC~3766, providing thereby a stronger evidence of their position in the color-magnitude diagram relative to SPB and $\delta$ Scuti stars.
A subsequent spectroscopic study of these objects enabled \citet{Mowlavi2016} to characterize them as fast-rotating pulsators.
They consequently named these new variables fast-rotating pulsating B (FaRPB) stars.
The characteristic signature of these objects is their short period (less than 0.55~d, clearly distinguishing them from SPB stars) while being fainter than $\beta$ Cephei stars.
In the sample of \citet{Mowlavi2013}, the periods range from 0.1~d to 0.55~d, with typical amplitudes between 1~mmag and 4~mmag.
These authors found many of them between the SPB and $\delta$~Scuti instability strips, but they could lie in the SPB instability strip as well.

Given the properties mentioned above, we identify nine FaRPB candidates from their periods shorter than 0.5~d and with magnitudes fainter than $\beta$ Cephei stars:
\begin{itemize}
\item six of them (stars 127, 212, 305, 338, 360 and 414) have periods between 0.1~d and 0.4~d, and amplitudes between 1~mmag and 7~mmag, in agreement with the ranges mentioned above for FaRPB stars.
The spectral types of stars 127, 338 and 360 derived by \citet{Marco2001} are B6, B8 and B9, respectively;
\vskip 2mm
\item two other FaRPB candidates (stars 122 and 184) have periods much shorter than 0.1~d, 0.0609429~d for the mono-periodic star 122, and 0.03244~d and 0.02616~d for the bi-periodic star 184;
\vskip 2mm
\item the last one, star 182, is classified as a classical Be star by \citet[][their star ID S3R1N4]{Marco2002}, with spectral type B1.5Ve.
Its low-resolution spectrum shows a strong H$\alpha$ emission line and a moderately reddened continuum.
\end{itemize}
These FaRPB candidates are shown in yellow filled circles in Figs.~\ref{fig:A_P_V_members} to \ref{fig:HR_MS}, except the classical Be star which is shown in black triangle.
They are further discussed in Sect.~\ref{Sect:Discussion_variables}.

Apart from these stars, there are four stars (stars 190, 240, 262 and 5269) in the magnitude range $V=13.5$~--~14.5~mag for which we could not find any data in the $U$ band, and therefore could not estimate the positions in the HR diagram.
They all have periods below 0.5~d, and we consequently classify them as FaRPB candidates.
They are marked in orange open circles in Figs.~\ref{fig:A_P_V_members} and \ref{fig:A_P_members}.
We note that three of them (stars 190, 262 and 5269) have periods shorter than 0.1~d, similarly to stars 122 and 184.

The folded light curves of the 13 FaRPB candidates are shown in Fig.~\ref{fig:classical_Be} (for the Be star) and Fig.~\ref{fig:NEW} (for the other FaRPB candidates).

\subsection{Pre-main-sequence variable stars}
\label{Sect:PMS}

We find 37 PMS periodic variable stars in the cluster NGC~1893.
In the $V/B-V$, $U-B/B-V$ and $J-H/H-K$ diagrams, most of them are fainter than 16 mag with infrared excess.
The periods of these PMS variable stars range from 0.068 to 10.937~d and the amplitudes range from 1.3 to 177~mmag.
They are marked with red points in Figs.~\ref{fig:A_P_V_members} and \ref{fig:A_P_members}.
The PMS stars also include the two Herbig Be candidates mentioned in Sect.~\ref{Sect:CMD_TCD}, that are brighter than 16~mag.
They are marked with black filled squares in Figs.~\ref{fig:A_P_V_members} and \ref{fig:A_P_members}.
The folded light curves of the 35 PMS variable stars are shown in Fig.~\ref{fig:PMS} and the ones of the two Herbig Be candidates in Fig.~\ref{fig:HerbigAeBe}.


\section{Eclipsing binaries}

We discover 15 eclipsing binaries in the field of view of NGC~1893.
Seven of them are members of the cluster.
Based on the shape of their folded light curves, shown in Fig.~\ref{binary_PD}, we classify six of them as Algol-type (EA), four as $\beta$ Lyrae-type (EB) and five as W Ursae Majoris-type (EW) binaries.
Their periods, memberships and classifications are listed in Table~\ref{tab:cat_binary}.

\begin{figure}
\centering
\includegraphics[width=0.45\textwidth]{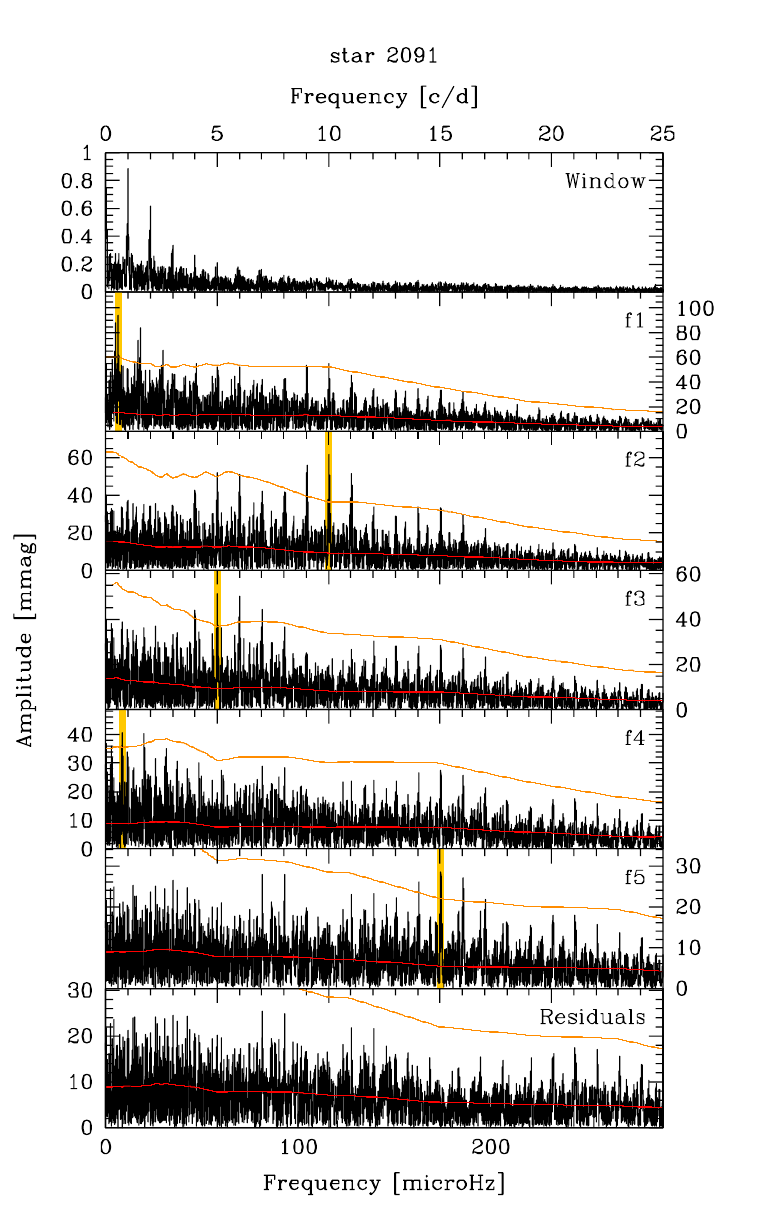}
\caption{Same as Fig.~\ref{fig:periodogram}, but for star 2091.
        }
\label{fig:star2091}
\end{figure}

One of the eclipsing binary candidates, star 2091 at $V=17.72$~mag has a rich periodogram, as shown in Fig.~\ref{fig:star2091}.
The frequencies extracted from the successive pre-whitening procedure reveals five significant frequencies which are listed in Table~\ref{tab:freq_star2091}, three of them being independent.
The first frequency ($f_1=0.5633$~c/d) is associated to the orbital period of the binary system, which is actually twice the period extracted from the periodogram.
The corresponding light curve folded with $P_\mathrm{orb}=3.55$~d, shown in Fig.~\ref{binary_PD}, supports the binary nature.
The additional clear signal at the high frequency of $f_3=5.0131$~c/d, with harmonics detected at $f_2=2\,f_3$ and $f_5=3\,f_3$, likely originates from the pulsation of one of the companions in the binary system, that could be a $\delta$~Scuti star.
We finally note a third independent frequency, at $f_4=0.7499$~c/d, detected with a S/N of 4.6.
The origin of this signal is unclear, and we would suggest additional photometric observations to conclude on it.

\begin{table}
\caption[]{\label{tab:freq_star2091} Multi-frequency solution of star 2091. The last-digit errors of the frequencies and amplitudes are given in parentheses.}
\begin{tabular}{lccc}
\hline \hline
No. & Frequency (c/d) & Amplitude (mmag) & S/N \\
\hline
$f_1$ & 0.5633(3) & 107(5) & 6.3 \\
$f_2$ &10.0281(5)&  61(6) & 6.7 \\
$f_3$ &5.0131(6) &  51(6) & 5.5 \\
$f_4$ &0.7499(8) &  42(5) & 4.6 \\
$f_5$ &15.040(1) &  26(6) & 5.2 \\
\hline
\end{tabular}
\end{table}


\section{Non-periodic variable stars}

From a visual check of the light curves, we identify 22 non-periodic variable stars.
They are listed in Table~\ref{tab:cat_unknown_P}.

We classify the non-periodic variable stars in two categories.
The first category, called irregular variables, comprises nine stars that show large-amplitude (above 1~mag) irregular variability over an observation time span of 90 days.
These variations are detected from the observations done during the 31~nights of the first observation run listed in Table~\ref{tab:obs_log1}, that span three months.
Their light curves are shown in Figure~\ref{fig:unknown_lc_1}.
Their large-amplitude irregular variations may be due to variable obscuration by circumstellar dust \citep{Herbst1994}.
Among them, stars 2697, 2882, 2936, 4162, 4671 and 5158 are members of the cluster, and consequently classified as PMS irregular variable stars.

In the second category of non-periodic variables, we gather 13 stars for which the periodic nature cannot be checked due to an insufficient number of measurements.
The stars are positioned far from the center of the cluster on the sky, and have been observed only during observation runs 6 and 7 (see Table~\ref{tab:obs_log1}).
Figure~\ref{fig:unknown_lc_2} shows their light curves obtained each night during observation run 6, from the Xinglong 85 cm telescope in 2014.
We label these stars as of unknown type.
Among them, only one (star 197) is classified as a MS star of the cluster.
The other ones (stars 441, 1069, 1098, 1301, 1961, 2640, 4056 and 4716) are classified as PMS stars of the cluster.


\section{Discussion}

\subsection{Comparison with previous works}

\begin{figure}
\centering
\includegraphics[width=0.45\textwidth]{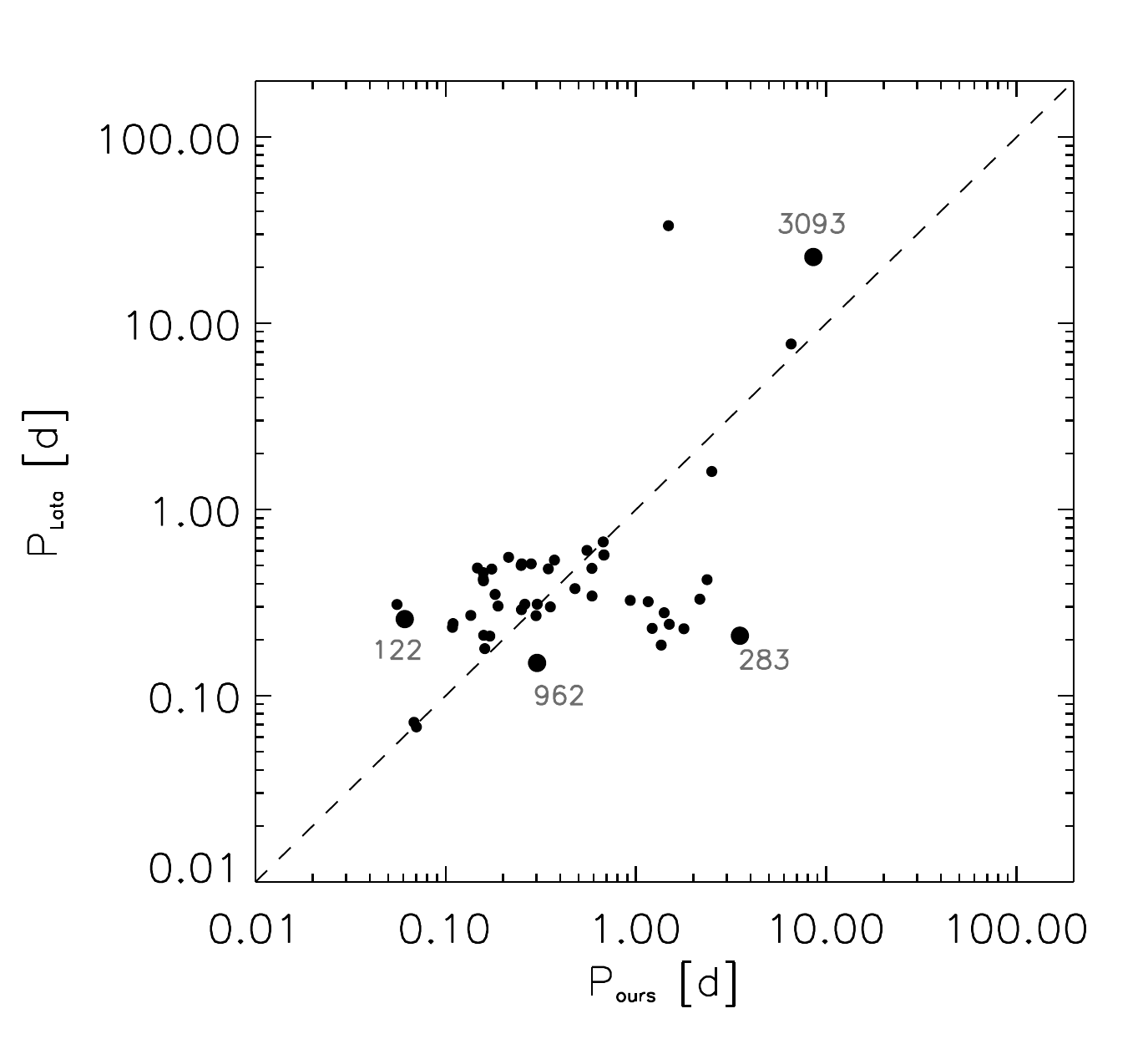}
\caption{Comparison of the periods of periodic variable stars between this work and \citet{Lata2012,Lata2014}. \label{fig:comparison_p}}
\end{figure}

\begin{figure}
\centering
\subfigure{\includegraphics[width=0.45\textwidth]{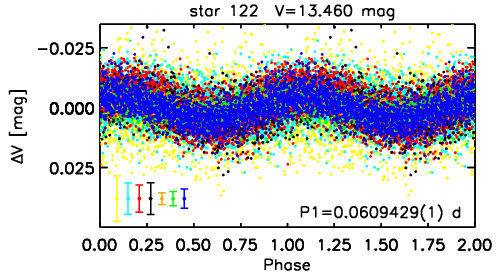}}
\subfigure{\includegraphics[width=0.45\textwidth]{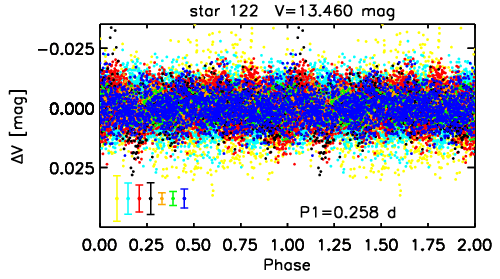}}
\caption{Folded light curves of star 122 with periods from this work (top panel) and from \citet{Lata2014} (bottom panel).
        }
\label{fig:star122_V154}
\end{figure}

Variable stars in NGC~1893 are mainly reported by \citet{Lata2012} and \citet{Lata2014}. These authors detected 157 variable stars in a $13 \times 13$ arcmin$^2$ FOV around the center of the cluster based on 16 nights of observations spread over about 5 years.
We therefore compare in this section our results with theirs.
In the same FOV, we find 99 variable stars, of which 62 stars are common to both works.

Among the 62 variable stars in common between our work and \citet{Lata2012, Lata2014}, 48 stars are found to be periodic in both works, one star is found to be periodic in this work only, and 13 stars are reported to be periodic in \citet{Lata2012, Lata2014} but not confirmed in this work.
The periods of the subsample found to be periodic in both works are compared in Fig.~\ref{fig:comparison_p}.
Some stars are seen to have consistent periods between the two works, but some do not.
It is not always easy to explain the differences, and additional observations may be required in some cases.
We nevertheless discuss here four examples, highlighted in Fig.~\ref{fig:comparison_p}.
\begin{itemize}
\item Star 122 ($V=13.460$~mag) has a period of 0.0609429~d in our work with an amplitude of 4.2~mmag.
The signal is very significant in our time series, with a S/N of 16.2, and is clearly visible in the folded light curve shown in the top panel of Fig.~\ref{fig:star122_V154}.
We however do not confirm the signal at 0.258~d found by \citet{Lata2014} (their star V154) with an amplitude of 8~mmag.
Our light curve folded with their period, shown in the bottom panel of Fig.~\ref{fig:star122_V154}, does not reveal a periodicity at this period.
\vskip 2mm
\item Star 283 (V=14.411~mag) has a period of 3.5162~d with an amplitude of 3.7~mmag (S/N=5.9) in our work, which is more than ten times longer than the period of 0.21~d found by \citet[][their star V142]{Lata2014}.
The analysis of our time series reveals the star to be multi-periodic (see Table~\ref{tab:cat_periodic}), with a additional short period at 0.385699~d ($A$=1.6~mmag, S/N=4.2) and a longer one at 5.989~d ($A$=2~mmag, S/N=4.1).
The folded light curves are shown in Fig.~\ref{fig:SPB}.
It may thus be that the multi-periodic nature of the star combined with the relatively sparse observations of \citet{Lata2014} made the detection difficult in their time series.
\vskip 2mm
\item A similar conclusion may hold for star 3093 (V=18.33~mag).
Two periods are detected in our time series, 8.5513~d ($A$=108~mmag, S/N=5.6) and 12.572~d ($A$=50~mmag, S/N=4.4), which are clearly visible in the folded light curves shown in Fig.~\ref{fig:PMS}.
\citet{Lata2012}, on the other hand, report only one period for this star (labelled V11 in their work) at 22.7~d.
Their folded light curve shown in their Fig.~10 indicates a relatively small number of observations, which may have prevented a correct identification of the periods of this multi-periodic star.
\vskip 2mm
\item Finally, the fourth example is star 962.
It is an eclipsing binary with a period of 0.30612~d (see its folded light curve in Fig.~\ref{binary_PD}).
The period found by \citet{Lata2012} is 0.15~d (their star V14), i.e. half this period.
\end{itemize}

Apart from the 62 variable stars in common between our work and \citet{Lata2012, Lata2014}, we find 37 variable stars in the FOV of \citet{Lata2012, Lata2014} that are not reported by these authors.
They are uniformly distributed across the FOV.

\begin{figure}
\centering
\includegraphics[width=0.45\textwidth]{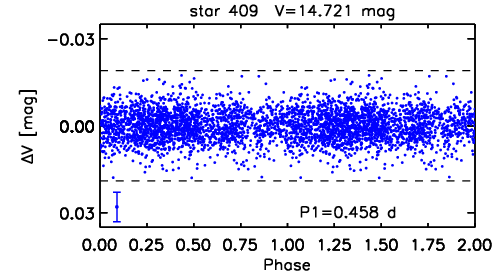}
\caption{Folded light curve of star 409 (labelled V98 in \citet{Lata2014}) with period from \citet{Lata2014}. The blue points represent the observations of group 1 which have higher photometric accuracy than other groups. The dashed lines represent the amplitude of 0.019 mag reported by \citet{Lata2014}. One can see that the photometric variability amplitudes in our time series are smaller than the amplitude reported by \citet{Lata2014}.
        }
\label{fig:star409_V98}
\end{figure}

But we do not confirm the variable nature of 95 other stars that are reported to be variable by these authors.
Some of them are too faint to secure a reliable variability detection in our time series (e.g. stars V5 and V27 in \citet{Lata2012, Lata2014} with $V$=20.21 and 19.77~mag, respectively), and we do not consider them in our comparison.
For many of the other ones, the  photometric variability amplitudes in our time series are smaller than the amplitudes reported by \citet{Lata2012, Lata2014}.
An example is given by star 409 (their star V98), of which our light curve folded with their period is shown in Fig.~\ref{fig:star409_V98}, with the variability amplitude reported by these authors indicated by the horizontal dashed lines.
We don't know what could be the origin of the variability reported by these authors, knowing that the epochs of their observations (2007-2013) overlap with ours (2008-2017).
We note that at least some of the variable stars reported by them but not by us are close to a bright star.
Additional precise observations would be required to further characterize the variability of these stars.

Outside the FOV of \citet{Lata2012,Lata2014}, we discovered 48 new variable stars.
In total, thus, we find 85 variable stars that have no counterparts in \citet{Lata2012, Lata2014}.
We consider these objects to be newly discovered variable stars, as indicated in the last columns of Table~\ref{tab:cat_periodic}, \ref{tab:cat_binary} and \ref{tab:cat_unknown_P}.

\subsection{Variable star content of the cluster}
\label{Sect:Discussion_variables}

Among the main sequence B-type variable stars that are members of NGC~1893, we identified in Sect.~\ref{Sect:PeriodicMembers} five $\beta$ Cephei stars, seven SPB stars and thirteen FaRPB stars.
We highlight here some properties of the FaRPB stars detected in this study.
\begin{itemize}
\item Most of the FaRPB stars are close to the faint border of the SPB instability strip, with some of them inside the instability strip, but some outside, between the red edge of SPB stars and the blue edge of $\delta$ Scuti stars.
This is in agreement with the stars being fast rotators \citep[e.g.][]{SalmonMontalbanReese_etal14,SaioEkstroemMowlavi_etal17}.
\vskip 2mm
\item The majority of them have periods between 0.1~d and 0.5~d, as found by \citet{Mowlavi2013}.
However, five of them have periods smaller than this range, between 0.03~d and 0.07~d.
They are located close to the faint border of the SPB instability strip.
This is unexpectedly low compared to both previous observations and model predictions.
We exclude them being $\delta$ Scuti stars given their magnitudes, because low-mass stars with an age of 1-5 Myr stars have not yet evolved to the MS stage \citep[see Figure~10 of][]{Palla1993}.
Additional observations, including spectral characterization, would be required to further characterize these short period FaRPB candidates.
\vskip 2mm
\item One of the FaRPB star, star 182, which lies well inside the SPB instability strip, is a Be star, as certified by the presence of emission lines in its spectrum \citep{Marco2002}.
We find it to be pulsating with a period of 0.1467416~d, in agreement with it being a FaRPB star rather than an SPB star.
This supports the suggestion by \citet{Mowlavi2016} of a link between FaRPB and Be stars.
\end{itemize}

The young open cluster NGC~1893 is a challenging case to study the nature of variable stars, due not only to the high differential reddening observed in its FOV, but also because the cluster may have had more than one birth episodes, or a continuous, still on-going, star formation history.
This was already noticed by \citet{MarcoNegueruelaSteele_02} and \citet{MarcoNegueruela_03} from the presence of massive Herbig Ae/Be stars in the cluster.
Our suggestion of the existence of an additional Herbig Be star, star 62 (see Sects.~\ref{Sect:CMD_TCD} and \ref{Sect:PMS}), would further support this view, if the PMS and Be natures of this star are confirmed by future spectra.


\section{Conclusions}

The young open cluster NGC~1893 is an active region which contains a lot of variable stars.
In this study we analyzed more than 20\,000 images taken over nine years, we detected 147 variable stars, including 110 periodic variable stars, 15 eclipsing binaries and 22 non-period variable stars.
We used kinematic membership probabilities and CMD+TCDs to explore the cluster membership of these variable stars, resulting in 84 member candidates.
These members are classified into periodic and non-period variable stars.
The periodic variable stars cover main-sequence variables ($\beta$ Cep, SPB, FaRPB stars) and pre-main-sequence variables.
We potentially find a new Herbig Be candidate, that is to be confirmed by spectroscopic observations, and a binary candidate harboring a pulsating star, that would deserve additional observations.

The thirteen FaRPB candidates have, for most of them, properties similar to the ones highlighted by \citet{Mowlavi2013} in the open cluster NGC~3766.
Moreover, one of the thirteen candidates in NGC~1893 is a known classical Be star, confirming its fast-rotating nature.
However, five of them are found with periods smaller than 0.1~d, down to 0.03~d, which would raise new challenges for their understanding.

The detection and analysis of multi-type variable stars in NGC~1893 yield valuable samples for the study of cluster asteroseismology, the properties of B type pulsators, stellar parameters of eclipsing binaries and the star formation process.
Future efforts of studying these samples, including multi-site time-series photometric observation campaigns and spectroscopic observations, may help to reveal the features of NGC~1893 and the nature of these variable members.


\section*{Acknowledgements}

JNF acknowledges the support from the National Natural Science Foundation of China (NSFC) through the grants 11673003 and 11833002 and the National Basic Research Program of China (973 Program 2014CB845700).
This work has been carried out within the framework of the National Centre for Competence in Research PlanetS supported by the Swiss National Science Foundation.
The authors acknowledge the financial support of the SNSF. We acknowledge the support of the staff of the Xinglong 85cm/50cm telescopes.
This work was partially supported by the Open Project Program of the Key Laboratory of Optical Astronomy, National Astronomical Observatories, Chinese Academy of Sciences.
We are thankful to all people who contributed to the monitoring campaign while performing the observation runs at the telescopes.
This publication makes use of data products from the Two Micron All Sky Survey, which is a joint project of the University of Massachusetts and the Infrared Processing and Analysis Center/California Institute of Technology, funded by the National Aeronautics and Space Administration and the National Science Foundation.



\appendix
\section{The photometric and kinematic data of variable stars}

\begin{figure*}
\centering
\subfigure{\includegraphics[width=0.45\textwidth]{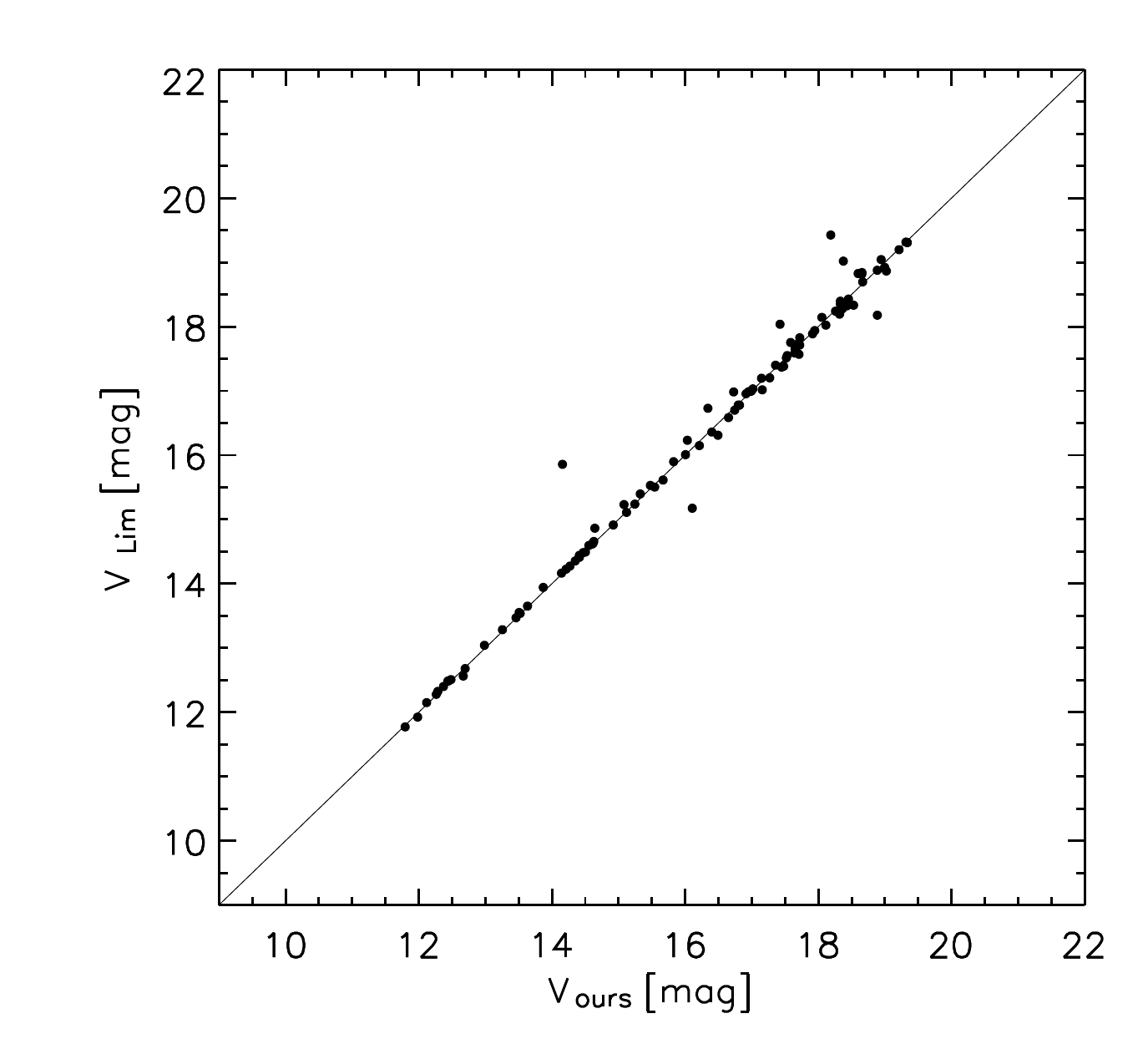}}
\subfigure{\includegraphics[width=0.45\textwidth]{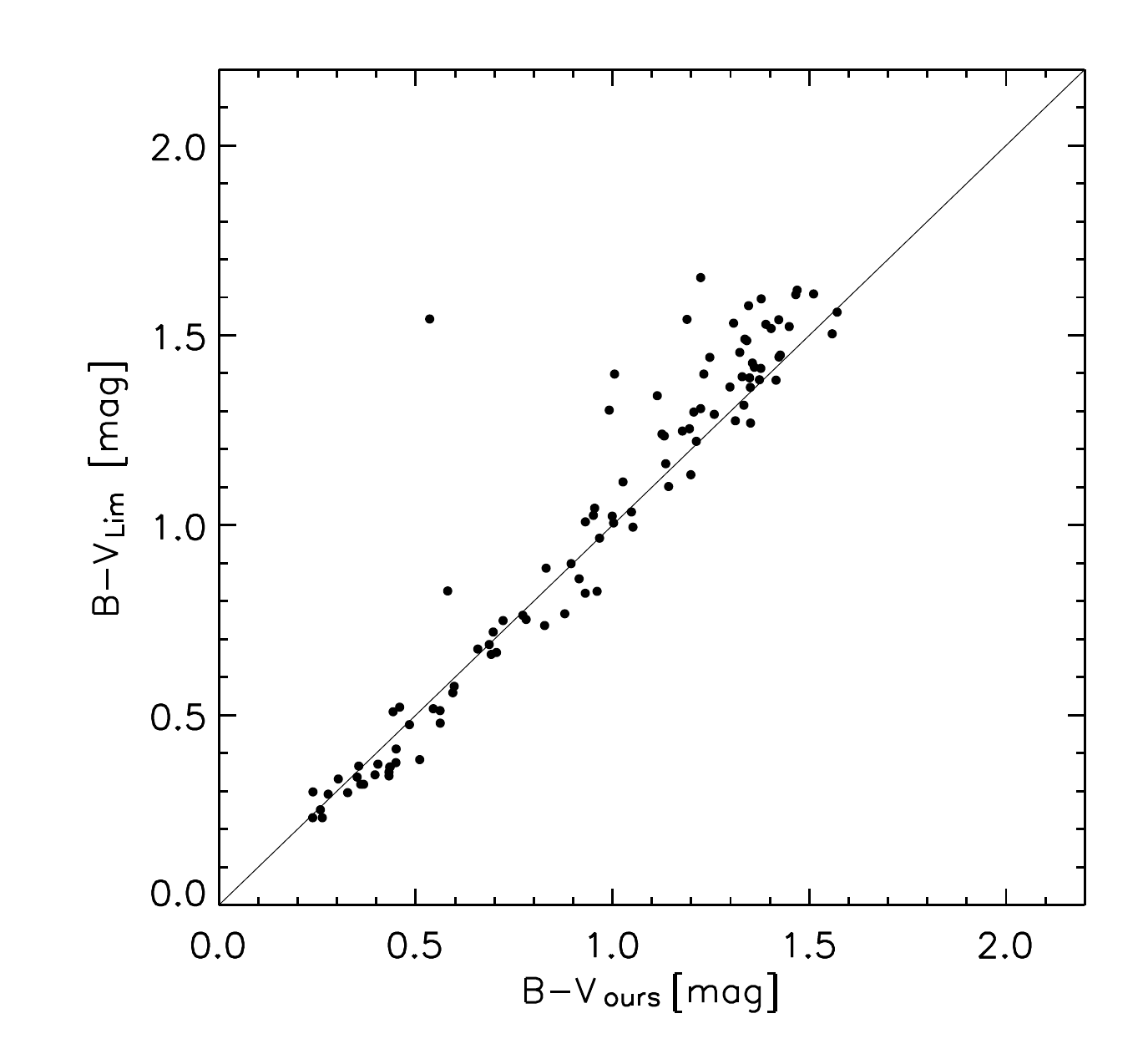}}
\caption{Comparison of our $V$, $B-V$ data with the ones from \citet{Lim2014} for 101 variable stars.}
\label{fig:comparison_photometry}
\end{figure*}

\begin{figure*}
\centering
\includegraphics[width=0.5\textwidth]{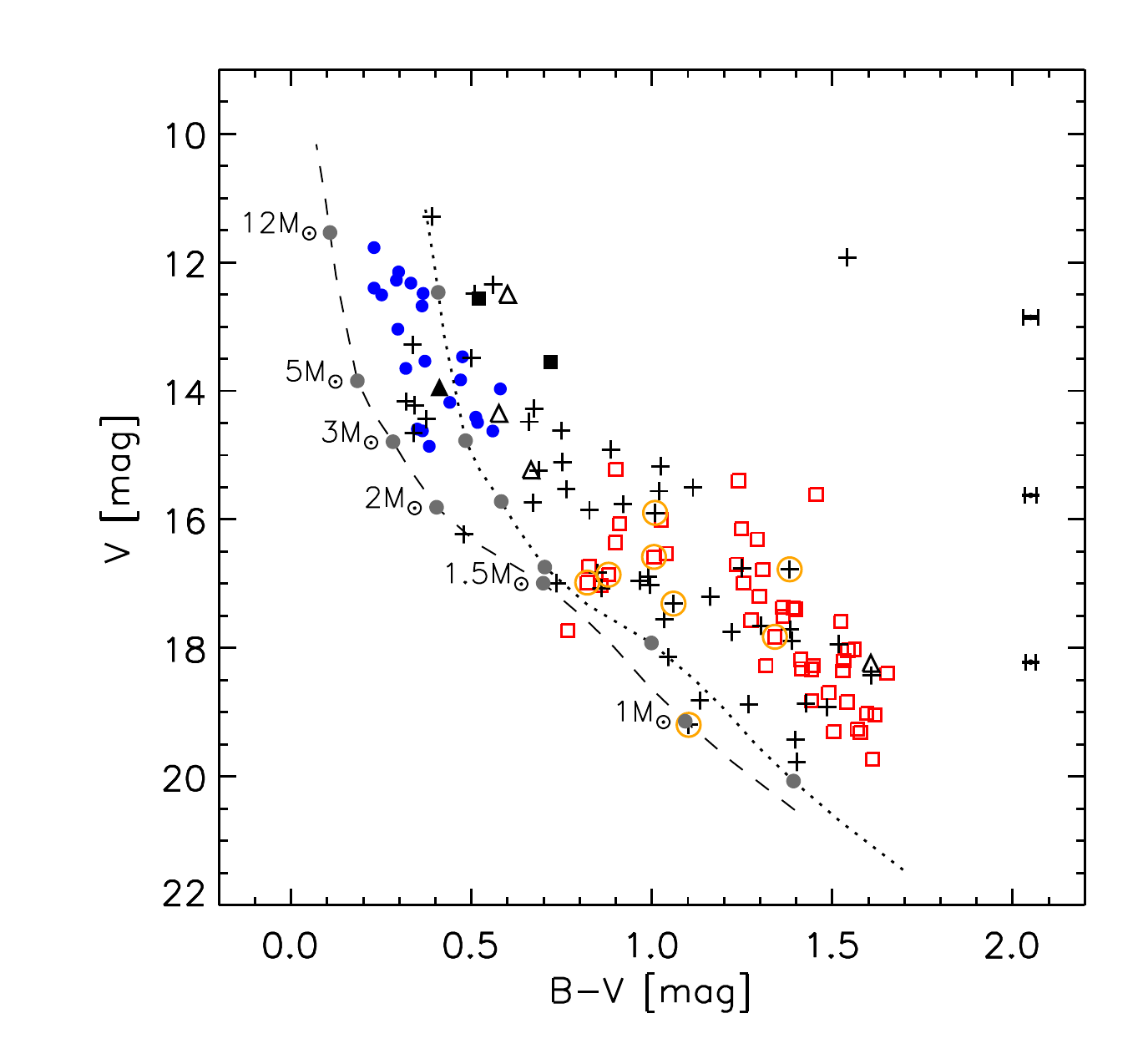}
\caption{Same as Fig.~\ref{fig:V_BV} in the main body of the text, but using the $V$ and $B-V$ data from \citet{Lim2014}.
         }
\label{fig:V_BV_Lim}
\end{figure*}

\clearpage
\onecolumn
\begin{landscape}\scriptsize
\begin{longtable}{ccccccccccccccc}
\caption{\label{tab:cat_variables} The photometric and kinematic data of variable stars in the field of view of NGC~1893. The second column denotes the WEBDA number (http://www.univie.ac.at/webda/). The third column denotes the identification number of variable stars given by \citet{Lata2012,Lata2014}.
The standardised $B$ and $V$ data are taken from the present work,
the $\mu_\alpha$ and $\mu_\delta$ data from the PPMXL catalog \citep{Roeser2010}, the radial velocity data from \citet{Xiang2017LAMOST}, the $B-V$ and $U-B$ data from \citet{Lim2014}, and the $J$, $H$, $K$ data from the 2MASS point source catalogue \citep{Cutri2003}.
}\\
\hline
\hline
Star ID &ID1 & ID2 & RA  & Dec & $B$   & $V$   & $\mu_\alpha$   & $\mu_\delta$   & Vr   & $U-B$ & $B-V$ & $J$   &  $H$ & $K$ \\
     {} &    &     & deg & deg & mag   & mag   & $mas\ yr^{-1}$ & $mas\ yr^{-1}$ & km/s & mag   & mag   & mag   & mag  & mag\\
\hline
\endfirsthead
\caption{continued.}\\
\hline
\hline
Star ID &ID1& ID2 & RA  & Dec & $B$   & $V$   & $\mu_\alpha$   & $\mu_\delta$   & Vr   & $U-B$ & $B-V$ & $J$   &  $H$ & $K$ \\
     {} &   &     &deg & deg & mag   & mag   & $mas\ yr^{-1}$ & $mas\ yr^{-1}$ & km/s & mag   & mag   & mag   & mag  & mag\\
\hline
\endhead
\hline
\endfoot
22 & 197  & V77 & 80.78103 & 33.49782 & 13.173$\pm$0.010 & 11.984$\pm$0.008 & 13.6$\pm$3.2 & -10.2$\pm$3.2 & $-$ & 1.574$\pm$0.011 & 1.542$\pm$0.015 & 9.197$\pm$0.022 & 8.457$\pm$0.026 & 8.295$\pm$0.018 \\
23 & 256  & V53 & 80.57688 & 33.47259 & 12.058$\pm$0.010 & 11.796$\pm$0.007 &  2.9$\pm$1.4 & -3.9$\pm$1.4 & $-$ & -0.450 & 0.230 & 11.128$\pm$0.019 & 11.035$\pm$0.021 & 10.978$\pm$0.019 \\
25 & 2372 & $-$ & 80.92893 & 33.38503 & 12.936$\pm$0.012 & 12.408$\pm$0.014 &  3.6$\pm$1.8 & -2.6$\pm$1.7 & $-$ & 0.270 & 0.600 & 11.179$\pm$0.022 & 10.950$\pm$0.024 & 10.870$\pm$0.021 \\
40 & 59   & V94 & 80.74172 & 33.44358 & 12.357$\pm$0.009 & 12.118$\pm$0.007 &  2.1$\pm$1.4 & -1.4$\pm$1.4 & $-$ & -0.545$\pm$0.029 & 0.298$\pm$0.046 & 11.460$\pm$0.022 & 11.320$\pm$0.025 & 11.261$\pm$0.023 \\
43 & 168  & V156& 80.64312 & 33.48914 & 12.589$\pm$0.009 & 12.285$\pm$0.007 &  1.7$\pm$1.4 & -1.9$\pm$1.4 & $-$ & -0.444$\pm$0.010 & 0.332$\pm$0.011 & 11.526$\pm$0.021 & 11.433$\pm$0.021 & 11.395$\pm$0.019 \\
50 & 33   & V52 & 80.68803 & 33.40670 & 12.541$\pm$0.010 & 12.263$\pm$0.007 &  1.0$\pm$1.4 & -7.3$\pm$1.4 & $-$ & -0.489$\pm$0.010 & 0.292$\pm$0.015 & 11.519$\pm$0.022 & 11.388$\pm$0.022 & 11.321$\pm$0.018 \\
52 & 228  & V97 & 80.74118 & 33.36895 & 12.741$\pm$0.010 & 12.483$\pm$0.007 & -0.6$\pm$1.4 & -4.2$\pm$1.4 & $-$ & -0.479$\pm$0.015 & 0.251$\pm$0.017 & 11.911$\pm$0.022 & 11.816$\pm$0.022 & 11.812$\pm$0.020 \\
54 & 38   & V41 & 80.66208 & 33.43215 & 12.893$\pm$0.010 & 12.451$\pm$0.007 & -0.3$\pm$1.4 & -5.6$\pm$1.4 & $-$ & 0.095$\pm$0.014 & 0.509$\pm$0.015 & 11.411$\pm$0.022 & 11.228$\pm$0.023 & 11.139$\pm$0.020 \\
55 & 140  & V112& 80.71899 & 33.38707 & 12.610$\pm$0.010 & 12.372$\pm$0.007 &  1.5$\pm$1.4 & -4.9$\pm$1.5 & $-$ & -0.463$\pm$0.008 & 0.230$\pm$0.017 & 11.822$\pm$0.022 & 11.760$\pm$0.022 & 11.713$\pm$0.020 \\
58 & 13   & V132& 80.69231 & 33.42260 & 12.792$\pm$0.019 & 12.437$\pm$0.015 & -1.0$\pm$1.8 & -4.1$\pm$1.8 & $-$ & -0.256$\pm$0.013 & 0.366$\pm$0.017 & 11.569$\pm$0.026 & 11.436$\pm$0.028 & 11.356$\pm$0.025 \\
62 & 196  & V73 & 80.78867 & 33.50070 & 13.127$\pm$0.010 & 12.668$\pm$0.008 &  4.2$\pm$1.7 & -4.3$\pm$1.7 & $-$ & -0.352$\pm$0.010 & 0.521$\pm$0.012 & 10.785$\pm$0.022 & 10.405$\pm$0.019 & 10.013$\pm$0.017 \\
65 & 62   & $-$ & 80.73958 & 33.43347 & 13.130$\pm$0.016 & 12.696$\pm$0.009 &  1.6$\pm$1.6 & -3.1$\pm$1.6 & $-$ & -0.312$\pm$0.030 & 0.363$\pm$0.033 & 11.609$\pm$0.023 & 11.435$\pm$0.025 & 11.291$\pm$0.022 \\
80 & 15   & V126& 80.69682 & 33.41887 & 13.313$\pm$0.021 & 12.986$\pm$0.022 & -5.7$\pm$2.3 & -7.8$\pm$2.3 & $-$ & -0.475$\pm$0.014 & 0.296$\pm$0.015 & 12.330$\pm$0.025 & 12.269$\pm$0.026 & 12.242$\pm$0.026 \\
96 & $-$  & $-$ & 80.38320 & 33.45217 & 13.809$\pm$0.009 & 13.231$\pm$0.012 &  8.2$\pm$4.2 &  2.1$\pm$4.2 & $-$ & $-$ & $-$ & 12.004$\pm$0.022 & 11.764$\pm$0.022 & 11.659$\pm$0.019 \\
101 & 123 & V78 & 80.78214 & 33.46735 & 13.607$\pm$0.006 & 13.256$\pm$0.008 & -9.9$\pm$5.6 & -20.6$\pm$5.6 & $-$ & -0.414$\pm$0.015 & 0.337$\pm$0.024 & 12.485$\pm$0.023 & 12.308$\pm$0.026 & 12.235$\pm$0.022 \\
122 & 4350& V154& 80.65286 & 33.41109 & 13.944$\pm$0.006 & 13.460$\pm$0.008 & 3.5$\pm$4.2 & -2.1$\pm$4.2 & -31$\pm$8 & 0.265$\pm$0.008 & 0.475$\pm$0.012 & 12.427$\pm$0.022 & 12.263$\pm$0.021 & 12.183$\pm$0.023 \\
127 & 77  & V140& 80.68340 & 33.38389 & 13.926$\pm$0.006 & 13.522$\pm$0.008 &-11.5$\pm$3.2 & -4.9$\pm$3.4 & $-$ & -0.314$\pm$0.015 & 0.371$\pm$0.018 & 12.619$\pm$0.022 & 12.446$\pm$0.023 & 12.356$\pm$0.025 \\
128 & 1463& $-$ & 80.49650 & 33.47517 & 13.933$\pm$0.005 & 13.430$\pm$0.008 & 4.4$\pm$5.6 & -8.5$\pm$5.6 & 5$\pm$4 & 0.16 & 0.5 & 12.422$\pm$0.022 & 12.268$\pm$0.021 & 12.164$\pm$0.020 \\
130 & 35  & V40 & 80.67923 & 33.41831 & 14.204$\pm$0.005 & 13.507$\pm$0.007 &-10.8$\pm$4.2 & -6.0$\pm$4.2 & $-$ & -0.125$\pm$0.009 & 0.719$\pm$0.011 & 11.649$\pm$0.022 & 11.312$\pm$0.022 & 10.922$\pm$0.020 \\
149 & 34  & V39 & 80.68460 & 33.41223 &14.001$\pm$0.006   &13.633$\pm$0.005  & -9.7$\pm$4.2& -2.3$\pm$4.2 & $-$  & -0.307$\pm$0.011 & 0.318$\pm$0.010  & 12.797$\pm$0.022 & 12.639$\pm$0.022 & 12.536$\pm$0.023\\
182 & 32  & V133& 80.69201 & 33.41621 & 14.319$\pm$0.014 & 13.869$\pm$0.011 & -3.0$\pm$20.9 & 10.0$\pm$13.1 & $-$ & -0.180$\pm$0.010 & 0.411$\pm$0.013 & 12.806$\pm$0.027 & 12.524 & 12.380 \\
184 & 2280& $-$ & 81.02486 & 33.44609 & 14.247$\pm$0.012 & 13.742$\pm$0.014 & 2.0$\pm$4.1 & -6.5$\pm$4.1 & $-$ & 0.160 & 0.470 & 12.842$\pm$0.023 & 12.682$\pm$0.024 & 12.587$\pm$0.022 \\
190 & $-$ & $-$ & 80.55212 & 33.61116 & 14.136$\pm$0.010 & 13.672$\pm$0.009 & 2.9$\pm$4.2 & -4.7$\pm$4.2 & $-$ & $-$ & $-$ & 12.738$\pm$0.022 & 12.539$\pm$0.022 & 12.443$\pm$0.023 \\
197 & 3264& $-$ & 81.02847 & 33.30689 & 14.552$\pm$0.008 & 13.966$\pm$0.009 & 8.5$\pm$4.1 & -4.4$\pm$4.1 & $-$ & 0.16 & 0.580 & 12.752$\pm$0.021 & 12.491$\pm$0.021 & 12.429$\pm$0.026 \\
212 & 1494& $-$ & 80.45012 & 33.45708 & 14.498$\pm$0.005 & 14.033$\pm$0.005 & -1.7$\pm$4.2 & 0.8$\pm$4.2 & $-$ & 0.23 & 0.440 & 13.114$\pm$0.023 & 12.960$\pm$0.023 & 12.893$\pm$0.030 \\
218 & 112 & V6  & 80.76758 & 33.47963 & 14.739$\pm$0.005 & 14.157$\pm$0.004 & 13.2$\pm$4.2 & -1.1$\pm$4.2 & $-$ & -0.161$\pm$0.002 & 0.827$\pm$0.008 & 12.010$\pm$0.028 & 11.120$\pm$0.022 & 10.249$\pm$0.034 \\
233 & 216 & V72 & 80.79527 & 33.39117 & 14.609$\pm$0.006 & 14.212$\pm$0.006 & -2.4$\pm$4.2 & -17.8$\pm$4.2 & $-$ & $-$ & $-$ & 13.258$\pm$0.027 & 13.105$\pm$0.032 & 13.060$\pm$0.036 \\
235 & 8   & V51 & 80.71320 & 33.42673 & 14.503$\pm$0.005 & 14.143$\pm$0.004 & 17.3$\pm$4.2 & 3.7$\pm$4.2 & $-$ & -0.308$\pm$0.014 & 0.318$\pm$0.019 & 13.443$\pm$0.030 & 13.327$\pm$0.035 & 13.223$\pm$0.037 \\
240 & $-$ & $-$ & 80.35389 & 33.59228 & 14.541$\pm$0.035 & 13.925$\pm$0.032 & 2.5$\pm$4.2 & -3.1$\pm$4.2 & $-$ & $-$ & $-$ & 12.580$\pm$0.023 & 12.334$\pm$0.023 & 12.250$\pm$0.022 \\
262 & $-$ & $-$ & 80.57247 & 33.58869 & 14.607$\pm$0.006 & 14.066$\pm$0.006 & 4.4$\pm$4.2 & -6.3$\pm$4.2 & $-$ & $-$ & $-$ & 13.015$\pm$0.023 & 12.848$\pm$0.023 & 12.721$\pm$0.027 \\
269 & 163 & V45 & 80.60909 & 33.47391 & 14.928$\pm$0.005 & 14.270$\pm$0.004 & 3.9$\pm$4.2 & -20.2$\pm$4.2 & $-$ & 0.456$\pm$0.014 & 0.674$\pm$0.015 & 12.746$\pm$0.060 & 12.508$\pm$0.025 & 12.423 \\
271 & 30  & V129& 80.69415 & 33.40377 & 14.947$\pm$0.006 & 14.349$\pm$0.005 & -9.3$\pm$5.4 & -14.1$\pm$5.4 & $-$ & 0.384$\pm$0.009 & 0.576$\pm$0.015 & 12.938$\pm$0.023 & 12.719$\pm$0.028 & 12.604$\pm$0.025 \\
283 & 81  & V142& 80.67296 & 33.38462 & 14.973$\pm$0.005 & 14.411$\pm$0.004 & 4.9$\pm$4.2 & -3.5$\pm$4.2 & $-$ & -0.057$\pm$0.022 & 0.512$\pm$0.023 & 13.179$\pm$0.022 & 13.001$\pm$0.023 & 12.859$\pm$0.027 \\
305 & 213 & V70 & 80.79815 & 33.40250 &15.047$\pm$0.006   &14.503$\pm$0.005  & 0.1$\pm$4.2 & -5.8$\pm$4.2 & $-$  & 0.329$\pm$0.012  & 0.517$\pm$0.013  & 13.183$\pm$0.024 & 12.902$\pm$0.025 & 12.786$\pm$0.026\\
313 & 204 & V65 & 80.80596 & 33.44969 & 15.158$\pm$0.005 & 14.466$\pm$0.005 & -2.0$\pm$4.2 & -13.7$\pm$4.2 & $-$ & 0.224$\pm$0.017 & 0.660$\pm$0.025 & 13.008$\pm$0.023 & 12.716$\pm$0.026 & 12.617$\pm$0.022 \\
325 & 292 & $-$ & 80.79351 & 33.52916 & 14.860$\pm$0.005 & 14.410$\pm$0.005 & 2.8$\pm$4.2 & -3.1$\pm$4.2 & $-$ & -0.174$\pm$0.015 & 0.375$\pm$0.018 & 13.412$\pm$0.024 & 13.252$\pm$0.026 & 13.185$\pm$0.030 \\
327 & 227 & V90 & 80.74676 & 33.36755 & 15.209$\pm$0.005 & 14.615$\pm$0.004 & 6.9$\pm$4.2 & -7.0$\pm$4.2 & $-$ & 0.168$\pm$0.016 & 0.559$\pm$0.021 & 13.411$\pm$0.023 & 13.209$\pm$0.025 & 13.125$\pm$0.029 \\
338 & 54  & V91 & 80.74239 & 33.45843 & 14.986$\pm$0.005 & 14.554$\pm$0.004 & 0.3$\pm$4.2 & -7.4$\pm$4.2 & $-$ & -0.137$\pm$0.019 & 0.350$\pm$0.026 & 13.700$\pm$0.027 & 13.529$\pm$0.030 & 13.432$\pm$0.036 \\
341 &$-$  & $-$ & 80.62017 & 33.63055 & 15.086$\pm$0.008 & 14.335$\pm$0.007 & 3.8$\pm$4.2 & 2.4$\pm$4.2 & $-$ & $-$ & $-$ & 12.748$\pm$0.022 & 12.446$\pm$0.022 & 12.303$\pm$0.023 \\
342 & 131 & V84 & 80.76639 & 33.39888 & 15.060$\pm$0.005 & 14.628$\pm$0.005 & 11.4$\pm$4.2 & -28.3$\pm$4.2 & $-$ & -0.079$\pm$0.011 & 0.340$\pm$0.018 & 13.763$\pm$0.028 & 13.663$\pm$0.038 & 13.640$\pm$0.041 \\
360 & 4   & V109& 80.72064 & 33.45624 & 15.046$\pm$0.005 & 14.611$\pm$0.004 & 2.2$\pm$4.2 & -8.9$\pm$4.2 & $-$ & -0.096$\pm$0.016 & 0.364$\pm$0.024 & 13.681$\pm$0.025 & 13.376$\pm$0.028 & 13.232$\pm$0.029 \\
368 & 91  & V149& 80.66039 & 33.49655 & 15.326$\pm$0.005 & 14.604$\pm$0.004 & 6.6$\pm$4.2 & -7.9$\pm$4.2 & $-$ & 0.243$\pm$0.010 & 0.749$\pm$0.012 & 13.100$\pm$0.022 & 12.757$\pm$0.023 & 12.636$\pm$0.023 \\
414 & 195 & V79 & 80.78025 & 33.51257 & 15.153$\pm$0.014 & 14.643$\pm$0.019 &-10.4$\pm$4.2 & 1.9$\pm$4.2 & $-$ & -0.024$\pm$0.012 & 0.383$\pm$0.018 & 13.519$\pm$0.030 & 13.416$\pm$0.044 & 13.271$\pm$0.047 \\
441 & $-$ & $-$ & 80.49708 & 33.12144 & 15.735$\pm$0.008 & 14.971$\pm$0.010 & 4.5$\pm$4.2 & -7.9$\pm$4.2 & -19$\pm$11 & $-$ & $-$ & 13.455$\pm$0.022 & 13.097$\pm$0.025 & 13.027$\pm$0.028 \\
472 & 20  & $-$ & 80.72466 & 33.45060 & 15.752$\pm$0.005 & 14.920$\pm$0.004 & 28.5$\pm$4.2 & -40.2$\pm$4.2 & $-$ & 0.355$\pm$0.016 & 0.887$\pm$0.022 & 13.260$\pm$0.024 & 12.852$\pm$0.024 & 12.723$\pm$0.023 \\
509  &249 & V59 & 80.60244 & 33.35910 &15.900$\pm$0.006   &15.120$\pm$0.005  & 7.2$\pm$4.2 & -10.3$\pm$4.2& $-$  & 0.278$\pm$0.030  & 0.752$\pm$0.029  & 13.576$\pm$0.026 & 13.218$\pm$0.028 & 13.128$\pm$0.032\\
528 & 201 & V1  & 80.79927 & 33.47937 & 15.787$\pm$0.006  & 15.081$\pm$0.006 & 7.7$\pm$4.2 & -4.8$\pm$4.2 & $-$  & 0.483$\pm$0.005  & 0.665$\pm$0.019  & 13.332$\pm$0.024 & 12.919$\pm$0.025 & 12.711$\pm$0.026\\
534 & 1131& $-$ & 80.79918 & 33.25738 & 16.127$\pm$0.006 & 15.225$\pm$0.005 & 5.4$\pm$4.2 & -3.9$\pm$4.2 & $-$ & 0.410 & 0.900 & 13.412$\pm$0.022 & 12.958$\pm$0.022 & 12.840$\pm$0.025 \\
591 & 120 & $-$ & 80.77458 & 33.46494 & 15.931$\pm$0.006 & 15.244$\pm$0.005 & -2.1$\pm$4.2 & -10.1$\pm$4.2 & $-$ & 0.217$\pm$0.014 & 0.686$\pm$0.025 & 13.815$\pm$0.026 & 13.551$\pm$0.031 & 13.430$\pm$0.033 \\
599 & 16  & V124& 80.70089 & 33.41688 & 16.451$\pm$0.015 & 15.325$\pm$0.008 & -3.3$\pm$4.2 & -2.1$\pm$4.2 & $-$ & 0.512$\pm$0.020 & 1.240$\pm$0.021 & 12.771 & 12.147$\pm$0.026 & 11.971 \\
638 & $-$ & $-$ & 80.54939 & 33.57767 & 15.882$\pm$0.007 & 15.204$\pm$0.007 & 4.2$\pm$4.2 & -9.9$\pm$4.2 & $-$ & $-$ & $-$ & 13.772$\pm$0.023 & 13.495$\pm$0.026 & 13.381$\pm$0.032 \\
660 & 232 & V103& 80.73177 & 33.35004 & 16.569$\pm$0.012 & 15.542$\pm$0.007 & 16.0$\pm$4.2 & -20.3$\pm$4.2 & $-$ & 0.901$\pm$0.008 & 1.114$\pm$0.007 & 13.418$\pm$0.038 & 12.844 & 12.708 \\
666 & 1468& $-$ & 80.49009 & 33.28063 & 16.568$\pm$0.010 & 15.565$\pm$0.008 & 8.2$\pm$4.2 & -9.4$\pm$4.2 & $-$ & 0.650 & 1.020 & 13.213$\pm$0.021 & 12.825$\pm$0.024 & 12.621$\pm$0.025 \\
706 & $-$ & $-$ & 80.38970 & 33.37950 & 16.518$\pm$0.014 & 15.619$\pm$0.011 & 4.9$\pm$4.2 & -3.4$\pm$4.2 & 5$\pm$15 & $-$ & $-$ & 13.581$\pm$0.025 & 13.068$\pm$0.025 & 12.939$\pm$0.027 \\
710 & 265 & V54 & 80.61602 & 33.50862 & 16.250$\pm$0.008 & 15.478$\pm$0.007 & -1.2$\pm$4.2 & -8.5$\pm$4.2 & $-$ & 0.217$\pm$0.003 & 0.763$\pm$0.002 & 14.006$\pm$0.023 & 13.622$\pm$0.025 & 13.524$\pm$0.032 \\
721 & 1697& V80 & 80.78675 & 33.32696 & 16.991$\pm$0.010 & 15.668$\pm$0.006 & 0.8$\pm$4.2 & -2.5$\pm$4.2 & $-$ & 0.973$\pm$0.002 & 1.455$\pm$0.004 & 12.689$\pm$0.019 & 11.995$\pm$0.018 & 11.784$\pm$0.021 \\
738 & 1492& $-$ & 80.45336 & 33.26993 & 16.633$\pm$0.012 & 15.759$\pm$0.010 & 3.2$\pm$4.2 & -8.6$\pm$4.2 & -26$\pm$12 & 0.280 & 0.920 & 13.745$\pm$0.022 & 13.344$\pm$0.025 & 13.236$\pm$0.033 \\
741 & $-$ & $-$ & 80.49465 & 33.18248 & 16.729$\pm$0.010 & 15.829$\pm$0.009 & -2.7$\pm$4.2 & -3.9$\pm$4.2 & $-$ & $-$ & $-$ & 13.950$\pm$0.024 & 13.434$\pm$0.031 & 13.301$\pm$0.035 \\
824 & $-$ & $-$ & 80.47082 & 33.08112 & 16.644$\pm$0.016 & 15.847$\pm$0.019 & 5.1$\pm$4.2 & -5.5$\pm$4.2 & $-$ & $-$ & $-$ & 14.200$\pm$0.029 & 13.850$\pm$0.031 & 13.810$\pm$0.047 \\
844 & 2351& $-$ & 80.96228 & 33.50493 & 16.440$\pm$0.010 & 15.693$\pm$0.010 & 4.7$\pm$4.1 & -4.6$\pm$4.1 & -6$\pm$14 & 0.33 & 0.670 & 14.272$\pm$0.031 & 14.071$\pm$0.037 & 13.815$\pm$0.049 \\
861 & 1847& $-$ & 80.65413 & 33.28915 & 16.956$\pm$0.011 & 16.005$\pm$0.007 & 0.1$\pm$4.2 & -5.8$\pm$4.2 & $-$ & 0.459$\pm$0.022 & 1.026$\pm$0.020 & 13.946$\pm$0.024 & 13.415$\pm$0.026 & 13.213$\pm$0.028 \\
922 & 7   & V22 & 80.71259 & 33.42986 & 17.106$\pm$0.015 & 16.106$\pm$0.015 & 2.3$\pm$4.2 & 8.7$\pm$4.2 & $-$ & 0.378$\pm$0.003 & 1.024$\pm$0.010 & 13.079$\pm$0.032 & 12.339$\pm$0.032 & 11.747$\pm$0.026 \\
926 & 69  & $-$ & 80.71622 & 33.40000 & 16.595$\pm$0.008 & 16.033$\pm$0.006 & -1.1$\pm$4.2 & -5.3$\pm$4.2 & $-$ & 0.315$\pm$0.007 & 0.479$\pm$0.006 & 14.607$\pm$0.033 & 14.214$\pm$0.047 & 13.654$\pm$0.037 \\
962 & 55  & V14 & 80.74806 & 33.45908 & 16.757$\pm$0.010 & 15.826$\pm$0.007 & 4.4$\pm$4.2 & -4.2$\pm$4.2 & $-$ & 0.552$\pm$0.009 & 1.009$\pm$0.010 & 13.917$\pm$0.030 & 13.468$\pm$0.030 & 13.289$\pm$0.035 \\
986 & 400 & V23 & 80.70495 & 33.43023 & 17.390$\pm$0.030 & 16.213$\pm$0.015 & 8.7$\pm$8.4 & -6.1$\pm$8.4 & $-$ & 0.631$\pm$0.003 & 1.248$\pm$0.004 & 13.564$\pm$0.031 & 12.982$\pm$0.033 & 12.838$\pm$0.035 \\
1049 &1771& V110& 80.72223 & 33.39394 & 17.302$\pm$0.048 & 16.341$\pm$0.062 & 0.0$\pm$0.0 & $-$ & $-$ & 0.429$\pm$0.006 & 0.826$\pm$0.002 & 14.721$\pm$0.056 & 14.355$\pm$0.068 & 14.285$\pm$0.068 \\
1069 & $-$& $-$ & 80.68839 & 33.14482 & 17.289$\pm$0.014 & 16.380$\pm$0.013 & 4.3$\pm$4.2 & -9.3$\pm$4.2 & $-$ & $-$ & $-$ & 14.707$\pm$0.032 & 14.302$\pm$0.043 & 14.262$\pm$0.065 \\
1098 &1988& $-$ & 80.41248 & 33.44180 & 17.191$\pm$0.013 & 16.233$\pm$0.011 & 5.5$\pm$4.2 & -8.4$\pm$4.2 & $-$ & 0.340 & 0.91 & 14.106$\pm$0.024 & 13.708$\pm$0.032 & 13.566$\pm$0.037 \\
1101 & $-$& $-$ & 80.63457 & 33.61266 & 16.906$\pm$0.012 & 16.109$\pm$0.009 & 0.1$\pm$4.2 & -3.0$\pm$4.2 & $-$ & $-$ & $-$ & 14.452$\pm$0.030 & 14.138$\pm$0.038 & 14.079$\pm$0.051 \\
1114 & 18 & $-$ & 80.73024 & 33.42574 & 17.295$\pm$0.011 & 16.400$\pm$0.009 & 1.3$\pm$4.2 & -2.8$\pm$4.2 & $-$ & 0.387$\pm$0.003 & 0.899$\pm$0.009 & 14.683$\pm$0.036 & 14.252$\pm$0.044 & 14.139$\pm$0.056 \\
1202 & 183& $-$ & 80.72136 & 33.52127 & 17.751$\pm$0.018 & 16.492$\pm$0.011 & 0.3$\pm$4.2 & -7.6$\pm$4.2 & $-$ & 0.697$\pm$0.005 & 1.292$\pm$0.011 & 13.559$\pm$0.024 & 12.832$\pm$0.025 & 12.647$\pm$0.025 \\
1208 & 322& V36 & 80.78607 & 33.34782 & 17.654$\pm$0.015 & 16.651$\pm$0.010 & 10.0$\pm$4.2 & -6.1$\pm$4.2 & $-$ & 0.492$\pm$0.006 & 1.006$\pm$0.004 & 14.511$\pm$0.030 & 13.946$\pm$0.031 & 13.783$\pm$0.038 \\
1277 & 64 & V12 & 80.75554 & 33.43008 & 17.659$\pm$0.017 & 16.728$\pm$0.012 & 2.8$\pm$4.2 & -2.4$\pm$4.2 & $-$ & 0.247$\pm$0.009 & 0.821$\pm$0.008 & 14.770$\pm$0.047 & 14.357$\pm$0.048 & 14.239$\pm$0.056 \\
1287 &1604& $-$ & 80.86589 & 33.21299 & 17.709$\pm$0.018 & 16.814$\pm$0.014 & 1.4$\pm$4.2 & -4.7$\pm$4.2 & $-$ & 0.39 & 0.880 & 15.238$\pm$0.045 & 14.803$\pm$0.053 & 14.614$\pm$0.096 \\
1301 &1885& $-$ & 80.57849 & 33.23121 & 17.677$\pm$0.015 & 16.685$\pm$0.011 & 8.3$\pm$4.2 & -2.8$\pm$4.2 & $-$ & 0.34 & 1.040 & 14.552$\pm$0.030 & 14.074$\pm$0.037 & 13.920$\pm$0.044 \\
1326 & $-$& $-$ & 80.36099 & 33.50592 & 17.584$\pm$0.023 & 16.522$\pm$0.015 & 0.8$\pm$4.2 & -0.2$\pm$4.2 & $-$ & $-$ & $-$ & 14.297$\pm$0.029 & 13.729$\pm$0.032 & 13.546$\pm$0.037 \\
1376 &5462& $-$ & 80.67035 & 33.47260 & 18.231$\pm$0.057 & 16.815$\pm$0.026 & 31.7$\pm$4.3 & 5.3$\pm$4.3 & $-$ & 0.842$\pm$0.014 & 1.382$\pm$0.002 & 13.795$\pm$0.044 & 13.070$\pm$0.039 & 12.890$\pm$0.042 \\
1380 & 398& $-$ & 80.70840 & 33.43412 & 17.874$\pm$0.020 & 16.743$\pm$0.012 & 2.8$\pm$4.2 & -3.4$\pm$4.2 & $-$ & 0.600$\pm$0.005 & 1.235$\pm$0.005 & 14.122$\pm$0.027 & 13.460$\pm$0.028 & 13.267$\pm$0.029 \\
1388 &1773& $-$ & 80.72079 & 33.25965 & 17.672$\pm$0.015 & 16.849$\pm$0.012 & -8.5$\pm$4.3 & -2.7$\pm$4.3 & $-$ & 0.200 & 0.850 & 15.229$\pm$0.043 & 14.882$\pm$0.062 & 14.780$\pm$0.092 \\
1390 &2827& $-$ & 80.99255 & 33.45909 & 17.746$\pm$0.021 & 16.565$\pm$0.015 & -2.3$\pm$4.1 &-14.2$\pm$4.1 & $-$ & 1.100 & 1.250 & 14.216$\pm$0.029 & 13.588$\pm$0.030 & 13.348$\pm$0.035 \\
1425 & 111 &V44 & 80.75760 & 33.47390 & 18.021$\pm$0.020 & 16.797$\pm$0.013 & -1.8$\pm$4.2 & -5.6$\pm$4.2 & $-$ & 0.667$\pm$0.005 & 1.307$\pm$0.001 & 13.947$\pm$0.032 & 13.292$\pm$0.033 & 13.063$\pm$0.034 \\
1572 & 118 &V43 & 80.76350 & 33.46478 & 18.147$\pm$0.023 & 16.952$\pm$0.015 & 1.4$\pm$4.2 & -9.2$\pm$4.2 & $-$ & 0.650$\pm$0.004 & 1.254$\pm$0.004 & 14.066$\pm$0.029 & 13.407$\pm$0.029 & 13.204$\pm$0.030 \\
1577 & 393 &$-$ & 80.71161 & 33.47999 & 17.881$\pm$0.019 & 16.914$\pm$0.014 & -0.3$\pm$4.2 &-10.6$\pm$4.2 & $-$ & 0.495$\pm$0.020 & 0.966$\pm$0.019 & 14.979 & 14.469 & 14.527$\pm$0.084 \\
1590 & $-$& $-$ & 80.74907 & 33.62620 & 17.817$\pm$0.024 & 16.741$\pm$0.014 & -1.1$\pm$4.2 & -2.8$\pm$4.2 & $-$ & $-$ & $-$ & 14.218$\pm$0.030 & 13.763$\pm$0.034 & 13.616$\pm$0.040 \\
1620 &1688& V68 & 80.79996 & 33.44704 & 17.931$\pm$0.019 & 17.016$\pm$0.015 & 1.8$\pm$4.2 & -6.8$\pm$4.2 & $-$ & 0.470$\pm$0.010 & 0.859$\pm$0.013 & 15.068$\pm$0.047 & 14.587$\pm$0.052 & 14.510$\pm$0.072 \\
1630 &2875& $-$ & 80.94002 & 33.35492 & 17.988$\pm$0.021 & 17.070$\pm$0.018 & -0.5$\pm$4.2 & 2.3$\pm$4.2 & $-$ & 0.450 & 0.860 & 15.088$\pm$0.047 & 14.779$\pm$0.070 & 14.496$\pm$0.085 \\
1657 &2856& $-$ & 80.96145 & 33.51507 & 17.808$\pm$0.021 & 16.867$\pm$0.018 & 2.9$\pm$4.2 & -1.3$\pm$4.2 & $-$ & 0.130 & 0.990 & 14.823$\pm$0.037 & 14.438$\pm$0.048 & 14.346$\pm$0.080 \\
1664 & 454& $-$ & 80.82959 & 33.49251 & 17.819$\pm$0.021 & 16.991$\pm$0.017 & 0.6$\pm$4.2 & -0.9$\pm$4.2 & $-$ & 0.337$\pm$0.008 & 0.736$\pm$0.011 & 15.330$\pm$0.048 & 14.957$\pm$0.062 & 14.764$\pm$0.105 \\
1665 &5237& $-$ & 80.61248 & 33.32712 & 18.209$\pm$0.024 & 17.157$\pm$0.017 & 7.2$\pm$4.2 & -2.9$\pm$4.2 & $-$ & 0.454$\pm$0.014 & 0.995$\pm$0.014 & 14.695$\pm$0.036 & 14.313$\pm$0.049 & 14.084$\pm$0.049 \\
1810 &6271& $-$ & 80.76815 & 33.46161 & 18.352$\pm$0.028 & 17.146$\pm$0.017 & -0.5$\pm$4.2 &-10.7$\pm$4.2 & $-$ & 0.672$\pm$0.022 & 1.298$\pm$0.014 & 14.410$\pm$0.032 & 13.788$\pm$0.033 & 13.633$\pm$0.039 \\
1893 &$-$ & $-$ & 80.44761 & 33.47057 & 18.309$\pm$0.031 & 17.262$\pm$0.021 & 3.8$\pm$4.2 & -5.6$\pm$4.2 & $-$ & $-$ & $-$ & 14.870$\pm$0.037 & 14.504$\pm$0.052 & 14.161$\pm$0.053 \\
1908 &3836& $-$ & 81.05360 & 33.26864 & 18.255$\pm$0.032 & 17.322$\pm$0.027 & 4.5$\pm$4.3 & -2.5$\pm$4.3 & $-$ & 0.08 & 1.060 & 14.743$\pm$0.036 & 14.377$\pm$0.050 & 14.476$\pm$0.091 \\
1913 &5176& $-$ & 80.59277 & 33.46169 & 18.405$\pm$0.029 & 17.269$\pm$0.019 & 0.6$\pm$4.2 & -5.4$\pm$4.2 & -4$\pm$10 & 0.862$\pm$0.018 & 1.162$\pm$0.012 & 14.951$\pm$0.036 & 14.285$\pm$0.041 & 14.236$\pm$0.058 \\
1921 &$-$ & $-$ & 81.08671 & 33.20631 & 18.375$\pm$0.048 & 17.397$\pm$0.035 & 5.0$\pm$4.2 & -6.4$\pm$4.2 & $-$ & $-$ & $-$ & 15.218$\pm$0.048 & 14.852$\pm$0.065 & 14.515$\pm$0.091 \\
1961 &6237& V10 & 80.76309 & 33.45141 & 18.588$\pm$0.032 & 17.356$\pm$0.020 & 2.5$\pm$4.2 &-11.5$\pm$4.2 & $-$ & 0.925$\pm$0.024 & 1.398$\pm$0.009 & 14.218$\pm$0.031 & 13.359$\pm$0.030 & 12.925$\pm$0.027 \\
1994 &4064& $-$ & 80.67710 & 33.38725 & 18.798$\pm$0.036 & 17.447$\pm$0.020 & 0.2$\pm$4.2 & -0.2$\pm$4.2 & $-$ & 0.692$\pm$0.008 & 1.363$\pm$0.007 & 14.387$\pm$0.030 & 13.718$\pm$0.028 & 13.443$\pm$0.034 \\
2005 &6124& $-$ & 80.75001 & 33.42573 & 18.807$\pm$0.036 & 17.478$\pm$0.021 & 6.8$\pm$4.2 & -5.9$\pm$4.2 & $-$ & 0.974$\pm$0.003 & 1.391$\pm$0.015 & 14.392$\pm$0.032 & 13.678$\pm$0.033 & 13.408$\pm$0.033 \\
2038 &$-$ & $-$ & 80.66655 & 33.28352 & 18.580$\pm$0.031 & 17.532$\pm$0.022 & -0.9$\pm$4.2 & -9.4$\pm$4.2 & -27$\pm$14 & 0.479$\pm$0.010 & 1.035$\pm$0.006 & 15.032$\pm$0.042 & 14.647$\pm$0.064 & 14.384$\pm$0.068 \\
2051 &4049& $-$ & 80.68953 & 33.39325 & 18.817$\pm$0.043 & 17.518$\pm$0.022 & -1.0$\pm$4.2 & -5.2$\pm$4.2 & $-$ & 0.763$\pm$0.016 & 1.364$\pm$0.013 & 14.748$\pm$0.036 & 13.930 & 13.566 \\
2074 &4005& $-$ & 80.70669 & 33.44446 & 18.643$\pm$0.076 & 17.652$\pm$0.037 & 4.3$\pm$5.6 &-29.6$\pm$5.6 & $-$ & 0.740$\pm$0.038 & 1.303$\pm$0.030 & 14.866$\pm$0.035 & 14.150$\pm$0.035 & 13.995$\pm$0.049 \\
2091 &4253& $-$ & 80.68932 & 33.41576 & 18.834$\pm$0.060 & 17.720$\pm$0.033 &-11.5$\pm$5.6 &-16.4$\pm$5.6 & $-$ & 0.769$\pm$0.037 & 1.341$\pm$0.009 & 15.069$\pm$0.045 & 14.367$\pm$0.048 & 14.181$\pm$0.058 \\
2186 &4012& V47 & 80.70561 & 33.43216 & 19.094$\pm$0.065 & 17.645$\pm$0.048 & 10.6$\pm$5.6 & 3.0$\pm$5.6 & $-$ & 1.121$\pm$0.026 & 1.523$\pm$0.003 & 14.322$\pm$0.034 & 13.558$\pm$0.035 & 13.293$\pm$0.036 \\
2302 &4592& V35 & 80.69177 & 33.42439 & 17.959$\pm$0.193 & 17.424$\pm$0.180 & $-$ & $-$ & $-$ & 0.687$\pm$0.053 & 1.543$\pm$0.011 & 14.295$\pm$0.036 & 13.350$\pm$0.036 & 12.690$\pm$0.029 \\
2347 &6169& $-$ & 80.75483 & 33.45615 & 18.796$\pm$0.039 & 17.583$\pm$0.028 & 2.9$\pm$4.2 & -7.2$\pm$4.2 & $-$ & 0.658$\pm$0.019 & 1.221$\pm$0.003 & 14.857 & 14.330$\pm$0.057 & 13.961 \\
2352 &1739& $-$ & 80.75288 & 33.43571 & 18.567$\pm$0.032 & 17.688$\pm$0.025 & 0.6$\pm$4.6 & -5.3$\pm$4.6 & $-$ & 0.452$\pm$0.006 & 0.767$\pm$0.004 & 15.800$\pm$0.082 & 15.416$\pm$0.106 & 15.315$\pm$0.148 \\
2450 &4244& $-$ & 80.69378 & 33.41084 & 19.021$\pm$0.048 & 17.709$\pm$0.028 & 5.5$\pm$4.2 &-11.0$\pm$4.2 & $-$ & 0.408$\pm$0.089 & 1.275$\pm$0.039 & 14.514$\pm$0.036 & 13.774$\pm$0.038 & 13.557$\pm$0.046 \\
2466 &4030& $-$ & 80.69711 & 33.42807 & 19.093$\pm$0.056 & 17.719$\pm$0.030 &-20.7$\pm$5.4 & 4.3$\pm$5.4 & $-$ & 0.780$\pm$0.001 & 1.383$\pm$0.004 & 14.845$\pm$0.045 & 14.130$\pm$0.047 & 13.885$\pm$0.050 \\
2611 &4251& V134& 80.68968 & 33.43261 & 19.260$\pm$0.047 & 17.912$\pm$0.032 & -0.4$\pm$4.2 &-11.8$\pm$4.2 & $-$ & 0.839$\pm$0.031 & 1.388$\pm$0.002 & 15.016$\pm$0.045 & 14.184$\pm$0.050 & 14.084$\pm$0.060 \\
2640 &4082& V30 & 80.66157 & 33.36836 & 19.555$\pm$0.083 & 18.331$\pm$0.058 & -4.7$\pm$5.6& -9.0$\pm$5.6 & $-$  & 0.963$\pm$0.059  & 1.652$\pm$0.004  & 14.682$\pm$0.035 & 13.701$\pm$0.034 & 13.08$\pm$0.026 \\
2697 &5691& V21 & 80.71266 & 33.42868 & 19.683$\pm$0.087 & 18.112$\pm$0.055 & $-$ & $-$ & $-$ & 1.161$\pm$0.057 & 1.561$\pm$0.008 & 14.607$\pm$0.040 & 13.780$\pm$0.037 & 13.522$\pm$0.040 \\
2773 &4074& $-$ & 80.66609 & 33.42473 & 19.348$\pm$0.056 & 17.945$\pm$0.031 &-36.7$\pm$4.2 &-15.9$\pm$4.2 & $-$ & 0.965$\pm$0.035 & 1.518$\pm$0.005 & 14.700$\pm$0.033 & 14.023$\pm$0.038 & 13.664$\pm$0.041 \\
2777 &6408& V2  & 80.79125 & 33.48579 & 19.191$\pm$0.054 & 18.186$\pm$0.043 & 10.4$\pm$4.2& -3.9$\pm$4.2 & $-$  & -0.018$\pm$0.046 & 1.398$\pm$0.004  & 15.395$\pm$0.053 & 14.106$\pm$0.038 & 13.196$\pm$0.028\\
2793 &$-$ & $-$ & 80.95059 & 33.61221 & 18.645$\pm$0.061 & 17.609$\pm$0.045 & 2.9$\pm$4.6 & -3.6$\pm$4.6 & $-$ & $-$ & $-$ & 15.788$\pm$0.084 & 14.979$\pm$0.083 & 15.341$\pm$0.183 \\
2882 &6269& V7  & 80.76744 & 33.48289 & 19.752$\pm$0.090 & 18.375$\pm$0.049 & 0.0$\pm$0.0 & $-$ & $-$ & 0.673$\pm$0.069 & 1.596$\pm$0.007 & 14.306$\pm$0.031 & 13.318$\pm$0.029 & 12.603$\pm$0.023 \\
2906 &4238& $-$ & 80.69411 & 33.44947 & 19.006$\pm$0.059 & 18.052$\pm$0.043 & 4.5$\pm$4.8 &-26.3$\pm$4.8 & $-$ & 0.405$\pm$0.025 & 1.045$\pm$0.017 & 15.952$\pm$0.079 & 15.788$\pm$0.143 & 15.356$\pm$0.155 \\
2936 &5576& V26 & 80.69513 & 33.49117 & 19.671$\pm$0.081 & 18.338$\pm$0.048 & 4.6$\pm$4.2 & -6.5$\pm$4.2 & $-$ & 0.535$\pm$0.039 & 1.316$\pm$0.012 & 14.831$\pm$0.037 & 13.900$\pm$0.034 & 13.359$\pm$0.034 \\
2975 &5035& $-$ & 80.53977 & 33.35309 & 19.626$\pm$0.071 & 18.318$\pm$0.043 & 2.1$\pm$4.2 & -3.3$\pm$4.2 & $-$ & 1.286$\pm$0.073 & 1.532$\pm$0.039 & 15.073$\pm$0.039 & 14.481$\pm$0.052 & 14.193$\pm$0.058 \\
2986 &5473& $-$ & 80.67348 & 33.48870 & 19.851$\pm$0.146 & 18.652$\pm$0.097 & -6.0$\pm$4.8 &-29.6$\pm$4.8 & $-$ & 0.797$\pm$0.064 & 1.133$\pm$0.007 & 16.119$\pm$0.089 & 15.762$\pm$0.141 & 15.489$\pm$0.167 \\
3093 &6179& V11 & 80.75606 & 33.45472 & 19.720$\pm$0.079 & 18.330$\pm$0.047 & 9.9$\pm$5.6 &-15.8$\pm$5.6 & $-$ & 1.174$\pm$0.056 & 1.529$\pm$0.010 & 14.959$\pm$0.040 & 14.149$\pm$0.038 & 13.919$\pm$0.044 \\
3214 &5853& V106& 80.72558 & 33.41627 & 19.789$\pm$0.081 & 18.428$\pm$0.050 & 2.4$\pm$4.2 & -7.7$\pm$4.2 & $-$ & 0.883$\pm$0.046 & 1.416$\pm$0.017 & 15.149$\pm$0.045 & 14.463$\pm$0.054 & 14.197$\pm$0.051 \\
3215 &$-$ & $-$ & 80.96517 & 33.26581 & 19.576$\pm$0.073 & 18.424$\pm$0.056 & -0.3$\pm$4.7 & -7.4$\pm$4.7 & $-$ & $-$ & $-$ & 15.916$\pm$0.077 & 15.524$\pm$0.094 & 15.399$\pm$0.198 \\
3219 &4242& $-$ & 80.69316 & 33.43692 & 19.723$\pm$0.080 & 18.257$\pm$0.044 & 12.5$\pm$4.2 & -4.7$\pm$4.2 & $-$ & 1.749$\pm$0.121 & 1.607$\pm$0.011 & 14.867$\pm$0.038 & 14.001$\pm$0.036 & 13.712$\pm$0.039 \\
3343 &4664& $-$ & 80.69810 & 33.43507 & 19.963$\pm$0.096 & 18.452$\pm$0.049 & 1.7$\pm$5.6 & -5.2$\pm$5.6 & $-$ & 1.382$\pm$0.067 & 1.609$\pm$0.009 & 14.944$\pm$0.038 & 13.993$\pm$0.046 & 13.791$\pm$0.046 \\
3349 &6186& $-$ & 80.75661 & 33.45178 & 19.785$\pm$0.064 & 18.359$\pm$0.047 & -0.4$\pm$4.2 & -2.8$\pm$4.2 & $-$ & 0.953$\pm$0.042 & 1.448$\pm$0.017 & 14.893$\pm$0.043 & 14.224$\pm$0.049 & 13.958$\pm$0.056 \\
3521 &5404& $-$ & 80.65383 & 33.48617 & 20.001$\pm$0.110 & 18.665$\pm$0.088 & -3.1$\pm$5.7 & -5.8$\pm$5.6 & $-$ & 1.071$\pm$0.066 & 1.490$\pm$0.013 & 15.291$\pm$0.059 & 14.467$\pm$0.060 & 14.161$\pm$0.064 \\
3627 &5626& $-$ & 80.70557 & 33.48396 & 19.951$\pm$0.102 & 18.529$\pm$0.056 & 2.0$\pm$4.2 & -6.0$\pm$4.2 & $-$ & 0.996$\pm$0.045 & 1.443$\pm$0.010 & 14.992$\pm$0.041 & 14.060$\pm$0.038 & 13.565$\pm$0.035 \\
3680 &5856& $-$ & 80.72576 & 33.40875 & 20.076$\pm$0.099 & 18.654$\pm$0.060 & 8.7$\pm$4.2 & -3.9$\pm$4.2 & $-$ & 1.295$\pm$0.078 & 1.541$\pm$0.010 & 15.405$\pm$0.054 & 14.557$\pm$0.052 & 14.055$\pm$0.051 \\
3695 &6230& $-$ & 80.76179 & 33.47169 & 19.844$\pm$0.089 & 18.597$\pm$0.059 & $-$ & $-$ & $-$ & 0.786$\pm$0.057 & 1.442$\pm$0.011 & 15.101$\pm$0.045 & 14.215$\pm$0.051 & 13.542$\pm$0.043 \\
4056 &6242& V9  & 80.76362 & 33.49035 & 20.262$\pm$0.124 & 18.885$\pm$0.090 & -3.7$\pm$4.2& -10.3$\pm$4.2& $-$  & 0.839$\pm$0.043  & 1.413$\pm$0.008  & 15.463$\pm$0.058 & 14.41$\pm$0.048  & 13.722$\pm$0.041\\
4082 &$-$ & $-$ & 80.80659 & 33.61557 & 20.047$\pm$0.117 & 18.755$\pm$0.079 & 7.3$\pm$4.5 &-13.0$\pm$4.5 & $-$ & $-$ & $-$ & 15.905$\pm$0.077 & 15.199$\pm$0.085 & 15.164$\pm$0.119 \\
4162 &5890& V17 & 80.72951 & 33.46505 & 20.660$\pm$0.143 & 19.315$\pm$0.125 & 5.7$\pm$5.6 &-11.1$\pm$5.6 & $-$ & 0.639$\pm$0.076 & 1.578$\pm$0.019 & 14.626$\pm$0.035 & 13.762$\pm$0.034 & 13.242$\pm$0.034 \\
4225 &5990& $-$ & 80.73762 & 33.46090 & 20.336$\pm$0.118 & 18.995$\pm$0.082 & -1.4$\pm$5.7 & -7.0$\pm$5.7 & $-$ & 1.086$\pm$0.086 & 1.486$\pm$0.017 & 15.963$\pm$0.082 & 14.990$\pm$0.065 & 14.842$\pm$0.093 \\
4495 &$-$ & $-$ & 80.54976 & 33.53785 & 20.234$\pm$0.126 & 18.884$\pm$0.083 & $-$ & $-$ & $-$ & 0.609$\pm$0.047 & 1.269$\pm$0.053 & 15.874$\pm$0.074 & 15.208$\pm$0.097 & 15.156$\pm$0.120 \\
4496 &6017& $-$ & 80.73990 & 33.42256 & 20.375$\pm$0.108 & 19.020$\pm$0.083 &-191.6$\pm$5.7 & 73.1$\pm$5.7 & $-$ & 0.857$\pm$0.060 & 1.427$\pm$0.008 & 15.750$\pm$0.068 & 14.966$\pm$0.072 & 14.484$\pm$0.073 \\
4671 &4559& V25 & 80.69805 & 33.45616 & 20.242$\pm$0.127 & 19.044$\pm$0.078 & 7.6$\pm$4.2 & -7.2$\pm$4.2 & $-$ & $-$ & $-$ & 15.617$\pm$0.060 & 14.551$\pm$0.051 & 14.184$\pm$0.058 \\
4716 &6265& V8  & 80.76672 & 33.49677 & 20.411$\pm$0.128 & 18.942$\pm$0.086 &  3.8$\pm$5.6&  2.5$\pm$5.6 & $-$ & 0.684$\pm$0.085  & 1.619$\pm$0.011  & 15.03$\pm$0.041  & 13.947$\pm$0.035 & 13.15$\pm$0.029 \\
4975 &$-$ & $-$ & 81.00633 & 33.20406 & 20.670$\pm$0.153 & 19.511$\pm$0.135 & 10.2$\pm$4.5 & -1.9$\pm$4.5 & $-$ & $-$ & $-$ & 16.223$\pm$0.102 & 16.196$\pm$0.217 & 15.149 \\
5048 &$-$ & $-$ & 80.79720 & 33.37944 & 20.354$\pm$0.129 & 19.211$\pm$0.102 & 4.3$\pm$5.4 & -2.9$\pm$5.4 & $-$ & 0.411$\pm$0.041 & 1.102$\pm$0.014 & $-$ & $-$ & $-$ \\
5130 &$-$ & $-$ & 80.55485 & 33.29024 & 21.274$\pm$0.205 & 19.900$\pm$0.168 & 0.6$\pm$4.9 & -0.3$\pm$4.9 & $-$ & $-$ & $-$ & 16.503$\pm$0.117 & 15.976$\pm$0.176 & 15.573$\pm$0.188 \\
5156 &5741& V48 & 80.71637 & 33.47757 & 20.961$\pm$0.180 & 19.648$\pm$0.138 & -8.9$\pm$5.7 & 2.0$\pm$5.7 & $-$ & $-$ & $-$ & 16.174$\pm$0.098 & 15.233$\pm$0.088 & 14.926$\pm$0.106 \\
5158 &6301& $-$ & 80.77197 & 33.47667 & 20.894$\pm$0.184 & 19.336$\pm$0.112 & $-$ & $-$ & $-$ & 0.571$\pm$0.086 & 1.504$\pm$0.009 & 15.975$\pm$0.084 & 14.785$\pm$0.066 & 14.316$\pm$0.067 \\
5262 &2393& $-$ & 80.89488 & 33.65379 & 12.759$\pm$0.031 & 12.263$\pm$0.023 & 4.8$\pm$1.4 & -5.1$\pm$1.4 & -34$\pm$2 & 0.230 & 0.560 & 11.122$\pm$0.023 & 10.890$\pm$0.023 & 10.827$\pm$0.019 \\
5269 &$-$ & $-$ & 80.82678 & 33.16976 & 13.462$\pm$0.012 & 13.055$\pm$0.013 & 5.9$\pm$3.3 &-10.6$\pm$3.3 & $-$ & $-$ & $-$ & 12.222$\pm$0.021 & 12.078$\pm$0.021 & 11.999$\pm$0.022 \\
5282 &$-$ & $-$ & 80.45517 & 33.20046 & 12.755$\pm$0.013 & 12.231$\pm$0.012 & 22.9$\pm$9.5 & 60.7$\pm$9.5 & $-$ & $-$ & $-$ & 10.947$\pm$0.018 & 10.690$\pm$0.022 & 10.587$\pm$0.017 \\
5283 &1514& $-$ & 80.42102 & 33.28128 & 11.558$\pm$0.014 & 11.272$\pm$0.013 & 3.4$\pm$1.2 & -9.3$\pm$1.2 & $-$ & 0.270 & 0.390 & 10.347$\pm$0.018 & 10.194$\pm$0.021 & 10.131$\pm$0.017 \\
5284 &$-$ & $-$ & 80.63320 & 33.11755 & 13.119$\pm$0.015 & 12.671$\pm$0.019 &-11.0$\pm$3.2 &-11.9$\pm$3.2 & $-$ & $-$ & $-$ & 11.639$\pm$0.022 & 11.435$\pm$0.021 & 11.338$\pm$0.020 \\
5569 &$-$ & $-$ & 80.83760 & 33.13245 & 19.749$\pm$0.112 & 18.461$\pm$0.060 & 7.5$\pm$4.2 & -4.2$\pm$4.2 & $-$ & $-$ & $-$ & 15.346$\pm$0.051 & 15.085$\pm$0.081 & 14.537$\pm$0.080 \\
\end{longtable}
\end{landscape}

\section{Periodic variables}

\begin{center}
\begin{longtable}{llllcccc}
\caption{\label{tab:cat_periodic} Periodic variable stars in the field of view of NGC~1893. $P$ is the period in day. $A$ is the amplitude in milli-magnitude. The last-digit errors of the frequencies and amplitudes are given in parentheses. For each star, only independent frequencies are listed. Z08 indicates that the variable star was detected in \citet{Zhang2008}, while L12 and L14 mean \citet{Lata2012} and \citet{Lata2014}, respectively. pp indicates that the variable star is detected in the present paper. Note: the eclipsing binary star 2091 hosting a pulsating companion is not reported here, but listed in Table~\ref{tab:freq_star2091}.}\\
\hline\hline
Star ID  & $P$(d) & $A$(mmag) & S/N & Membership & Sp.Type & Classification & Discovery \\
\hline
\endfirsthead
\caption{continued.}\\
\hline\hline
Star ID  & $P$(d) & $A$(mmag) & S/N & Membership & Sp.Type & Classification & Discovery \\
\hline
\endhead
\hline
\endfoot
22	&	1.7854(2)	    &	 0.86(3) 	&	4.3 	&	no  &  $-$  &  field     & L14\\
23	&	0.3548553(9)	&	14.88(9) 	&	6.7 	&	yes & 	B4$^a$  &	$\beta$ Cep & Z08 \\
	&	0.1800465(6)	&	 5.52(8) 	&	4.0 	&       &       &	         &    \\
25	&	0.132938(3)	    &	11.2(5) 	&	6.2 	&	$?$ &  $-$  &	$-$      & pp \\
	&	0.074472(1) 	&	 5.9(5)  	&	5.0 	&       &	    &	         &    \\
	&	0.122497(5) 	&	 5.5(5)  	&	4.8 	&       &	    &	         &    \\
	&	0.079010(2) 	&	 5.5(5)  	&	6.9 	&       &	    &	         &    \\
40	&	9.523(3)    	&	 2.05(4) 	&	4.6 	&	yes &	B2$^a$  &	SPB         & L14 \\
    &   6.008(3)        &    0.93(3)    &   4.7     &       &	    &	         &    \\
	&   0.55240(3)      &    0.79(4)    &   4.4     &       &	    &	         &    \\
43	&	0.174560(3) 	&	 0.60(3) 	&	6.5 	&	yes &	B3$^a$  &	$\beta$ Cep & L14 \\
    &   1.3661(2)       &    0.51(3)    &   4.0 	&       &	    &	         &    \\
50	&	0.1818344(8)	&	 2.56(4) 	&  11.4 	&	yes &	B2$^a$  &	$\beta$ Cep & Z08 \\
    &   1.45936(7)      &    1.75(3)    &   7.3     &       &	    &	         &    \\
52	&	0.58662(2)	    &	 1.13(4) 	&	4.1 	&	yes &	$-$ &	SPB      & L14\\
    &   0.50644(2)      &    0.85(4)    &   4.2     &       &	    &	         &    \\
54	&	0.672712(6)	    &	11.4(1) 	&	4.9 	&	no  &	F0\uppercase\expandafter{\romannumeral3}-\uppercase\expandafter{\romannumeral4}$^b$ 	&	field & Z08\\
	&	0.78317(1)	    &	 7.2(1) 	&	4.2 	&       &	    &	         &    \\
55	&	0.93309(2)	    &    1.63(4) 	&	4.2 	&	yes &	B3$^a$  &	SPB         & L14 \\
58	&	0.21407(9)  	&    2.5(1) 	&	6.0 	&	yes &	B4$^a$  &	$\beta$ Cep & L14 \\
62	&	0.298365(5) 	&	 1.00(4)	&	5.6 	&	yes &	$-$	&	Herbig Ae/Be& L14 \\
    &   2.5674(7)       &    0.69(4)    &   4.7     &       &       &            &    \\
65	&	1.2605(8)   	&	 3.74(6) 	&	4.0 	&	yes &	B5$^a$  &	SPB         & pp  \\
80	&	0.2492(1)   	&	 4.7(2) 	&	4.9 	&	yes	&	B2$^a$  &	$\beta$ Cep & L14 \\
96	&	0.146388(1) 	&  	58(1)   	&	6.9 	&	$?$	&	$-$ &	$-$      & pp \\
	&	0.140987(2) 	&	29(1)   	&	6.6 	&       &	    &	         &    \\
	&	0.143781(5) 	&	13(1)   	&	5.5 	&       &	    &	         &    \\
	&	0.083090(2) 	&	10(1)   	&	4.8 	&       &	    &	         &    \\
101	&	0.157490(2) 	&	 0.90(5) 	&	6.8 	&	no  &	$-$ &	field    & L14\\
122	&	0.0609429(1)	&	4.18(7) 	&	16.2 	&	yes &	$-$ &	FaRPB      & L14\\
127	&	0.157985(2) 	&	1.28(6) 	&	6.5 	&	yes &	B6$^a$  &	FaRPB         & L14 \\
	&	0.342568(6) 	&	1.59(5) 	&	5.9 	&       &	    &	         &    \\
	&	0.166082(2) 	&	1.36(6) 	&	7.1 	&       &	    &	         &    \\
	&	0.364430(1) 	&	1.04(5) 	&	4.3 	&       &	    &	         &    \\
128	&	0.06088143(5)	&	4.6(1)  	&	29.9	&	no  &	$-$ &	field    & pp \\
130	&	0.346131(2) 	&	5.78(8) 	&	8.5 	&	yes &	B0.5\uppercase\expandafter{\romannumeral4}e$^b$ 	&	Herbig Be & Z08\\
    &   8.469(2)        &   2.98(7)     &   4.9     &       &       &            &    \\
	&	0.369156(6) 	&	2.24(7) 	&	4.3 	&       &	    &	         &    \\
	&	0.356231(4) 	&	2.38(7) 	&	4.8 	&       &	    &	         &    \\
149 &   1.159(3)        &   1.36(7)     &   4.3     &   yes &  B5$^a$  &  SPB    & L12\\
182	&	0.1467416(8)	&	6.1(2)  	&	12.7	&	yes &	B1.5\uppercase\expandafter{\romannumeral5}e$^b$ 	&	classical Be & L14\\
184	&	0.03244(3)  	&	2.1(1)	    &	7.5	    &	yes &	$-$	&	FaRPB      & pp \\
    &	0.02616(2)      &	2.0(2)  	&	10.4	&       &	    &	         &    \\
190	&	0.03472643(4)	&	1.5(1)  	&	6.0 	&	yes &	$-$	&	FaRPB      & pp \\
	&	0.04525800(8)	&	1.6(1)  	&	5.8 	&       &	    &	         &    \\
	&	0.02750443(3)	&	1.3(1)  	&	6.7 	&       &	    &	         &    \\
	&	0.0465461(1)	&	1.0(1)  	&	6.5 	&       &	    &	         &    \\
	&	0.02241206(4)	&	0.8(1)  	&	5.0 	&       &	    &	         &    \\
212	&	0.231060(1)  	&	3.1(2)  	&	4.5 	&	yes &	$-$ &	FaRPB      & pp \\
233	&	0.17058(1)  	&	1.1(1)  	&	4.3 	&	no 	&	$-$ &	field    & L14\\
235	&	1.11273(1)  	&	26.2(2) 	&	12.7	&	no  &	B6$^a$ &	field       & Z08\\
240	&	0.1360(5)   	&	5.3(3)  	&	6.0   	&	yes &	$-$ &   FaRPB     & pp \\
262	&	0.05815637(4)	&	7.1(1)  	&	19.9	&	yes &	$-$ & 	FaRPB      & pp \\
	&	0.0634753(2)	&	1.4(1)  	&	5.3 	&       &	    &   	     &    \\
	&   0.89920(6)      &   1.0(1)      &   4.5     &       &       &            &    \\
	&	0.0567738(3)	&	0.9(1)  	&	4.0 	&       &	    &	         &    \\
269	&	0.1354645(6)	&	3.91(9) 	&	14.1	&	no  &	$-$ &	field    & L12\\
271	&	0.0554638(5)	&	0.88(9) 	&	6.3 	&	$?$ &	$-$ & 	$-$      & L14\\
283	&	3.5162(2)   	&	3.70(8) 	&	5.9 	&	yes &	B6$^a$ & SPB & L14 \\
    &   5.989(2)        &   1.99(7)     &   4.1     &       &	    &	         &    \\
	&   0.385699(7)     &   1.58(8)     &   4.2     &       &	    &	         &    \\
305 &   0.371(5)        &   1.2(1)      &   4.1     &   yes &   $-$ &   FaRPB      & L14\\
313	&	0.587530(8) 	&	1.13(9) 	&	5.1 	&	no  &	$-$ &	field    & L14\\
325	&	1.52798(3)  	&	25.1(1) 	&	18.9	&	no 	&	$-$ &	field    & pp \\
327	&	1.4063(1)   	&	2.5(1)  	&	4.6 	&	yes &	$-$ &	SPB      & L14\\
    &   1.9441(4)       &   1.9(1)      &   4.5     &       &       &            &    \\
338	&	0.281350(5) 	&	2.0(1)  	&	10.4	&	yes &	B8$^a$  &	FaRPB         & L14 \\
	&	0.37393(2)  	&	1.1(1)  	&	3.7 	&       &	    &	         &    \\
	&	0.28350(2)  	&	0.69(9) 	&	4.8 	&       &	    &	         &    \\
	&	0.18102(1)  	&	0.7(1)  	&	4.7 	&       &	    &	         &    \\
342	&	0.260471(2) 	&	4.8(1)  	&	11.4	&	no  &	$-$	&	field    & L14\\
	&	0.278360(7) 	&	1.8(1)  	&	5.1 	&       &	    &	         &    \\
	&	0.52163(6)  	&	0.9(1)  	&	4.1 	&       &	    &	         &    \\
360	&	0.250974(2) 	&	1.71(7) 	&	4.3 	&	yes &	B9$^a$  &	FaRPB         & L14 \\
368	&	0.160493(8) 	&	1.1(1)  	&	4.1 	&	no  &	$-$	&	field    & L14\\
414	&	0.250153(3) 	&	6.3(2)  	&	6.9 	&	yes &	$-$ &	FaRPB      & L14\\
472	&	5.3450(7)   	&	7.7(1)  	&	5.1 	&	no 	&	G3\uppercase\expandafter{\romannumeral5}$^b$ &	field & pp \\
509 &   0.1093(5)       &   1.5(2)      &   4.3     &   no  &   $-$ &   field      & L14\\ 
534	&	2.8168(4)   	&	7.1(2)  	&	6.2 	&	yes &	$-$	&	PMS      & pp \\
591	&	1.48126(7)  	&	6.7(2)  	&	7.4 	&	no  &	$-$ &	field    & pp \\
599	&	1.4963(1)   	&	7.8(2)  	&	6.5 	&	yes &	F9\uppercase\expandafter{\romannumeral5}$^b$ 	&	PMS & L14\\
638	&	0.03861865(5)	&	4.4(2)  	&   6.1 	&	yes &	$-$	&	PMS      & pp \\
    &   0.0794580(2)    &	3.9(2)      &	4.2 	&       &	    &	         &    \\
    &   0.0673194(2) 	&   2.6(2)      &   4.0 	&       &	    &	         &    \\
    &   0.0646079(3) 	&   2.6(2)      &	4.7 	&       &	    &	         &    \\
    &   0.0421728(1)    & 	2.5(2)      &	4.2 	&       &	    &	         &    \\
    &   0.0831568(4)    &   2.5(2)      &   4.8 	&       &	    &	         &    \\
660	&	0.18841(1)  	&	1.5(2)  	&	4.6 	&	no	&	$-$ &	field    & L14\\
666	&	0.0636964(1)	&	4.6(4)  	&	6.8 	&	no  &	$-$ &	field    & pp \\
	&	0.04426010(8)	&	2.9(3)  	&	4.6 	&       &	    &	         &    \\
	&	0.0926221(5)	&	2.3(4)  	&	4.0 	&       &	    &	         &    \\
710	&	0.1583(9)   	&	2.2(2)  	&	4.6 	&	no 	&	$-$ &	field    & L14\\
721	&	0.108524(5) 	&	1.3(2)  	&	4.9 	&	yes &	$-$ &	PMS      & L14\\
    &   0.33882(5)      &   1.2(2)      &   4.1     &       &       &            &    \\
738	&	0.572213(7) 	&	10.7(5) 	&	9.2 	&	no 	&	$-$ &	field    & pp \\
741	&	0.447(3)    	&	52(3)   	&	19.7	&	yes	&	$-$ &	PMS      & pp \\
	&	0.182(7)    	&	6(2)     	&	5.0 	&       &	    &	         &    \\
	&	0.36(4)     	&	4(2)     	&	4.2 	&       &	    &	         &    \\
824	&	0.263(1)    	&	28(1)   	&	9.3 	&	yes &	$-$	&	PMS      & pp \\
861	&	0.317314(2) 	&	23.9(7) 	&	8.9	    &	yes &	$-$	&	PMS      & pp \\
986	&	1.48152(5)  	&	35.5(6) 	&	7.6 	&	yes &	$-$	&	PMS      & L12\\
    &   1.4263(1)       &   15.3(5)     &   4.8     &       &       &            &    \\
    &   2.7763(6)       &   10.8(7)     &   4.4     &       &       &            &    \\	
1049 &	0.070152(8) 	&	29.7(9) 	&	9.8 	&	yes &	$-$ &	PMS      & L14\\
1101 &	0.0792482(2)	&	6.2(3)  	&	7.2 	&	yes &	$-$	&	PMS      & pp \\
	 &	0.0904108(4)	&	3.1(3)  	&	5.0 	&       &	    &	         &    \\
1114 &	11.257(2)   	&	10.7(3) 	&	7.2 	&	yes	&	$-$ &	PMS      & pp \\
1202 &	1.88693(7)  	&	62.2(7) 	&	8.6 	&	yes &	$-$ &	PMS      & pp \\
1326 &	0.2592(6)   	&	97(1)   	&	25.7	&	yes &	$-$	&	PMS      & pp \\
     &  0.332(6)   	    &   13(2)       &	5.0     &       &	    &	         &    \\
     &  0.40(2)         &	6(2)        &	8.4     &       &	    &	         &    \\
1380 &	0.68407(1)  	&	11.9(4) 	&	7.0 	&	yes &	$-$ &	PMS      & pp \\
1388 &	0.04135875(4)	&	9.8(5)  	&	15.8	&	no	&	$-$ &	field    & pp \\
1390 &	0.78970(2)  	&	59(1)   	&	6.1 	&	no  &	$-$	&	field    & pp \\
1425 &	2.1658(1)   	&	31.4(6) 	&	7.6 	&	yes &	$-$	&	PMS      & L12\\
1572 &	2.3642(1)   	&	40.1(6) 	&	8.5 	&	yes &	$-$	&	PMS      & L12\\
1577 &	6.409(3)    	&	13.3(7) 	&	5.4 	&	no  &	$-$	&	field    & pp \\
1590 &	0.097613(1) 	&	9.2(4)  	&	6.1 	&	yes &	$-$	&	PMS      & pp \\
	 &	0.098816(3) 	&	4.4(4)  	&	4.6 	&       &	    &	         &    \\
1620 &	0.0681402(7)	&	9.1(5)  	&	5.3 	&	yes &	$-$	&	PMS      & L14\\
1630 &	0.109075(1) 	&	14.4(6) 	&	7.6 	&	no  &	$-$	&	field    & pp \\
1664 &	0.405640(4) 	&	21.9(9) 	&	4.7 	&	no 	&	$-$  &	field    & pp \\
	 &	0.62303(1)  	&	20.0(9) 	&	4.0 	&       &	     &	         &    \\
	 &	0.81838(3)	    &	16.0(9) 	&	4.6  	&       &	     &	         &    \\
1665 &	3.8660(2)   	&	36.2(7) 	&	4.3 	&	no	&	$-$  &	field    & pp \\
	 &	1.56909(4)   	&	23.3(8) 	&	5.1 	&       &	     &	         &    \\
1810 &	3.7328(7)   	&	26.4(7) 	&	8.2 	&	yes &	$-$  &	PMS      & pp \\
1913 &	3.5650(6)   	&	24.3(7) 	&	5.3 	&	no  &	$-$  &	field    & pp \\
1921 &	0.2188(9)   	&	97(5)   	&	9.8 	&	no	&	$-$  &	field    & pp \\
	 &	0.188(2)    	&	30(5)    	&	12.5	&       &	     &	         &    \\
	 &	0.109(1)    	&	16(3)    	&	7.5 	&       &	     &        	 &    \\
	 &	0.44(2)     	&	16(4)   	&	4   	&       &	     &	         &    \\
	 &	0.35(3)     	&	9(4)    	&	6.9	    &       &	     &	         &    \\
1994 &	2.45302(7)   	&	24.8(6) 	&	4.6 	&	yes &	$-$  &	PMS      & pp \\
2005 &	9.99(2)     	&	36.3(7)  	&	7.0  	&	yes &	$-$  &	PMS      & pp \\
2038 &	0.0691036(1)	&  	14(1)   	&	7.9 	&	no  &	$-$  &	field    & pp \\
	 &	0.0951992(5)	&	7(1)    	&	4.1 	&       &	     &	         &    \\
2051 &	4.0024(3)   	&	24.5(7) 	&	7.0 	&	yes &	$-$	 &	PMS      & pp \\
2074 &	2.0031(2)   	&	39(1)   	&	6.7 	&	no 	&	$-$  &	field    & pp \\
2186 &	1.21670(6)  	&	38.7(1) 	&	4.7     &	yes	&	$-$  &	PMS      & L12\\
	 &	0.79301(3)  	&	27.5(1) 	&	5.0 	&       &	     &	         &    \\
	 &	0.64960(3)  	&	22.0(1) 	&	4.1 	&       &	     &	         &    \\
2302 &	6.537(6)    	&	349(6)  	&	6.7 	&	yes &	$-$  &	PMS      & L12\\
2347 &	2.2449(2)   	&	28.1(6) 	&	7.0 	&	no  &	$-$	 &	field    & pp \\
2352 &	0.06082(2)  	&	5.1(6)  	&	5.5 	&	yes	&	$-$  &	PMS      & pp \\
2450 &	1.7325(2)   	&	22(1)   	&	6.5 	&	yes &	$-$  &	PMS      & pp \\
2466 &	3.6148(3)   	&	21.0(9) 	&	6.5 	&	no  &	$-$  &	field    & pp \\
2611 &	0.478016(4) 	&	35.9(8) 	&	7.8 	&	no  &	$-$  &	field    & L14\\
2773 &	1.28221(3)  	&	46.2(8) 	&	8.8 	&	no 	&	$-$  &	field    & pp \\
2906 &	0.0819110(7)	&	5.5(7)  	&	4.6 	&	no  &	$-$  &	field    & pp \\
2975 &	0.628764(9) 	&	36(1)   	&	6.7 	&	yes &	$-$  &	PMS      & pp \\
2986 &	0.166209(3) 	&	27(3)   	&	4.4 	&	no  &	$-$	 &	field    & pp \\
	 &	0.110800(1) 	&	26(3)   	&	5.6 	&       &	     &	         &    \\
3093 &	8.5513(9)   	&	108(3)  	&	5.6 	&	yes &	$-$	 &	PMS      & L12\\
     &  12.572(6)       &    50(2)      &   4.4     &       &        &           &    \\
3214 &	1.35749(4)  	&	55(1)   	&	6.0 	&	yes &	$-$  &	PMS      & L14\\
3219 &	0.83266(3)  	&	34(1)	    &	4.9 	&	$?$ &	$-$  &	$-$      & pp \\
	 &	0.48605(1)  	&	28(1)   	&	5.8 	&       &	     &	         &    \\
3343 &	0.031073(6) 	&	8(1)    	&	4.7 	&	no  &	$-$  &	field    & pp \\
3349 &	2.7721(1)       &	45(1)   	&	8.8 	&	yes &	$-$  &	PMS      & pp \\
3521 &	3.8931(4)   	&	62(2)   	&	5.2 	&	yes &	$-$  &	PMS      & pp \\
3627 &	7.935(1)    	&	40(1)   	&	5.3 	&	yes	&	$-$  &	PMS      & pp \\
3680 &	10.937(2)   	&	56(2)   	&	5.0 	&	yes &	$-$	 &	PMS      & pp \\
3695 &	5.1109(5)   	&	58(2)   	&	5.1 	&	yes	&	$-$  &	PMS      & pp \\
4225 &	6.7303(9)   	&	79(2)   	&	7.1 	&	no  &	$-$  &	field    & pp \\
4495 &	0.1703063(6)	&	66(2)   	&	13.3	&	no  &	$-$  &	field    & pp \\
4496 &	7.9051(9)   	&	69(2)    	&	6.3 	&	no  &	$-$  &	field    & pp \\
4975 &	0.1321(3)   	&	193(11) 	&	16.8	&	no  &	$-$  &	field    & pp \\
5130 &	0.1871779(7)	&	153(6)  	&	14.2	&	no  &	$-$  &	field    & pp \\
5156 &	2.50218(7)  	&	177(3)  	&	11.9	&	yes &	$-$  &	PMS      & L12\\
5262 &	0.0841(2)   	&	4.6(3)	    &	4.7 	&	no  &	$-$  &	field    & pp \\
5269 &	0.0352494(2)	&	2.5(2)  	&	7.0 	&	yes &	$-$  &	FaRPB      & pp \\
5283 &	0.0899076(6)	&	16.9(4) 	&	7.8 	&	no  &	$-$  &	field    & pp \\
	 &	0.084353(1) 	&	6.7(4)  	&	7.2 	&       &	     &	         &    \\
	 &	0.089973(2) 	&	4.2(4)  	&	4.7 	&       &	     &       	 &    \\
5284 &	0.295(2)    	&	5.8(3)  	&	11.7	&	no  &	$-$  &	field    & pp \\
	 &	0.39(1)     	&	1.2(3)  	&	6.7 	&       &	     &	         &    \\
5569 &	0.07731(4)  	&	164(3)  	&	28.7	&	no  &	$-$  &	field    & pp \\
\hline
\multicolumn{8}{l}{$^a$ Spectral type from \citet{Marco2001}}\\
\multicolumn{8}{l}{$^b$ Spectral type from \citet{Marco2002}}\\
\end{longtable}
\end{center}
\clearpage

\begin{figure*}
\centering
\subfigure{\includegraphics[width=0.3\textwidth]{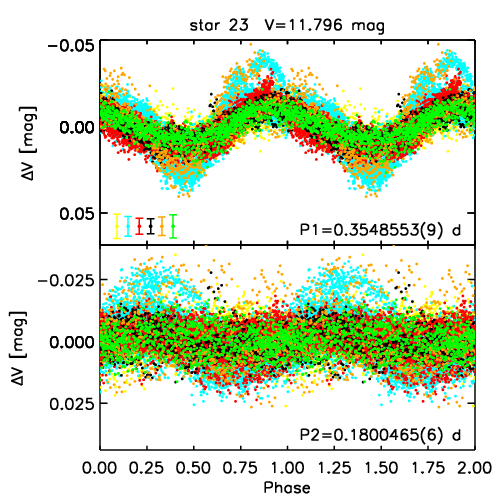}}
\subfigure{\includegraphics[width=0.3\textwidth]{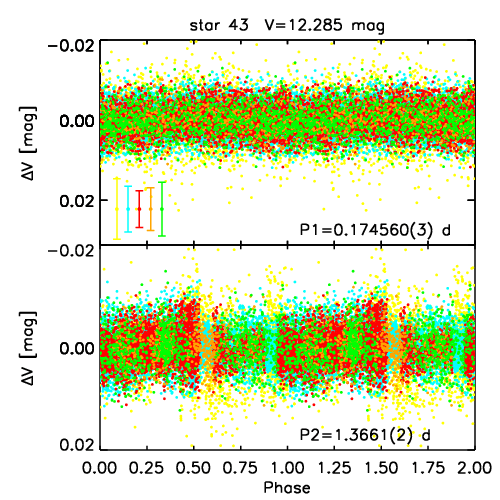}}
\subfigure{\includegraphics[width=0.3\textwidth]{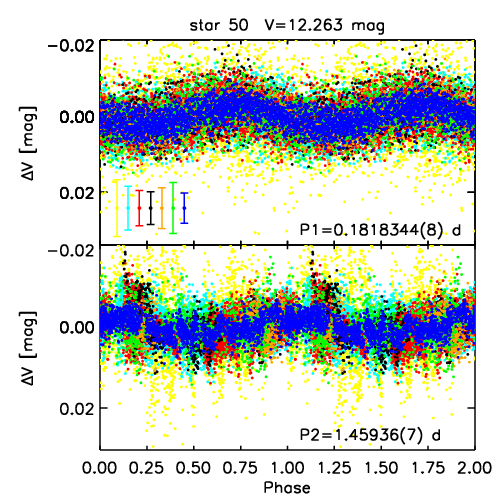}}
\subfigure{\includegraphics[width=0.3\textwidth]{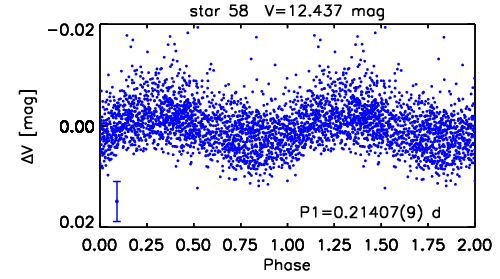}}
\subfigure{\includegraphics[width=0.3\textwidth]{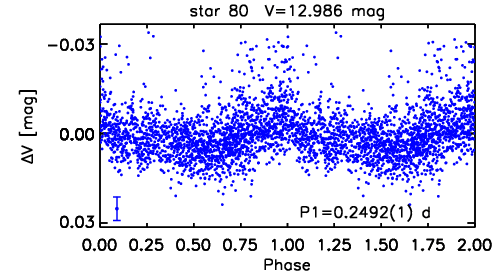}}
\caption{Folded light curves of the monoperiodic, biperiodic $\beta$ Cep stars.
         For the multiperodic stars, the top panel shows the light curve with the dominant frequency and the subsequent panels from second to bottom show the folded residual light curves after subtraction of all preceding frequencies.
         The period used to fold the light curves is written in the lower right corner of each panel.
         The observations of groups 1, 2, 3, 4, 5, 6 and 7 (the group number refers to the observation run IDs listed in Table~\ref{tab:obs_log1} of the main body of the text) are respectively represented in blue, black, red, cyan, yellow, green and orange.
         The error bars represent the mean errors of the measurements.
         The mean $V$ magnitude has been subtracted from the light curves.
         The title of each subfigure mentions the star number in the present work and its $V$ magnitude.}
\label{fig:Beta_Cep_PD1}
\end{figure*}

\begin{figure*}
\centering
\subfigure{\includegraphics[width=0.3\textwidth]{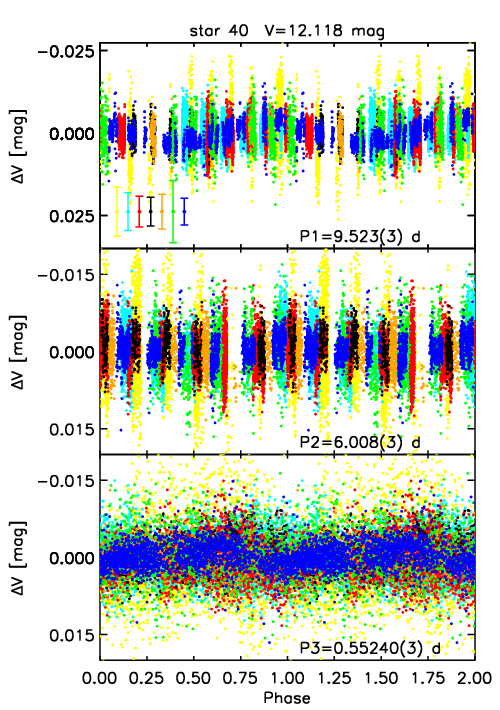}}
\subfigure{\includegraphics[width=0.3\textwidth]{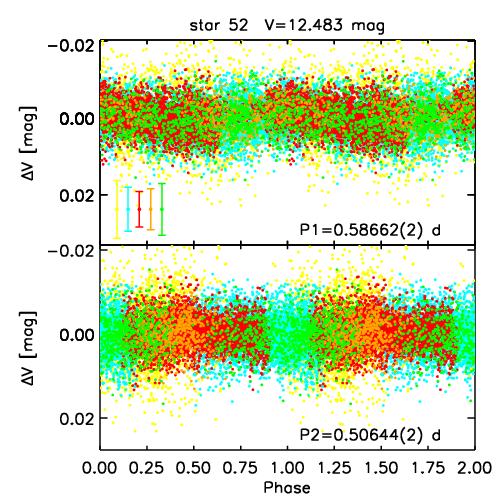}}
\subfigure{\includegraphics[width=0.3\textwidth]{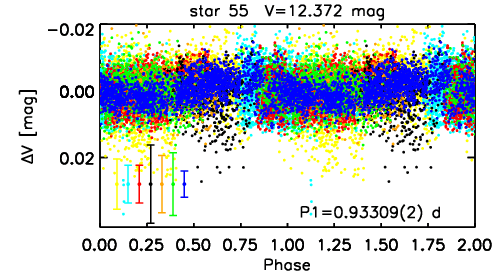}}
\caption{Same as Figure~\ref{fig:Beta_Cep_PD1}, but for the SPB stars.}
\label{fig:SPB}
\end{figure*}

\begin{figure*}
\centering
\setcounter{figure}{1}
\subfigure{\includegraphics[width=0.3\textwidth]{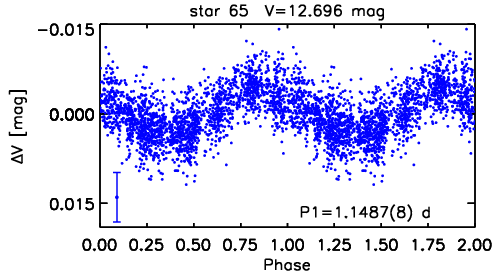}}
\subfigure{\includegraphics[width=0.3\textwidth]{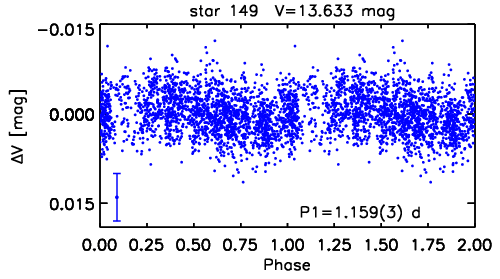}}
\subfigure{\includegraphics[width=0.3\textwidth]{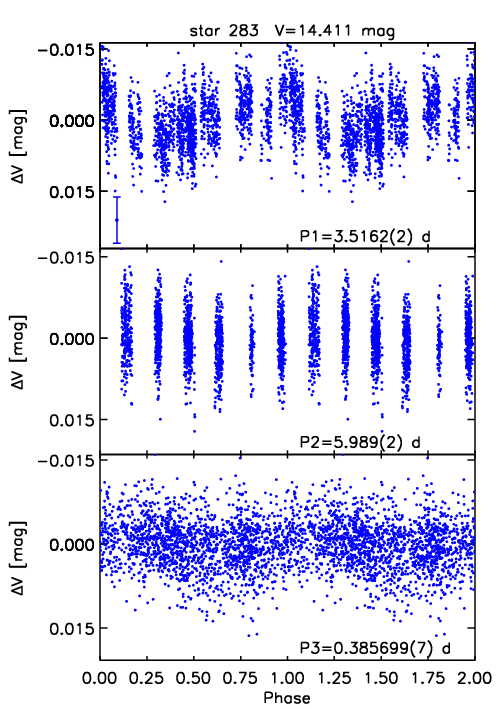}}
\subfigure{\includegraphics[width=0.3\textwidth]{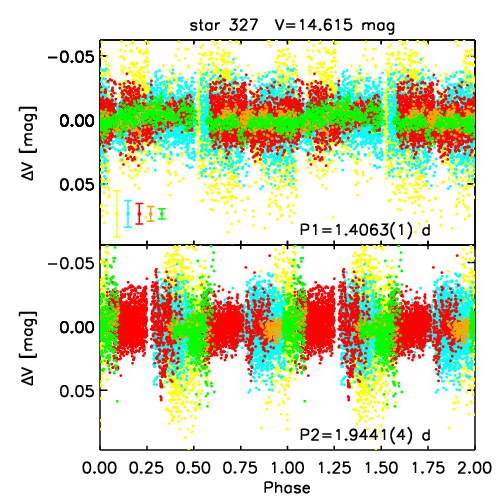}}
\caption{Continued.}
\end{figure*}

\begin{figure*}
\centering
\includegraphics[width=0.3\textwidth]{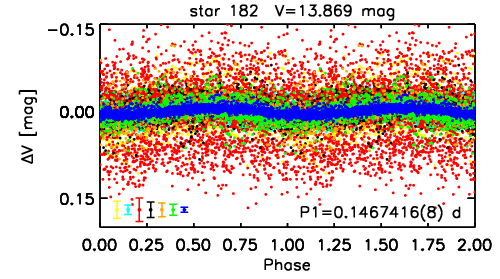}
\caption{Same as Figure~\ref{fig:Beta_Cep_PD1}, but for the classical Be star. \label{fig:classical_Be}}
\end{figure*}

\begin{figure*}
\centering
\subfigure{\includegraphics[width=0.3\textwidth]{figure/00122phase.png}}
\subfigure{\includegraphics[width=0.3\textwidth]{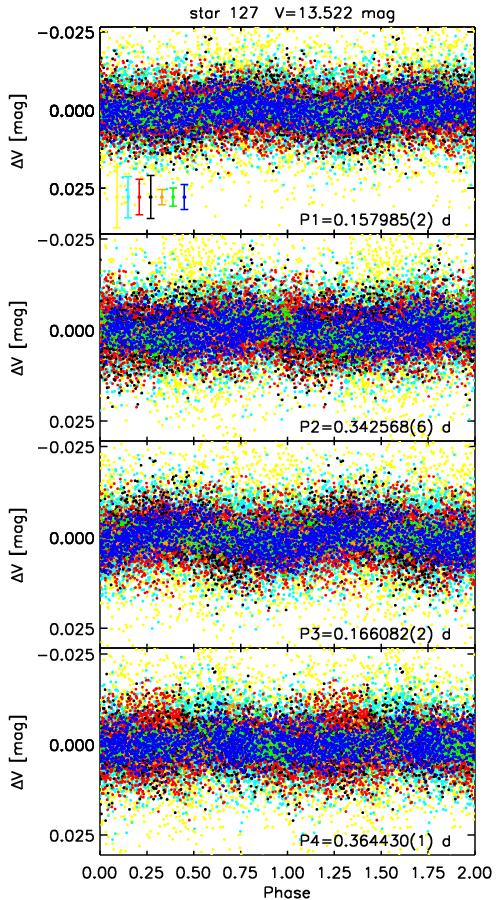}}
\subfigure{\includegraphics[width=0.3\textwidth]{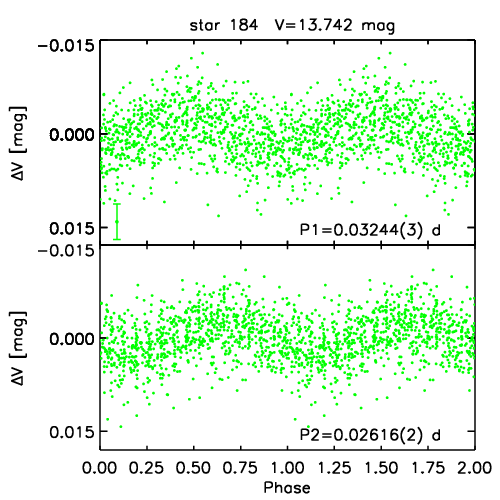}}
\subfigure{\includegraphics[width=0.3\textwidth]{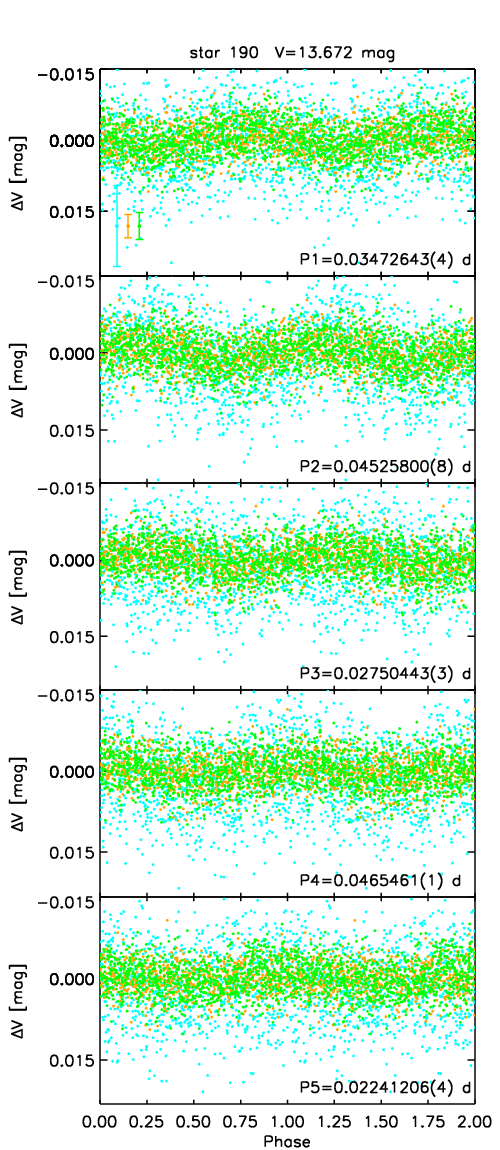}}
\subfigure{\includegraphics[width=0.3\textwidth]{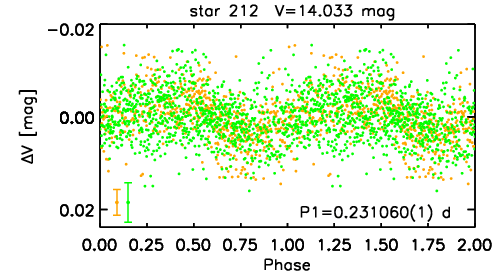}}
\subfigure{\includegraphics[width=0.3\textwidth]{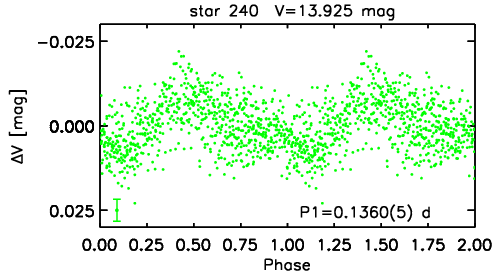}}
\caption{Same as Figure~\ref{fig:Beta_Cep_PD1}, but for the fast-rotating pulsating B stars.}
\label{fig:NEW}
\end{figure*}

\begin{figure*}
\centering
\setcounter{figure}{3}
\subfigure{\includegraphics[width=0.3\textwidth]{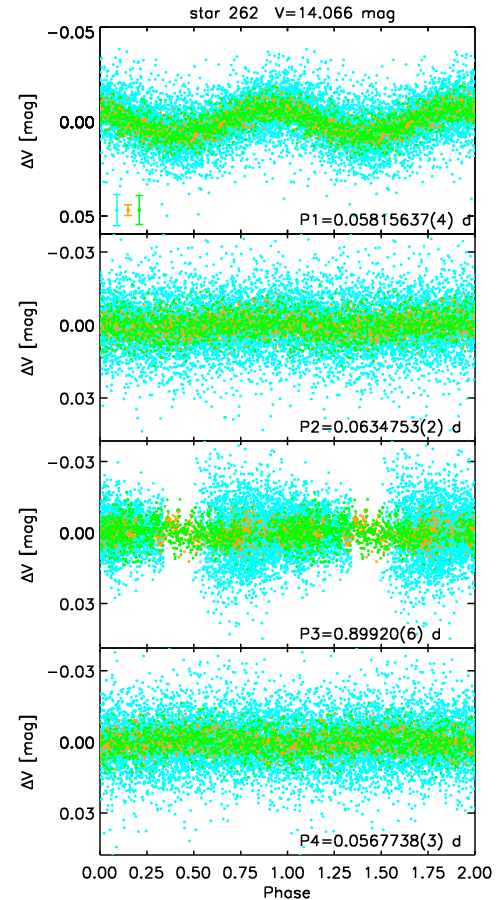}}
\subfigure{\includegraphics[width=0.3\textwidth]{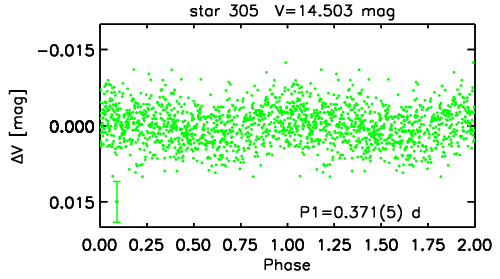}}
\subfigure{\includegraphics[width=0.3\textwidth]{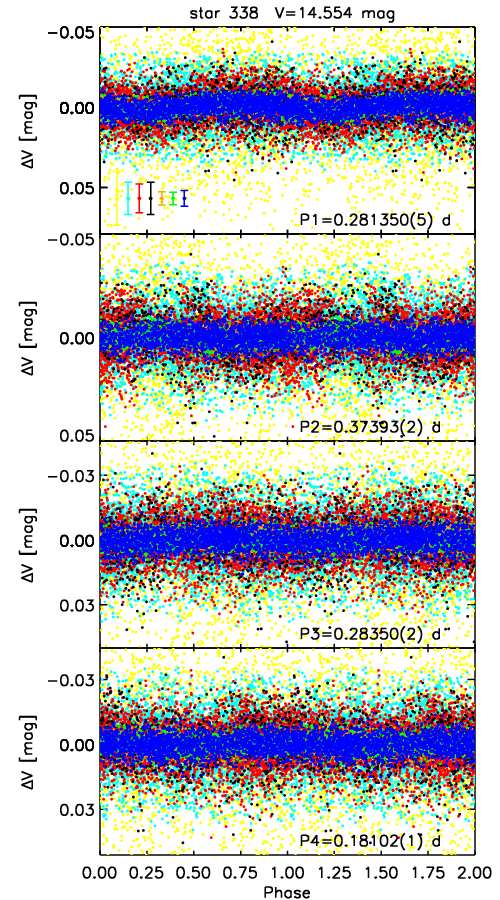}}
\subfigure{\includegraphics[width=0.3\textwidth]{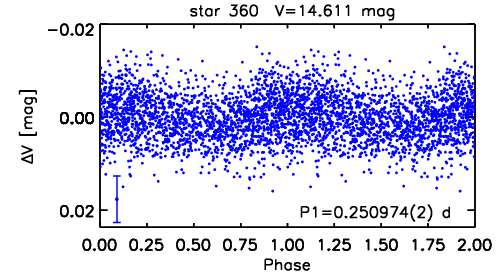}}
\subfigure{\includegraphics[width=0.3\textwidth]{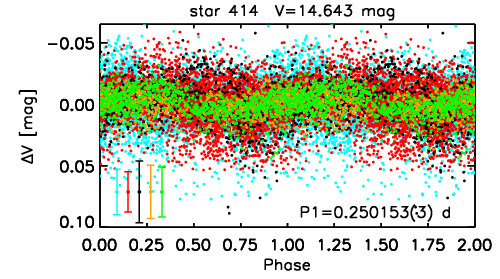}}
\subfigure{\includegraphics[width=0.3\textwidth]{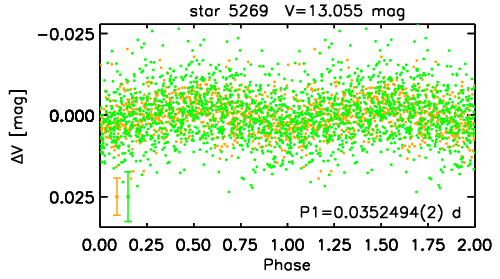}}
\caption{Continued.}
\end{figure*}

\begin{figure*}
\centering
\subfigure{\includegraphics[width=0.3\textwidth]{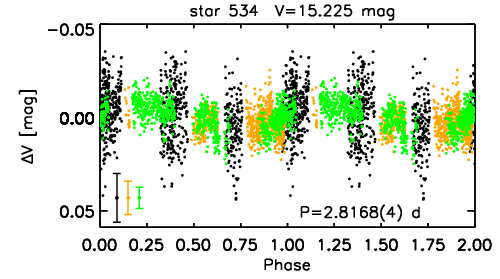}}
\subfigure{\includegraphics[width=0.3\textwidth]{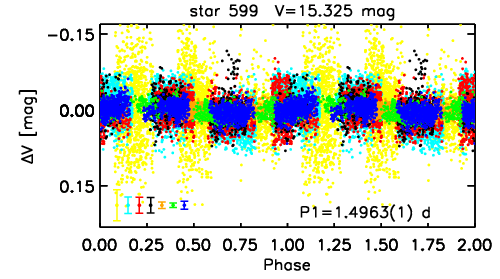}}
\subfigure{\includegraphics[width=0.3\textwidth]{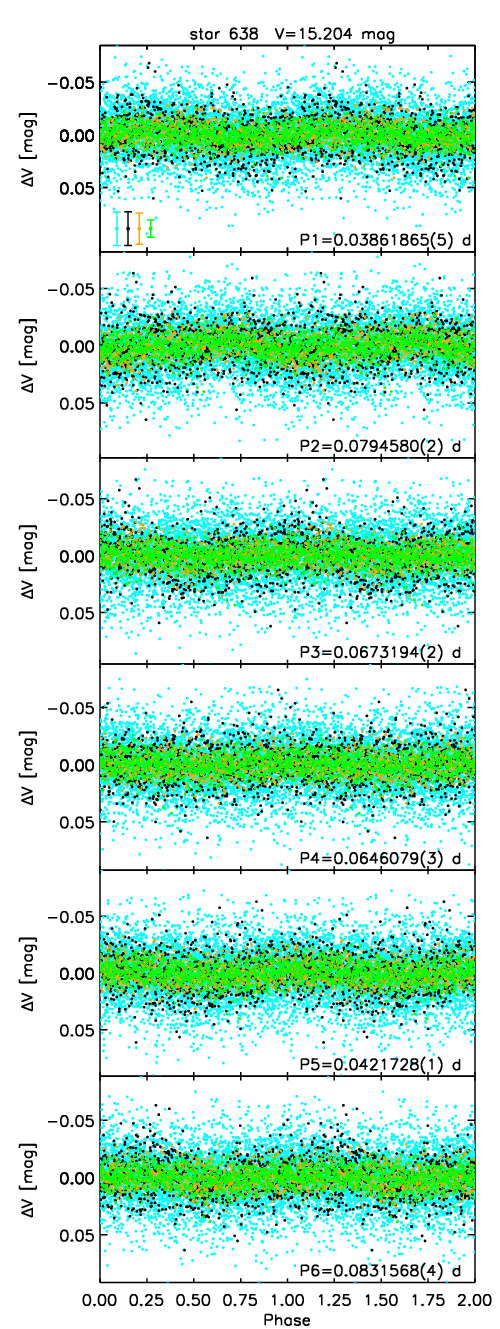}}
\subfigure{\includegraphics[width=0.3\textwidth]{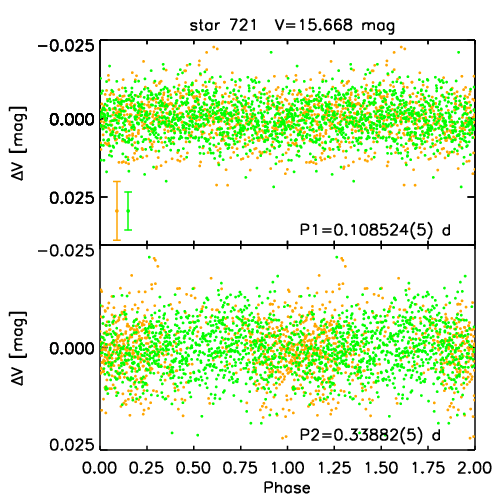}}
\subfigure{\includegraphics[width=0.3\textwidth]{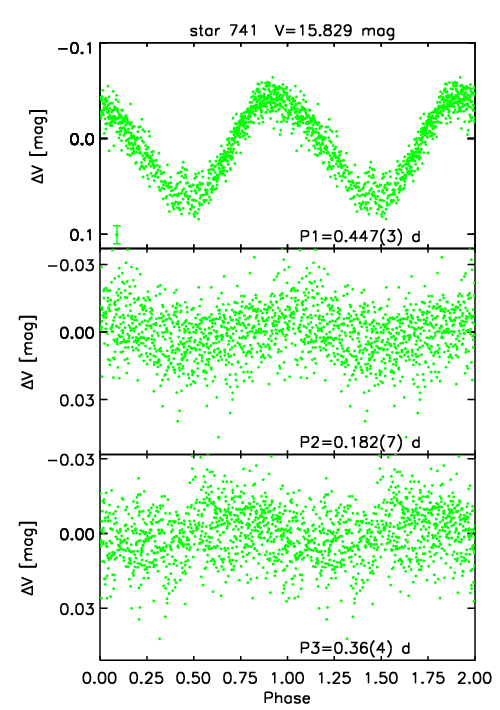}}
\subfigure{\includegraphics[width=0.3\textwidth]{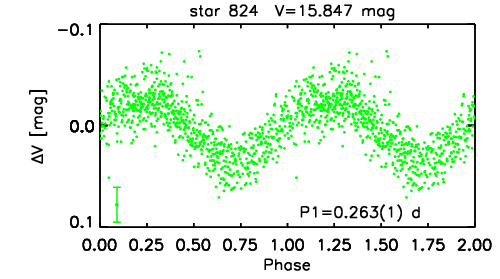}}
\caption{Same as Figure~\ref{fig:Beta_Cep_PD1}, but for the PMS variable stars.}
\label{fig:PMS}
\end{figure*}
\clearpage

\begin{figure*}
\centering
\setcounter{figure}{4}
\subfigure{\includegraphics[width=0.3\textwidth]{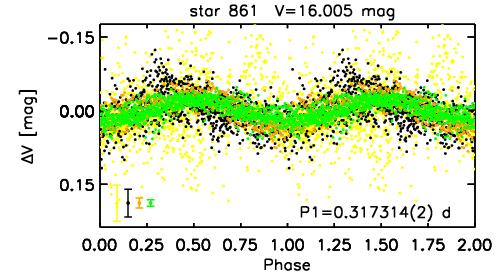}}
\subfigure{\includegraphics[width=0.3\textwidth]{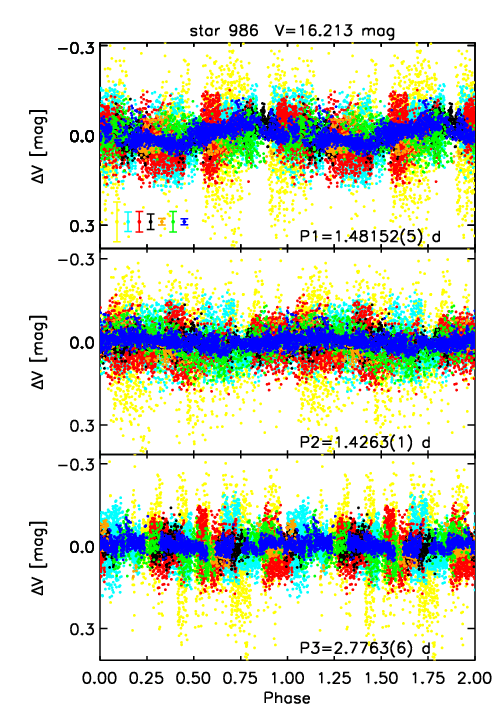}}
\subfigure{\includegraphics[width=0.3\textwidth]{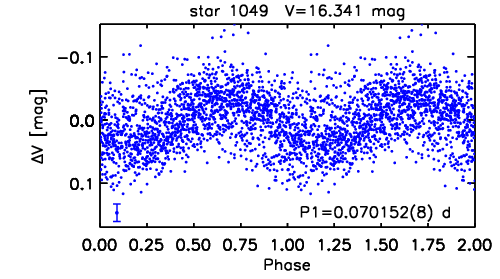}}
\subfigure{\includegraphics[width=0.3\textwidth]{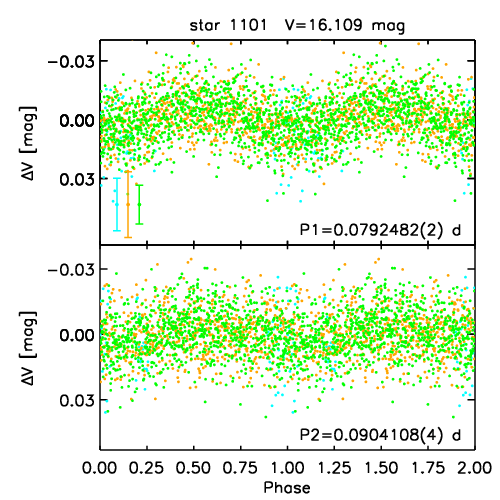}}
\subfigure{\includegraphics[width=0.3\textwidth]{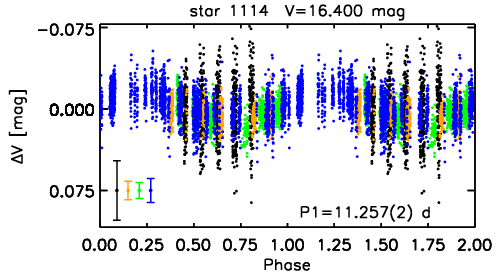}}
\subfigure{\includegraphics[width=0.3\textwidth]{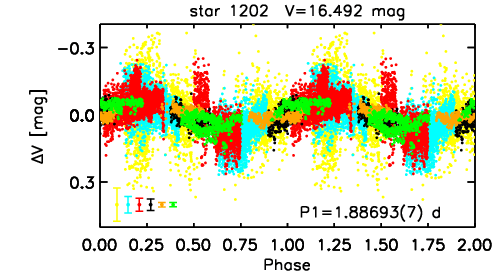}}
\subfigure{\includegraphics[width=0.3\textwidth]{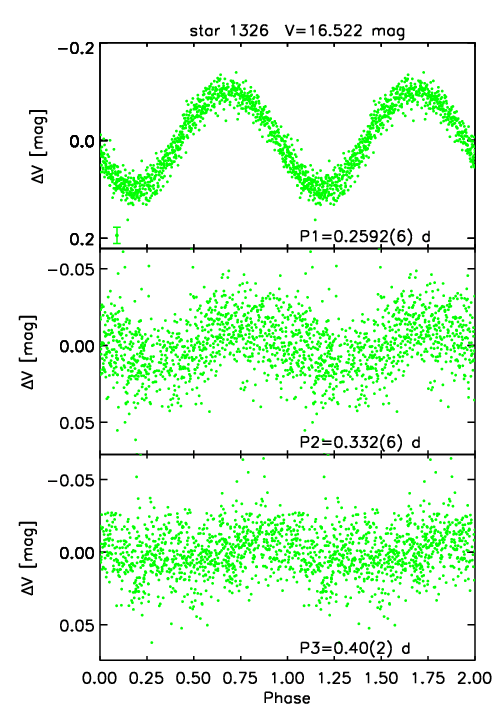}}
\subfigure{\includegraphics[width=0.3\textwidth]{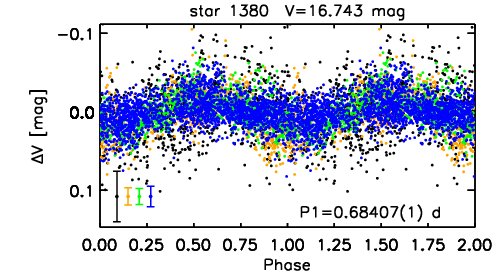}}
\subfigure{\includegraphics[width=0.3\textwidth]{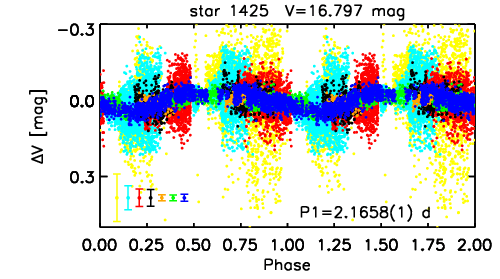}}
\caption{Continued.}
\end{figure*}

\begin{figure*}
\centering
\setcounter{figure}{4}
\subfigure{\includegraphics[width=0.3\textwidth]{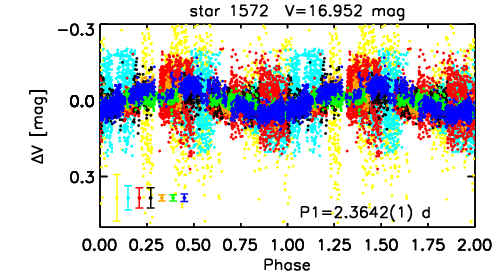}}
\subfigure{\includegraphics[width=0.3\textwidth]{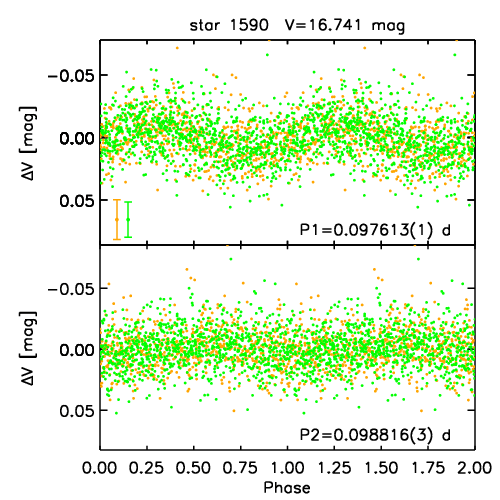}}
\subfigure{\includegraphics[width=0.3\textwidth]{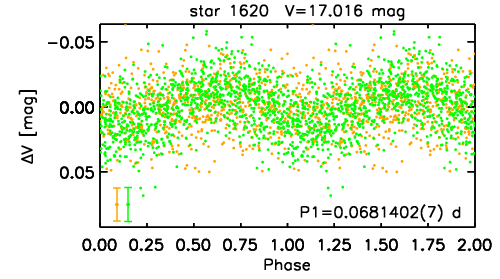}}
\subfigure{\includegraphics[width=0.3\textwidth]{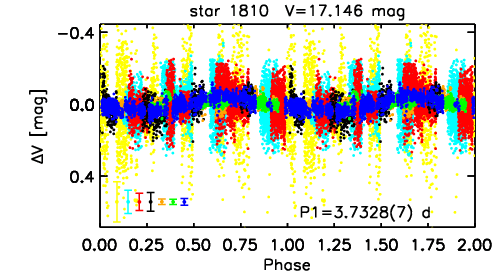}}
\subfigure{\includegraphics[width=0.3\textwidth]{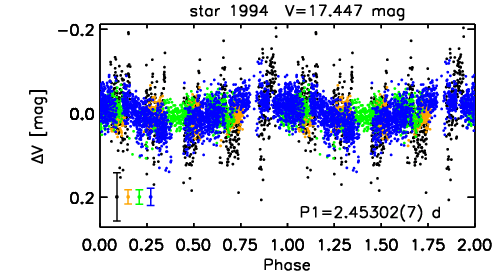}}
\subfigure{\includegraphics[width=0.3\textwidth]{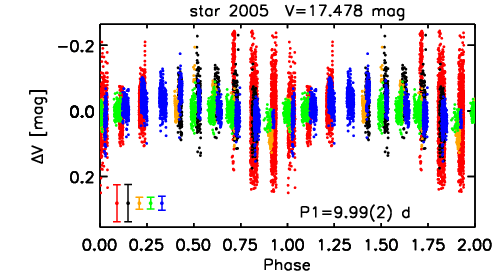}}
\subfigure{\includegraphics[width=0.3\textwidth]{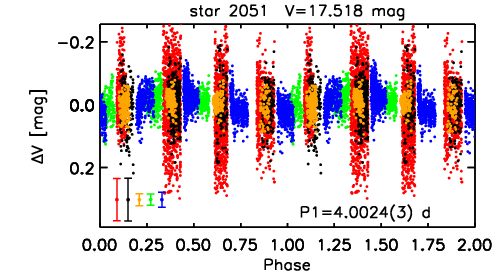}}
\subfigure{\includegraphics[width=0.3\textwidth]{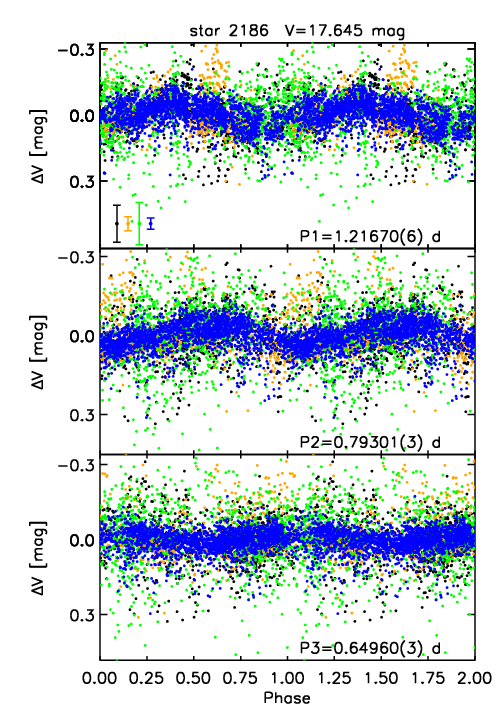}}
\subfigure{\includegraphics[width=0.3\textwidth]{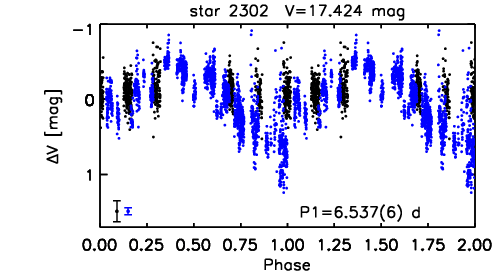}}
\subfigure{\includegraphics[width=0.3\textwidth]{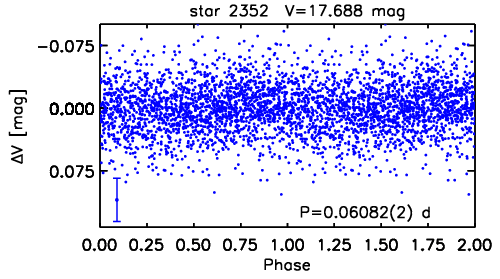}}
\subfigure{\includegraphics[width=0.3\textwidth]{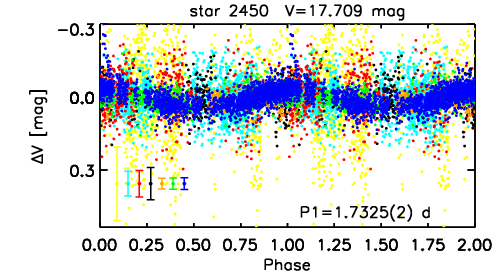}}
\subfigure{\includegraphics[width=0.3\textwidth]{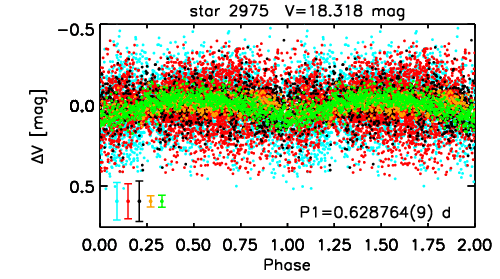}}
\caption{Continued.}
\end{figure*}

\begin{figure*}
\centering
\setcounter{figure}{4}
\subfigure{\includegraphics[width=0.3\textwidth]{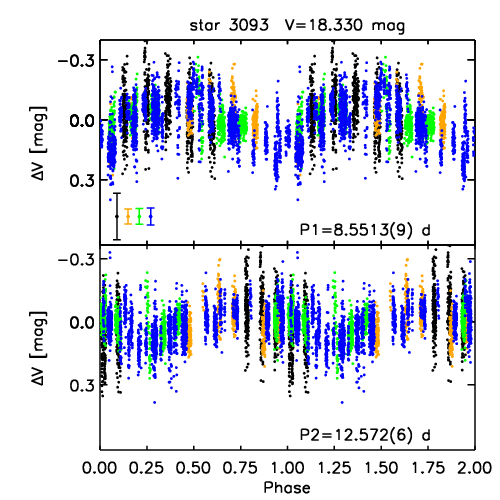}}
\subfigure{\includegraphics[width=0.3\textwidth]{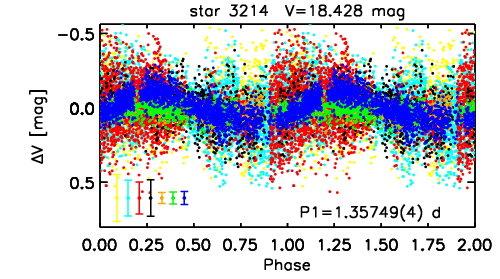}}
\subfigure{\includegraphics[width=0.3\textwidth]{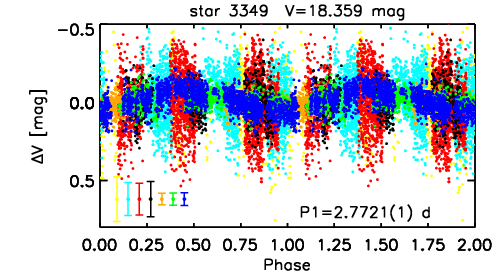}}
\subfigure{\includegraphics[width=0.3\textwidth]{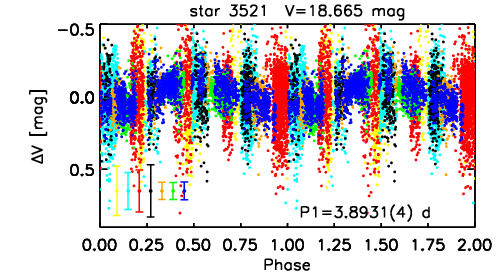}}
\subfigure{\includegraphics[width=0.3\textwidth]{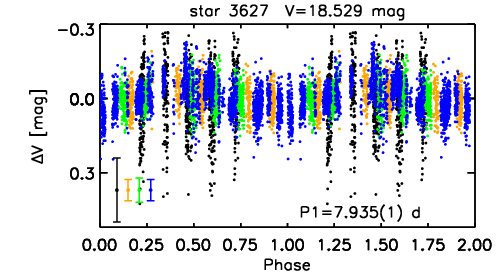}}
\subfigure{\includegraphics[width=0.3\textwidth]{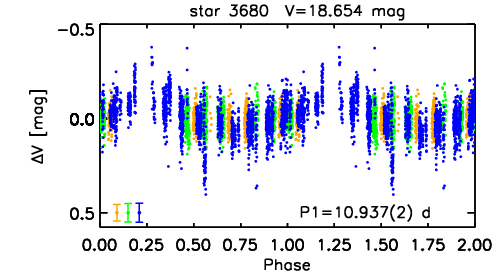}}
\subfigure{\includegraphics[width=0.3\textwidth]{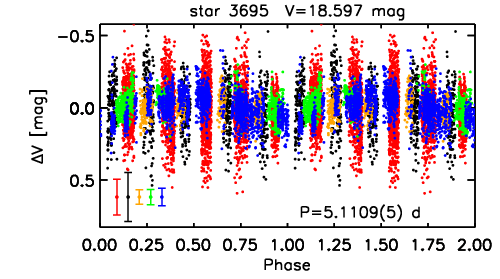}}
\subfigure{\includegraphics[width=0.3\textwidth]{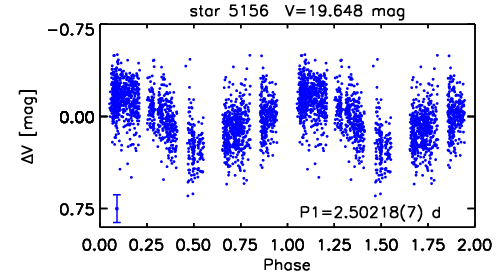}}
\caption{Continued.}
\end{figure*}

\begin{figure*}
\centering
\subfigure{\includegraphics[width=0.3\textwidth]{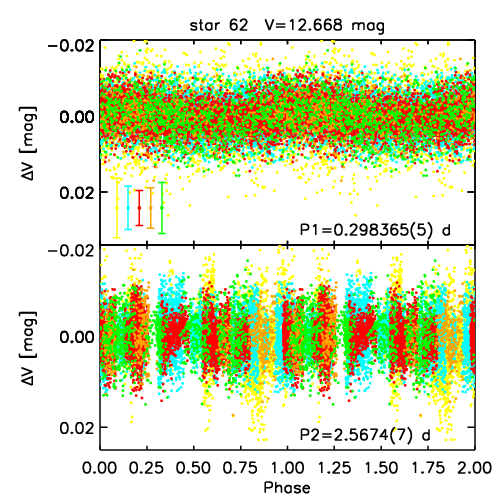}}
\subfigure{\includegraphics[width=0.3\textwidth]{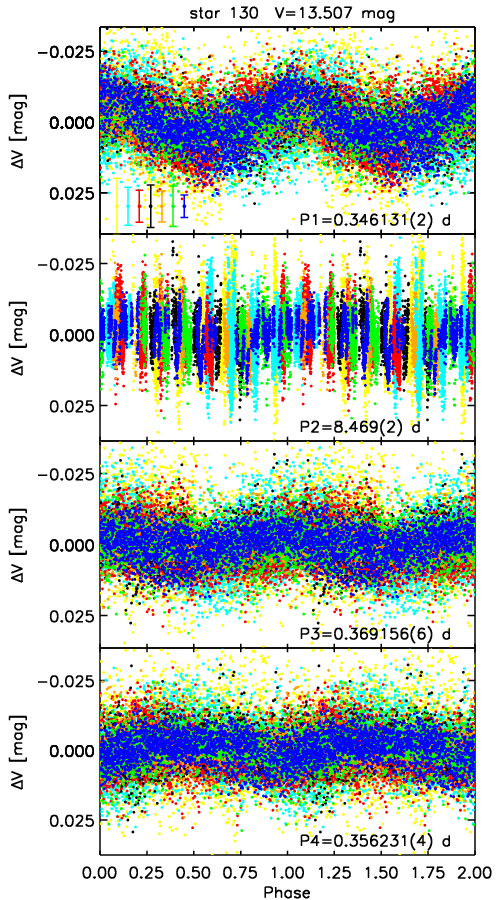}}
\caption{Same as Figure~\ref{fig:Beta_Cep_PD1}, but for the two Herbig Be candidates.}
\label{fig:HerbigAeBe}
\end{figure*}
\clearpage

\begin{figure*}
\centering
\subfigure{\includegraphics[width=0.3\textwidth]{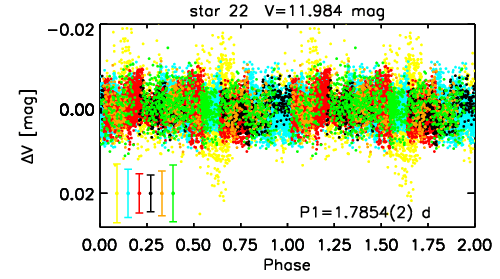}}
\subfigure{\includegraphics[width=0.3\textwidth]{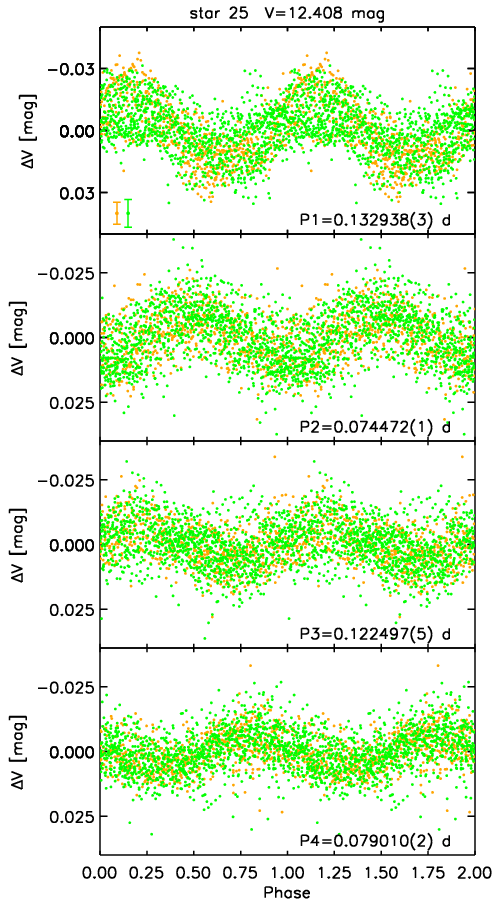}}
\subfigure{\includegraphics[width=0.3\textwidth]{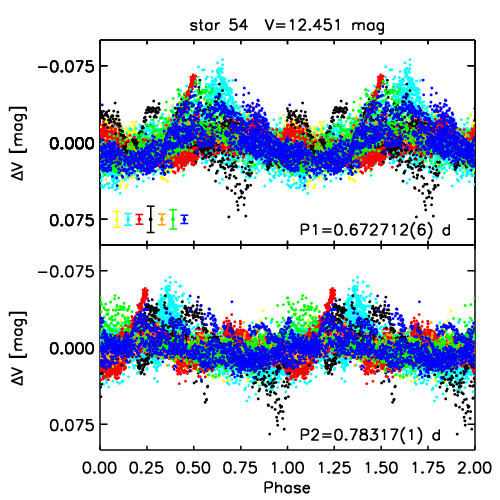}}
\subfigure{\includegraphics[width=0.3\textwidth]{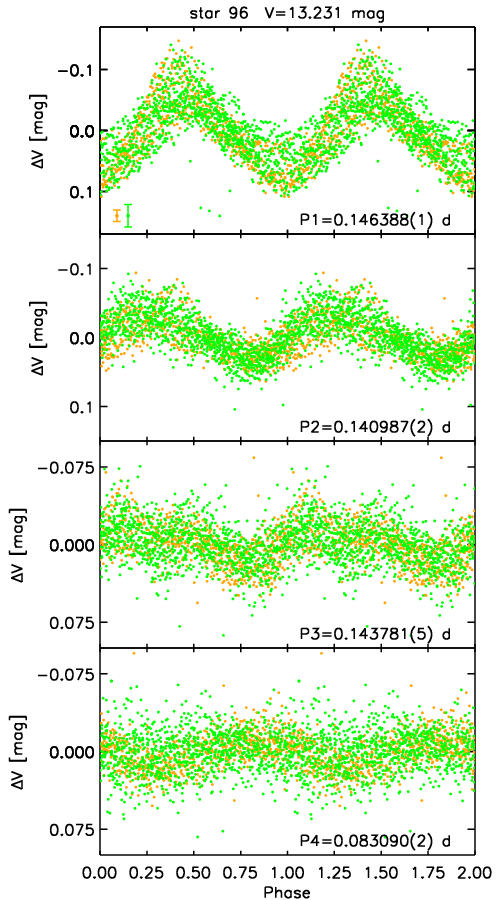}}
\subfigure{\includegraphics[width=0.3\textwidth]{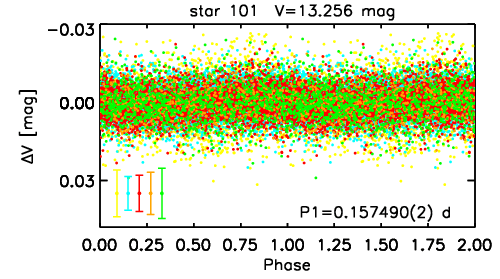}}
\subfigure{\includegraphics[width=0.3\textwidth]{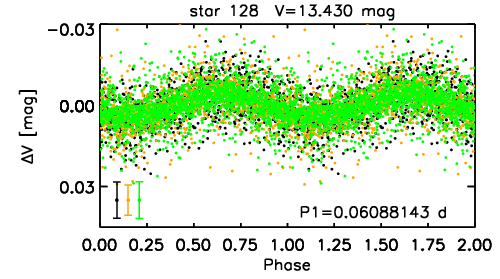}}
\subfigure{\includegraphics[width=0.3\textwidth]{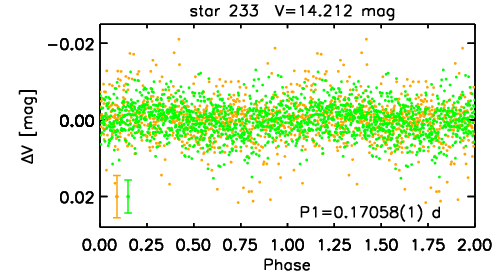}}
\subfigure{\includegraphics[width=0.3\textwidth]{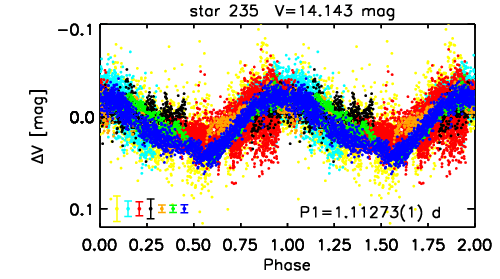}}
\subfigure{\includegraphics[width=0.3\textwidth]{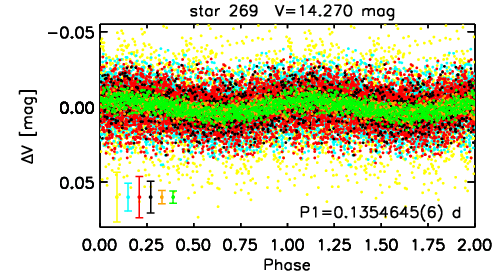}}
\caption{Same as Figure~\ref{fig:Beta_Cep_PD1}, but for the non-member variable stars, including field variable stars and undetermined-membership variable stars.}
\label{fig:non_member_star}
\end{figure*}

\begin{figure*}
\centering
\setcounter{figure}{6}
\subfigure{\includegraphics[width=0.3\textwidth]{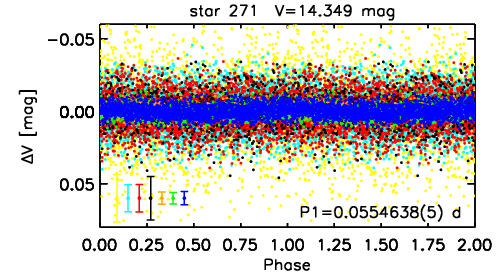}}
\subfigure{\includegraphics[width=0.3\textwidth]{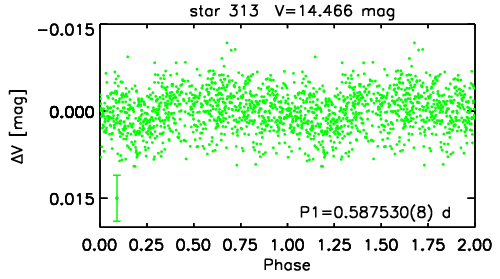}}
\subfigure{\includegraphics[width=0.3\textwidth]{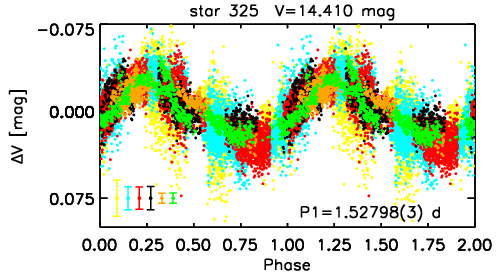}}
\subfigure{\includegraphics[width=0.3\textwidth]{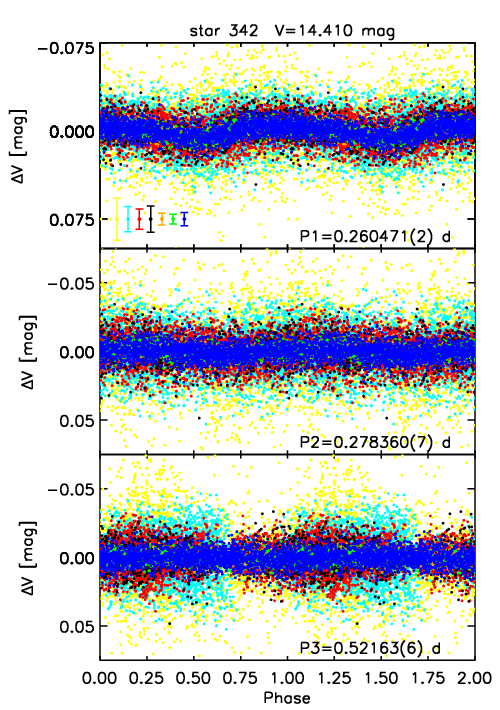}}
\subfigure{\includegraphics[width=0.3\textwidth]{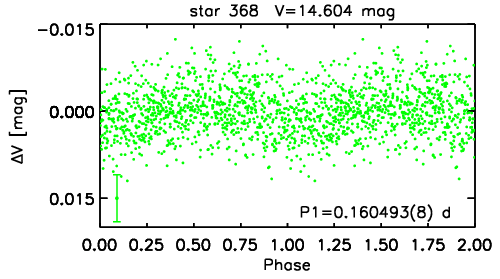}}
\subfigure{\includegraphics[width=0.3\textwidth]{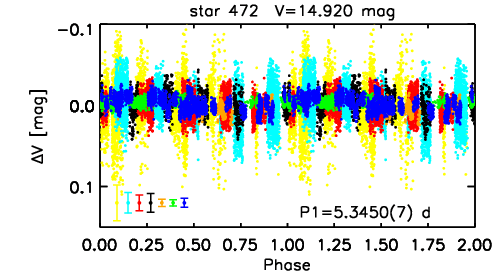}}
\subfigure{\includegraphics[width=0.3\textwidth]{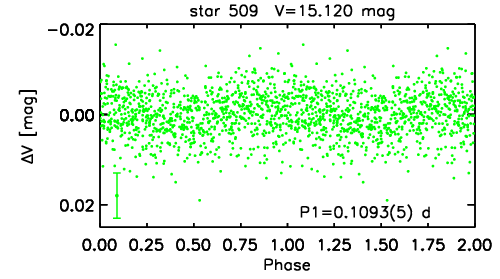}}
\subfigure{\includegraphics[width=0.3\textwidth]{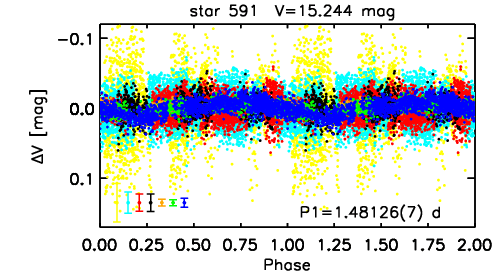}}
\subfigure{\includegraphics[width=0.3\textwidth]{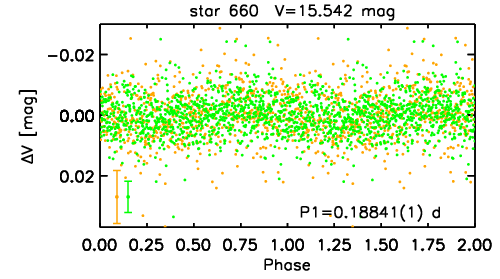}}
\subfigure{\includegraphics[width=0.3\textwidth]{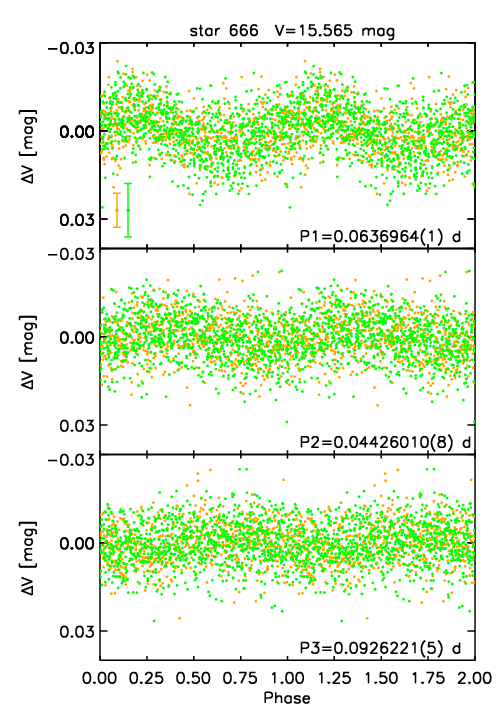}}
\subfigure{\includegraphics[width=0.3\textwidth]{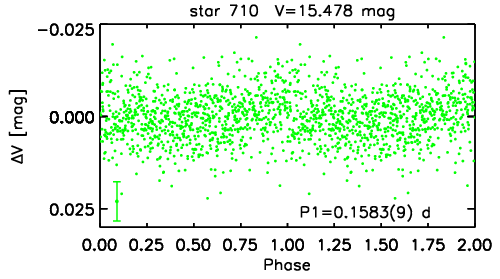}}
\subfigure{\includegraphics[width=0.3\textwidth]{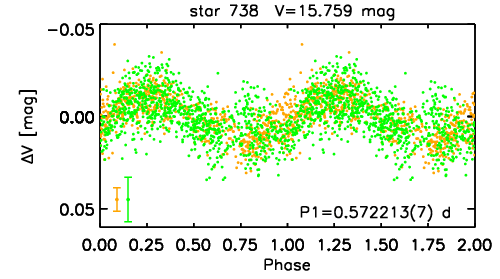}}
\caption{Continued.}
\end{figure*}

\begin{figure*}
\centering
\setcounter{figure}{6}
\subfigure{\includegraphics[width=0.3\textwidth]{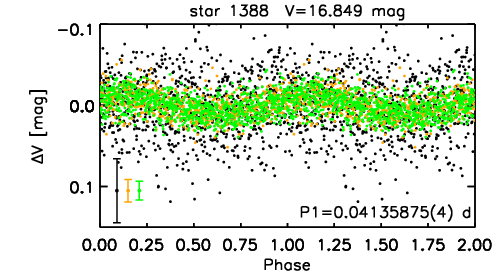}}
\subfigure{\includegraphics[width=0.3\textwidth]{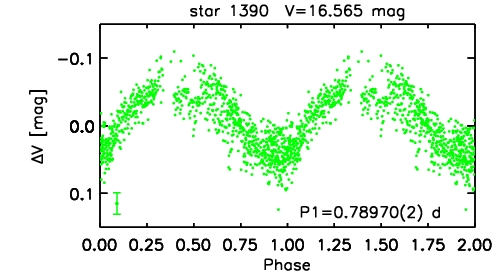}}
\subfigure{\includegraphics[width=0.3\textwidth]{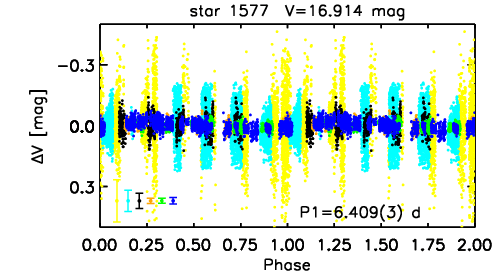}}
\subfigure{\includegraphics[width=0.3\textwidth]{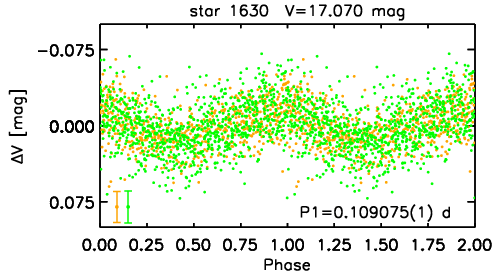}}
\subfigure{\includegraphics[width=0.3\textwidth]{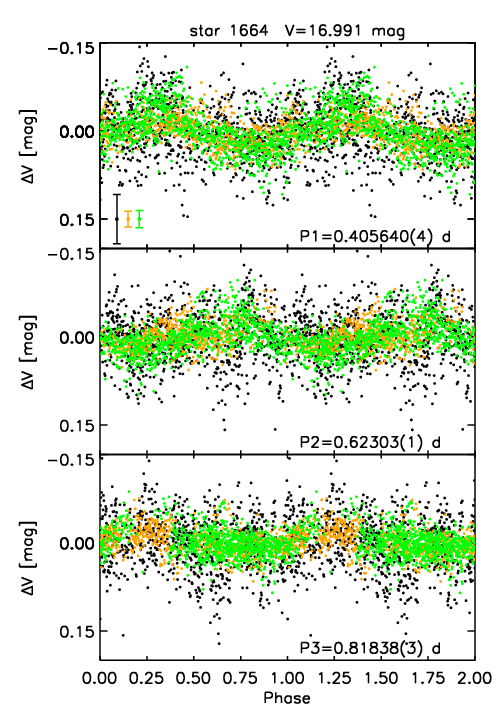}}
\subfigure{\includegraphics[width=0.3\textwidth]{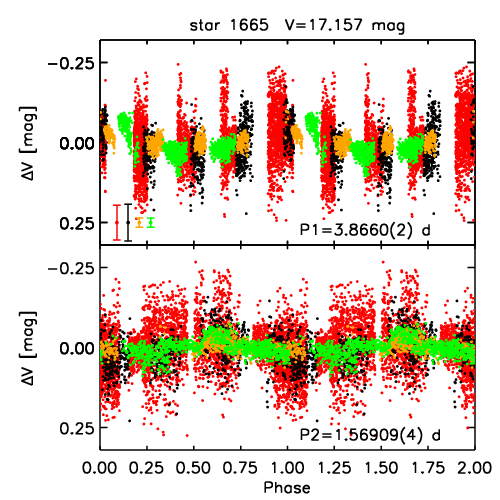}}
\subfigure{\includegraphics[width=0.3\textwidth]{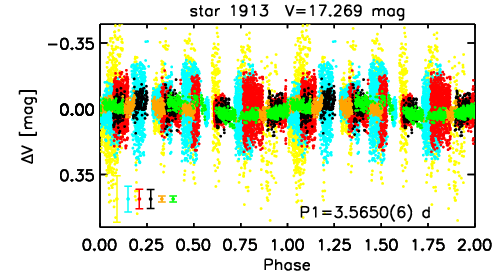}}
\subfigure{\includegraphics[width=0.3\textwidth]{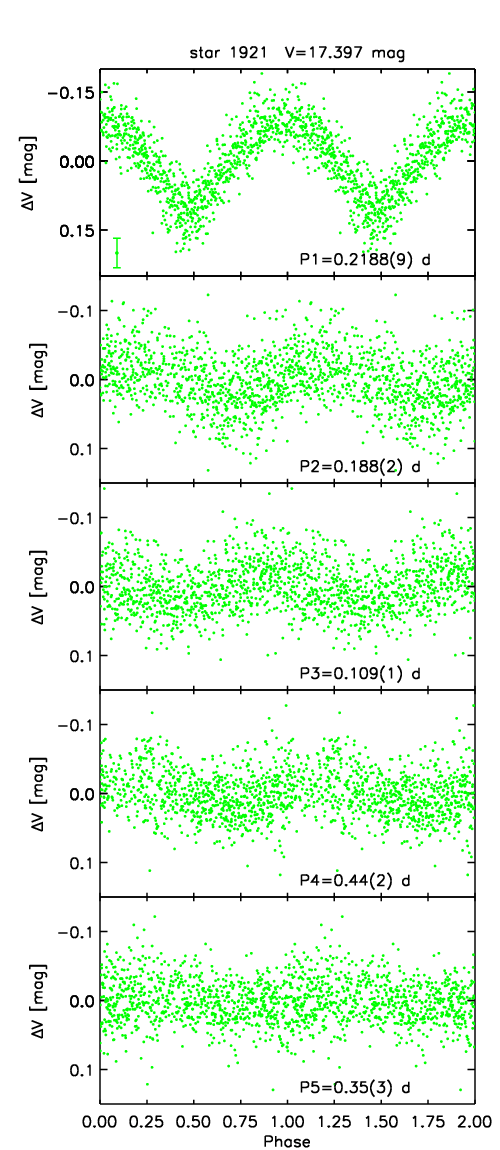}}
\subfigure{\includegraphics[width=0.3\textwidth]{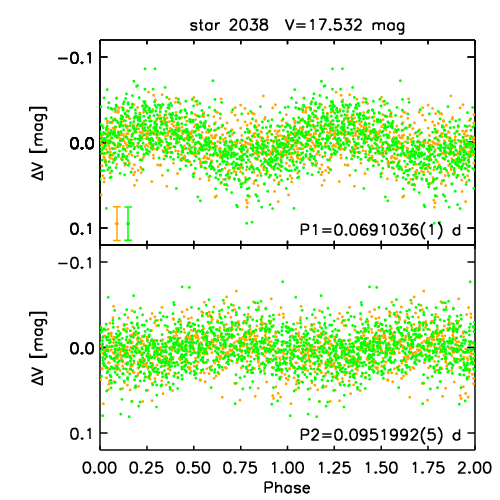}}
\caption{Continued.}
\end{figure*}

\begin{figure*}
\centering
\setcounter{figure}{6}
\subfigure{\includegraphics[width=0.3\textwidth]{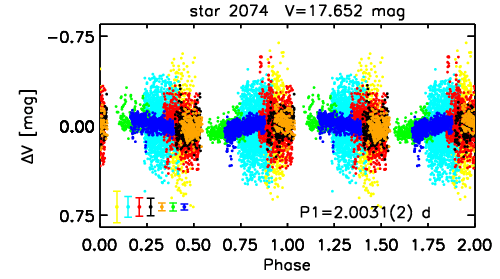}}
\subfigure{\includegraphics[width=0.3\textwidth]{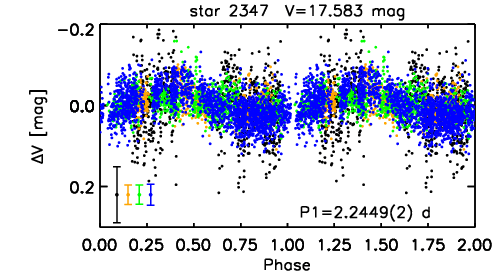}}
\subfigure{\includegraphics[width=0.3\textwidth]{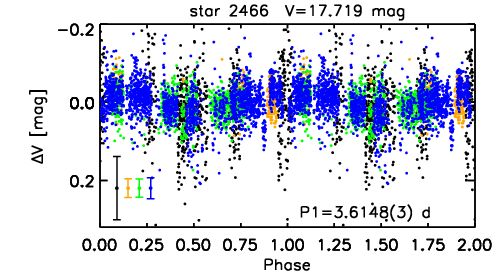}}
\subfigure{\includegraphics[width=0.3\textwidth]{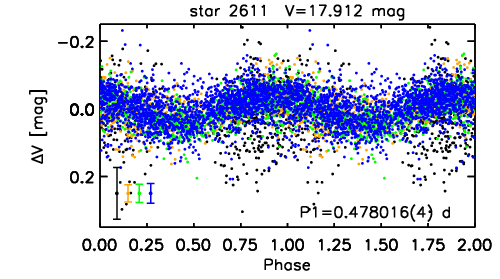}}
\subfigure{\includegraphics[width=0.3\textwidth]{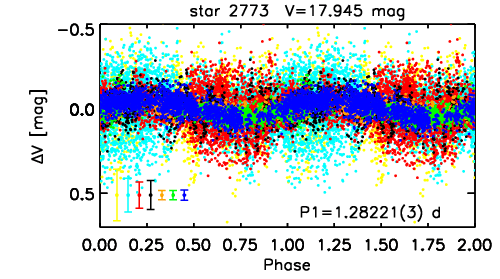}}
\subfigure{\includegraphics[width=0.3\textwidth]{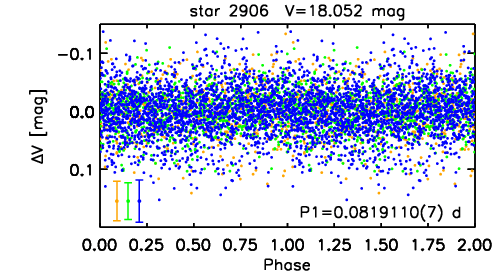}}
\subfigure{\includegraphics[width=0.3\textwidth]{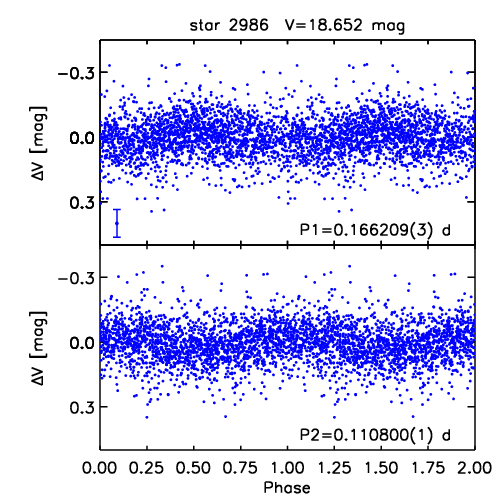}}
\subfigure{\includegraphics[width=0.3\textwidth]{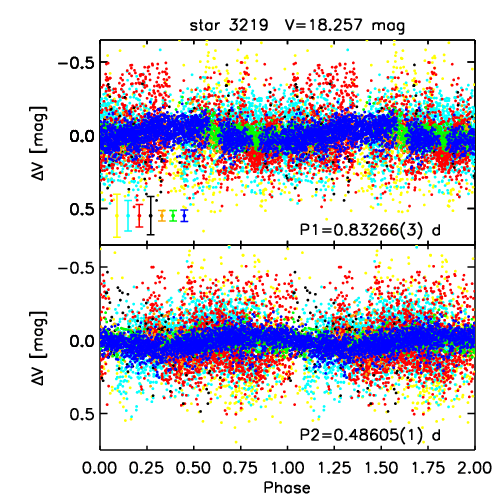}}
\subfigure{\includegraphics[width=0.3\textwidth]{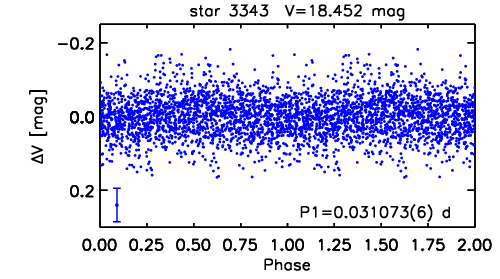}}
\subfigure{\includegraphics[width=0.3\textwidth]{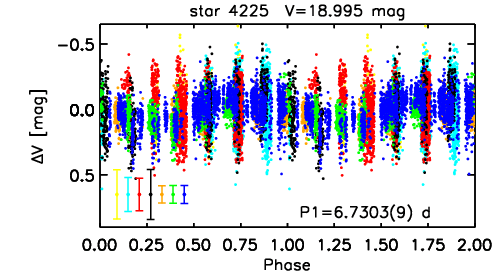}}
\subfigure{\includegraphics[width=0.3\textwidth]{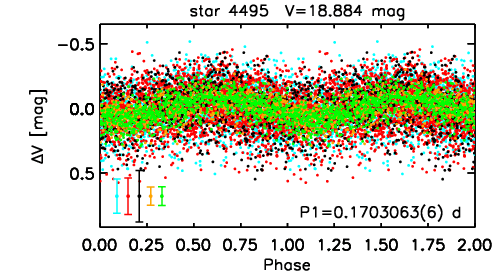}}
\subfigure{\includegraphics[width=0.3\textwidth]{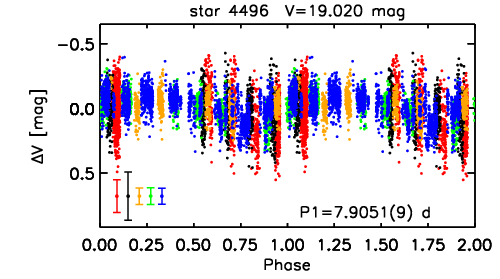}}
\subfigure{\includegraphics[width=0.3\textwidth]{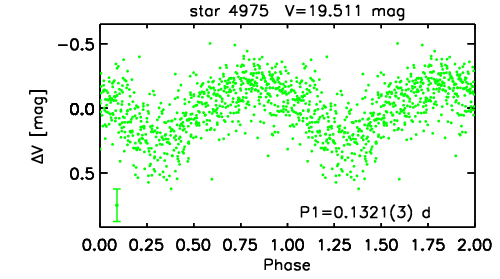}}
\subfigure{\includegraphics[width=0.3\textwidth]{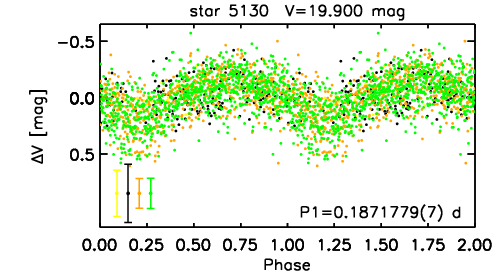}}
\subfigure{\includegraphics[width=0.3\textwidth]{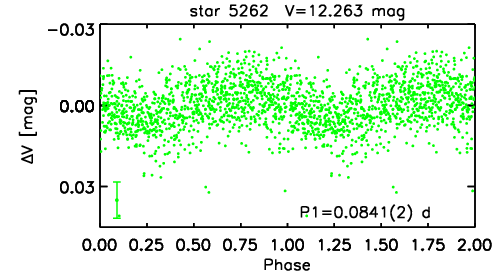}}
\caption{Continued.}
\end{figure*}

\begin{figure*}
\centering
\setcounter{figure}{6}
\subfigure{\includegraphics[width=0.3\textwidth]{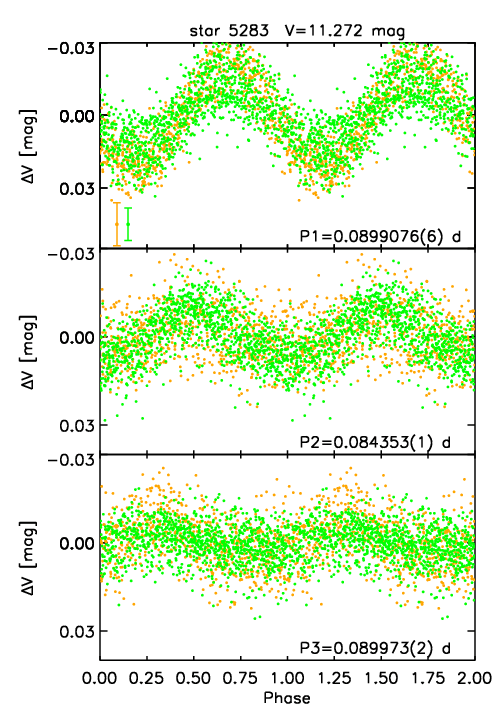}}
\subfigure{\includegraphics[width=0.3\textwidth]{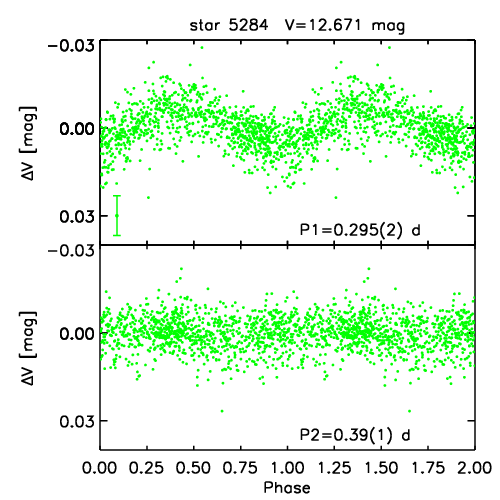}}
\subfigure{\includegraphics[width=0.3\textwidth]{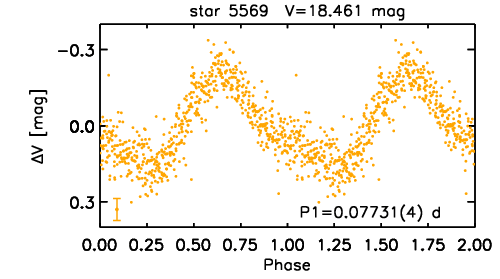}}
\caption{Continued.}
\end{figure*}

\onecolumn
\section{Eclipsing binaries}
\clearpage

\begin{table*}
\caption{\label{tab:cat_binary} Eclipsing binaries in the field of view of NGC~1893. The interpretation of L12 and pp are the same as in Table~\ref{tab:cat_periodic}.}
\centering
\begin{tabular}{lcccc}
\hline\hline
Star ID & P(d)  & Membership & Classification & Discovery\\
\hline
341     & 1.771411   &      yes  &      EB & pp\\
706     & 0.443540   &      yes  &      EB & pp\\
962     & 0.3061276  &      no   &      EW & L12\\
1208    & 0.678215   &      yes  &      EA & L12\\
1277    & 0.302597   &      yes  &      EW & L12\\
1287    & 2.330385   &      yes  &      EA & pp\\
1376    & $-$        &      no   &      EB & pp\\
1893    & 1.162861   &      no   &      EA & pp\\
1908    & 0.335324   &      no   &      EW & pp\\
2091    & 3.550000   &      yes  &      EB & pp\\
2793    & 0.255430   &      no   &      EW & pp\\
3215    & $-$        &      yes  &      EA & pp\\
4082    & 0.272121   &      no   &      EW & pp\\
5048    & 0.548099   &      no   &      EA & pp\\
5282    & 1.558758   &      no   &      EA & pp\\
\hline
\end{tabular}
\end{table*}

\begin{figure*}
\centering
\subfigure{\includegraphics[width=0.3\textwidth]{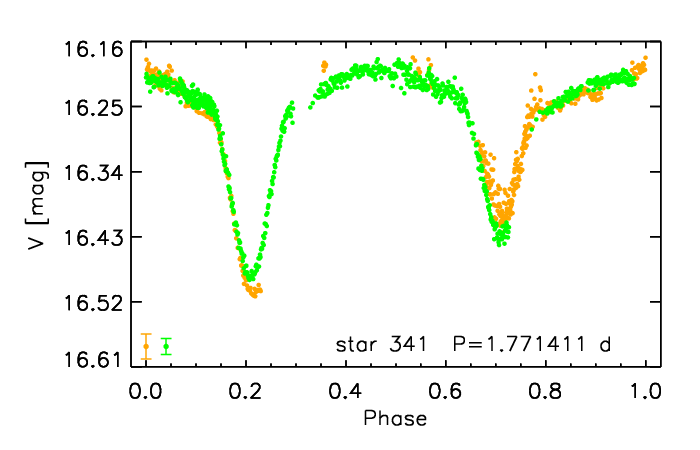}}
\subfigure{\includegraphics[width=0.3\textwidth]{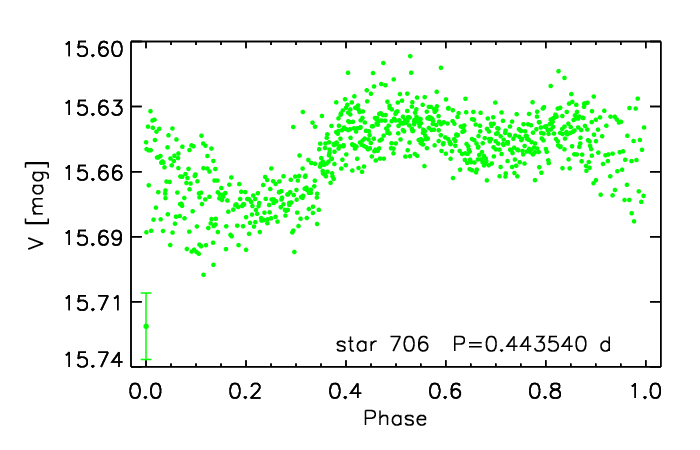}}
\subfigure{\includegraphics[width=0.3\textwidth]{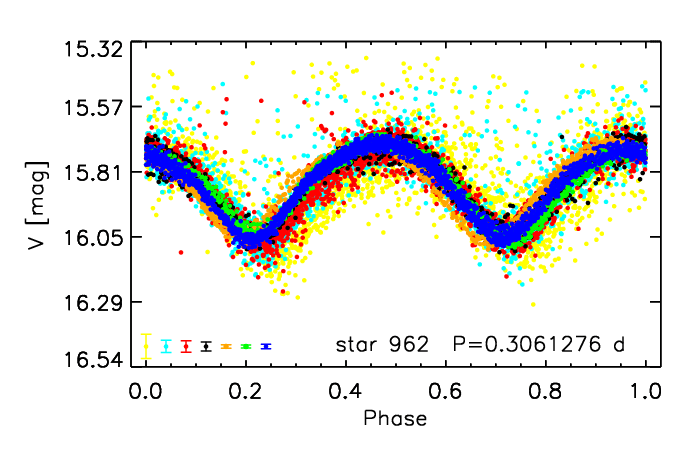}}
\subfigure{\includegraphics[width=0.3\textwidth]{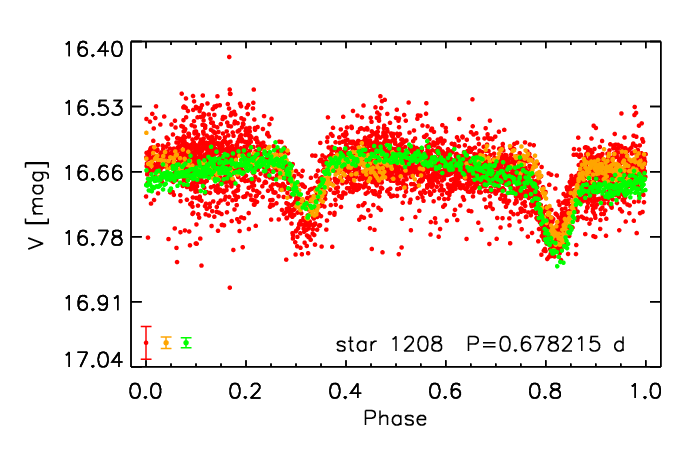}}
\subfigure{\includegraphics[width=0.3\textwidth]{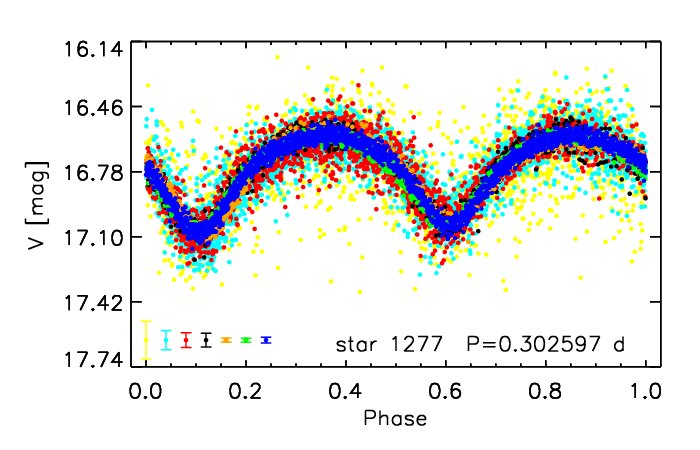}}
\subfigure{\includegraphics[width=0.3\textwidth]{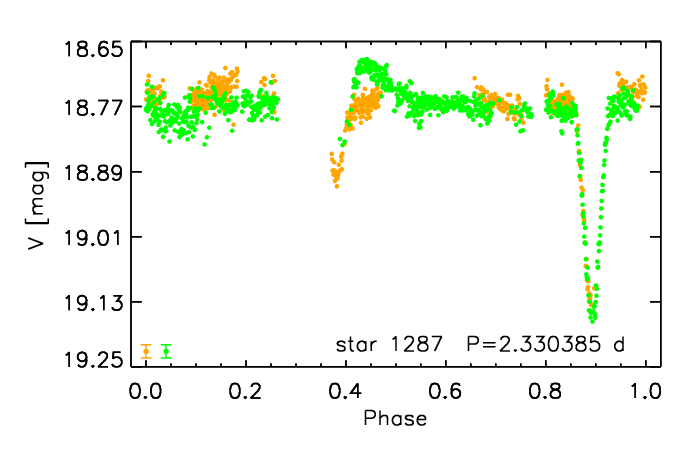}}
\subfigure{\includegraphics[width=0.3\textwidth]{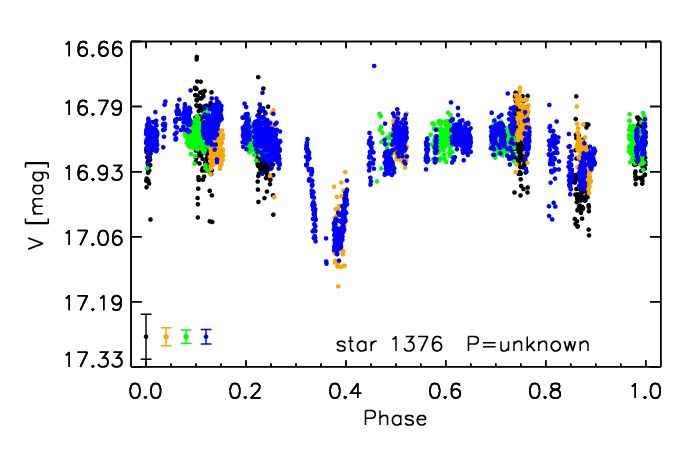}}
\subfigure{\includegraphics[width=0.3\textwidth]{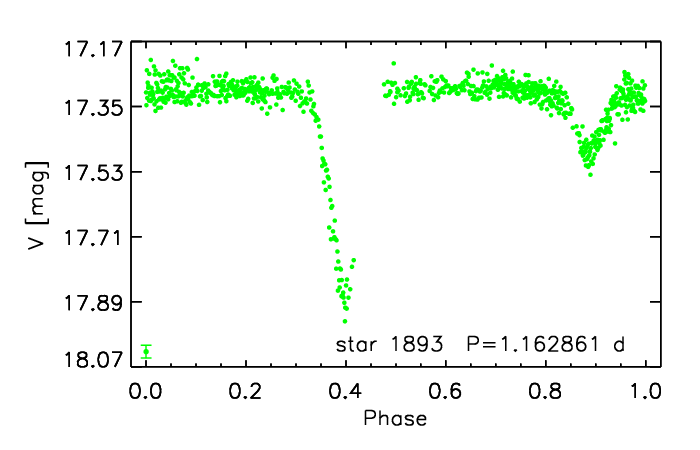}}
\subfigure{\includegraphics[width=0.3\textwidth]{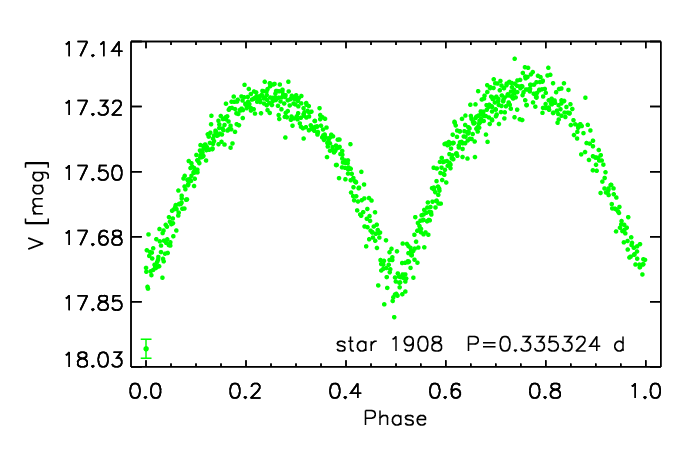}}
\subfigure{\includegraphics[width=0.3\textwidth]{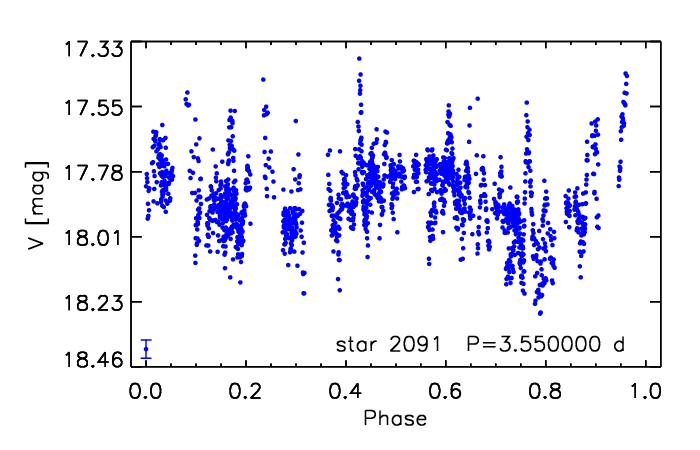}}
\subfigure{\includegraphics[width=0.3\textwidth]{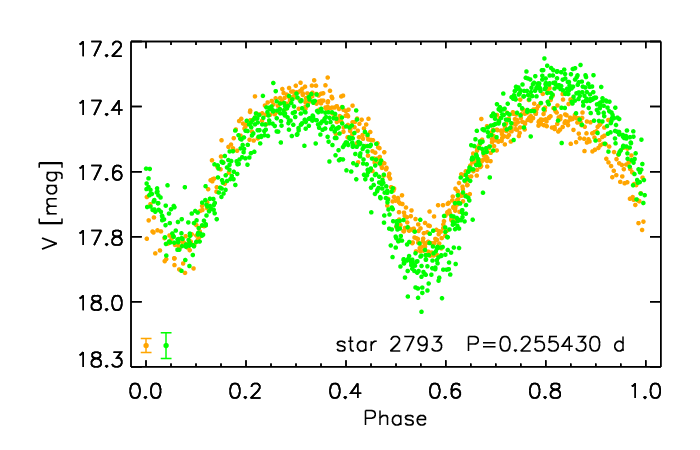}}
\subfigure{\includegraphics[width=0.3\textwidth]{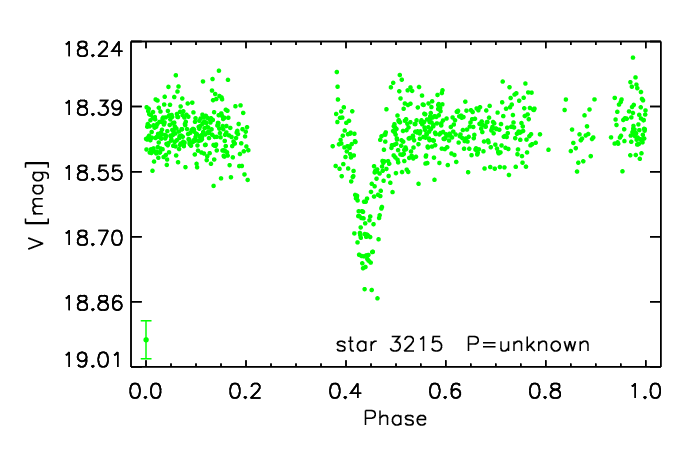}}
\subfigure{\includegraphics[width=0.3\textwidth]{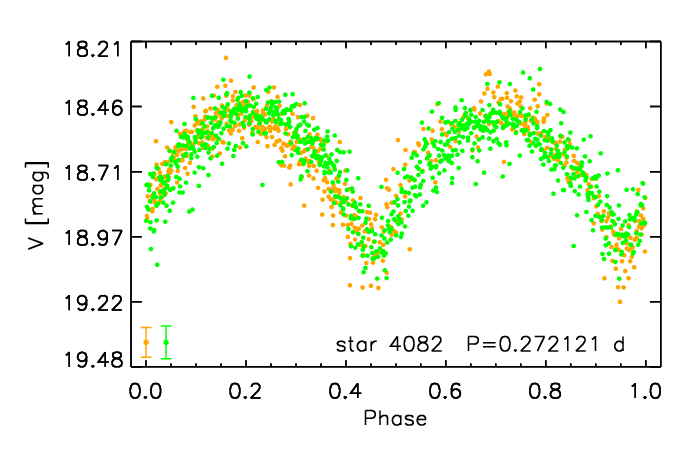}}
\subfigure{\includegraphics[width=0.3\textwidth]{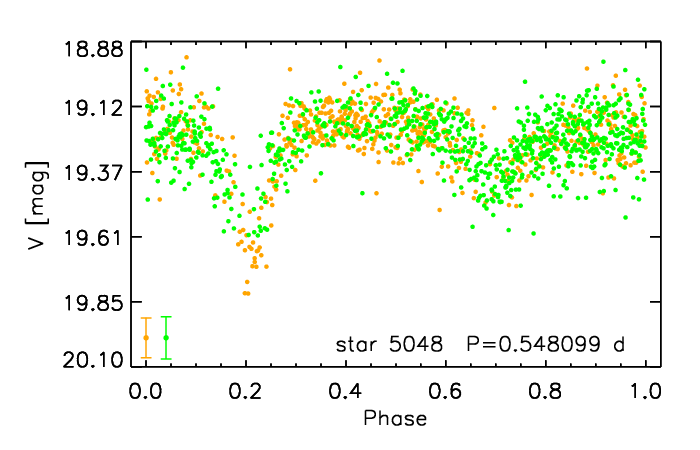}}
\subfigure{\includegraphics[width=0.3\textwidth]{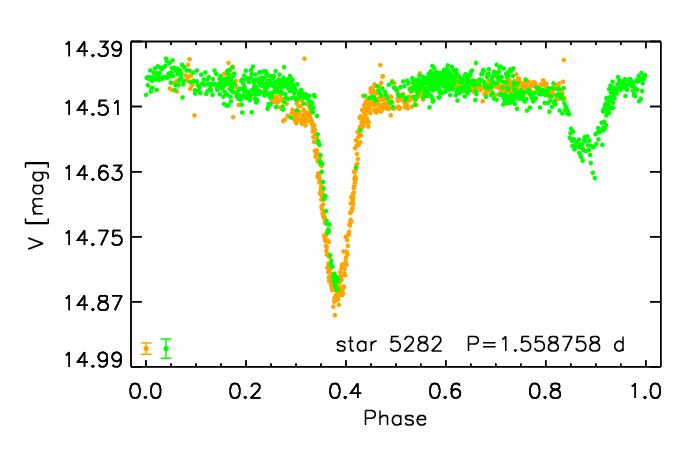}}
\caption{Folded light curves of the 15 eclipsing binaries listed in Tabel~\ref{tab:cat_binary}.
         The observations of groups 1, 2, 3, 4, 5, 6 and 7 (the group number refers to the observation run IDs listed in Table~\ref{tab:obs_log1} of the main body of the text) are respectively represented in blue, black, red, cyan, yellow, green and orange.
         The error bars represent the mean errors of the measurements.}
\label{binary_PD}
\end{figure*}
\clearpage

\onecolumn
\section{Non-period variable stars}
\clearpage

\begin{table*}
\caption{\label{tab:cat_unknown_P} Non-period variable stars in the field of view of NGC~1893. The interpretation of Z08 and L12 and pp is the same as in Table~\ref{tab:cat_periodic}.}
\centering
\begin{tabular}{lccc}
\hline\hline
Star ID  &  Membership  & Classification & Discovery\\
\hline
197    &   yes  &     MS/unknown-type               & pp \\
218    &   no   &     field/irregular                   & Z08\\
441    &   yes  &     PMS/unknown-type              & pp \\
528    &   $?$  &     unknown-type                & L12\\ 
844    &   no   &     field/unknown-type                & pp \\
922    &   no   &     field/irregular                   & L12\\
926    &   no   &     field/irregular                   & pp \\
1069   &   yes  &     PMS/unknown-type              & pp \\
1098   &   yes  &     PMS/unknown-type              & pp \\
1301   &   yes  &     PMS/unknown-type              & pp \\
1657   &   no   &     field/unknown-type                & pp \\
1961   &   yes  &     PMS/unknown-type              & L12\\
2640   &   yes  &     PMS/unknown-type              & L12\\
2697   &   yes  &     PMS/irregular                 & L12\\
2777   &    no  &     field/unknown-type              & L12\\  
2882   &   yes  &     PMS/irregular                 & L12\\
2936   &   yes  &     PMS/irregular                 & L12\\
4056   &   yes  &     PMS/unknown-type              & L12\\
4162   &   yes  &     PMS/irregular                 & L12\\
4671   &   yes  &     PMS/irregular                 & L12\\
4716   &   yes  &     PMS/unknown-type              & L12\\
5158   &   yes  &     PMS/irregular                 & pp \\
\hline
\end{tabular}
\end{table*}

\begin{figure*}
\centering
\subfigure{\includegraphics[width=0.9\textwidth]{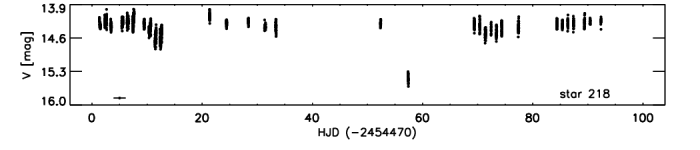}}
\subfigure{\includegraphics[width=0.9\textwidth]{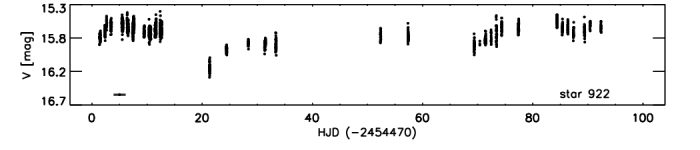}}
\subfigure{\includegraphics[width=0.9\textwidth]{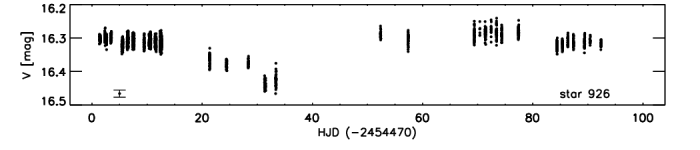}}
\subfigure{\includegraphics[width=0.9\textwidth]{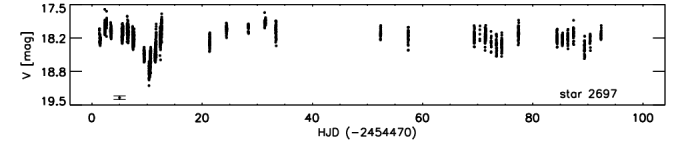}}
\subfigure{\includegraphics[width=0.9\textwidth]{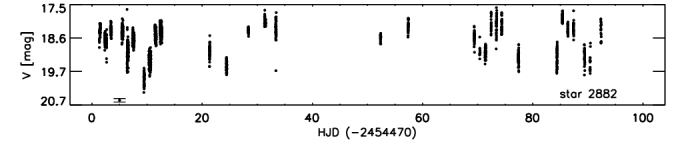}}
\subfigure{\includegraphics[width=0.9\textwidth]{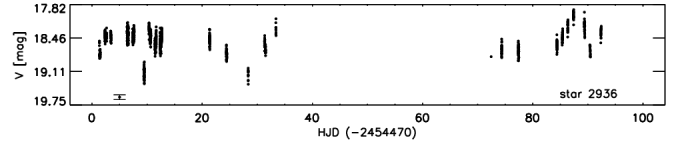}}
\caption{Light curves of the irregular variable stars in the field of view of NGC~1893 taken with Mercator in 2008. The mean error of the measurements is marked in the left bottom of each subfigure.}
\label{fig:unknown_lc_1}
\end{figure*}

\begin{figure*}
\centering
\setcounter{figure}{0}
\subfigure{\includegraphics[width=0.9\textwidth]{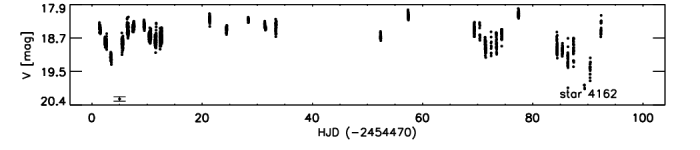}}
\subfigure{\includegraphics[width=0.9\textwidth]{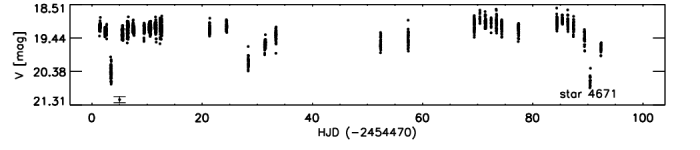}}
\subfigure{\includegraphics[width=0.9\textwidth]{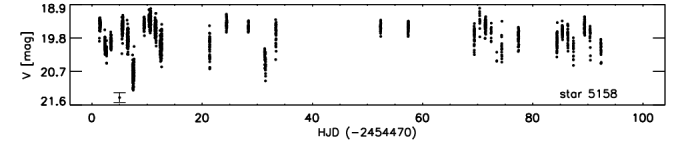}}
\caption{Continued.}
\end{figure*}

\begin{figure*}
\centering
\subfigure{\includegraphics[width=0.9\textwidth]{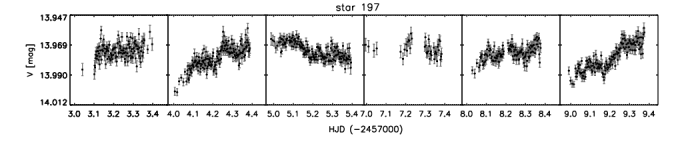}}
\subfigure{\includegraphics[width=0.9\textwidth]{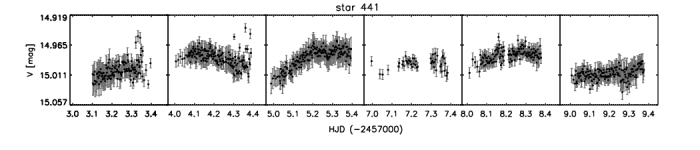}}
\caption{Light curves of variable stars for which a period could not be determined. The data were taken with the Xinglong 85 cm telescope in 2014.}
\label{fig:unknown_lc_2}
\end{figure*}

\begin{figure*}
\centering
\setcounter{figure}{1}
\subfigure{\includegraphics[width=0.9\textwidth]{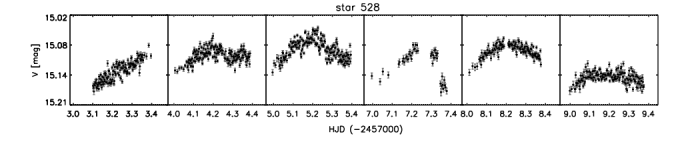}}
\subfigure{\includegraphics[width=0.9\textwidth]{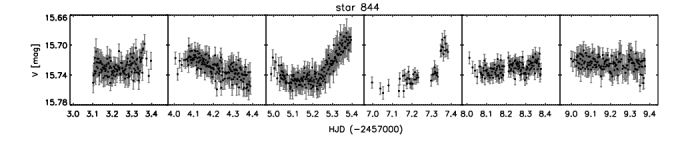}}
\subfigure{\includegraphics[width=0.9\textwidth]{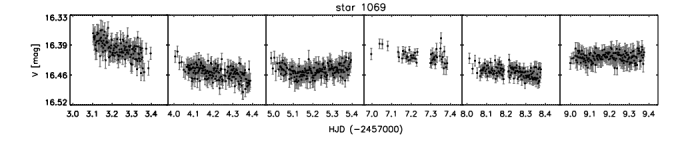}}
\subfigure{\includegraphics[width=0.9\textwidth]{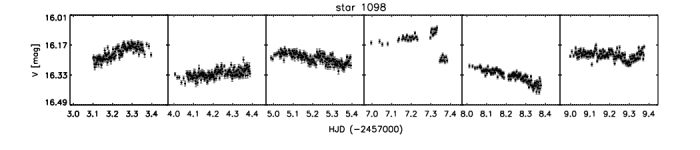}}
\subfigure{\includegraphics[width=0.9\textwidth]{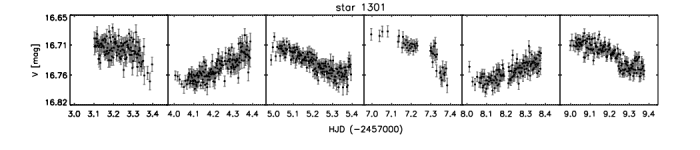}}
\subfigure{\includegraphics[width=0.9\textwidth]{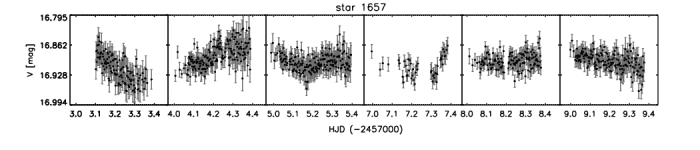}}
\caption{Continued.}
\end{figure*}

\begin{figure*}
\centering
\setcounter{figure}{1}
\subfigure{\includegraphics[width=0.9\textwidth]{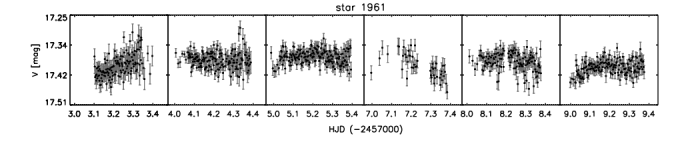}}
\subfigure{\includegraphics[width=0.9\textwidth]{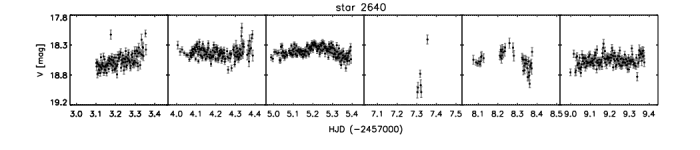}}
\subfigure{\includegraphics[width=0.9\textwidth]{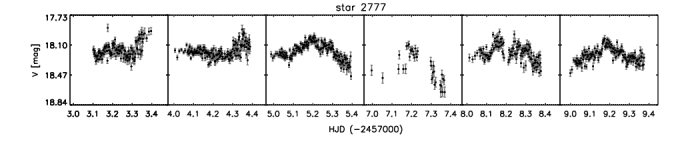}}
\subfigure{\includegraphics[width=0.9\textwidth]{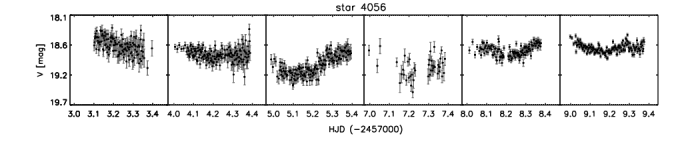}}
\subfigure{\includegraphics[width=0.9\textwidth]{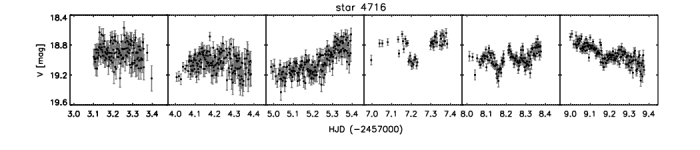}}
\caption{Continued.}
\end{figure*}

\bsp    
\label{lastpage}
\end{document}